\documentclass[reprint, amsmath, amssymb, aps]{revtex4-1}
\usepackage{amsmath}
\usepackage{amsfonts}
\usepackage{graphicx}
\usepackage{appendix}
\usepackage{dsfont}
\usepackage{amssymb}
\usepackage{bm}
\usepackage{color}
\usepackage{ulem}
\usepackage{soul}
\usepackage{xcolor}
\usepackage{natbib}
\usepackage{nccmath}
\usepackage{array}
\newcommand{\beq}{\begin{equation}}
\newcommand{\eneq}{\end{equation}}
\newcommand{\be}{\begin{equation}}
\newcommand{\ee}{\end{equation}}
\newcommand{\bea}{\begin{eqnarray}}
\newcommand{\eea}{\end{eqnarray}}

\usepackage{multirow}
\usepackage{wasysym}

\begin{document}
\title{Spin-Hall Current and Nonlocal Transport in Ferromagnet-Free Multi-band models for ${\rm Sr Ti O}_3$-Based 
Nanodevices in the presence of impurities} 
\author{Domenico Giuliano$^{(1,2)}$,  Andrea Nava$^{(1,2,3)}$,
Carmine Antonio Perroni$^{(4,5,6)}$, Manuel Bibes$^{(7)}$, 
Felix Trier$^{(8)}$,  and Marco Salluzzo$^{(5)}$  }

\affiliation{
$^{(1)}$ Dipartimento di Fisica, Universit\`a della Calabria Arcavacata di 
Rende I-87036, Cosenza, Italy \\
$^{(2)}$ I.N.F.N., Gruppo collegato di Cosenza, 
Arcavacata di Rende I-87036, Cosenza, Italy\\
$^{(3)}$Institut f\"ur Theoretische Physik IV - Heinrich-Heine Universit\"at, D-40225  D\"usseldorf, Germany\\
$^{(4)}$ Dipartimento di Fisica ``E. Pancini'', Complesso Universitario Monte S. Angelo, Via Cintia, I-80126 Napoli, Italy \\
$^{(5)}$ CNR-SPIN, Complesso Universitario Monte S. Angelo, Via Cintia, I-80126 Napoli, Italy \\
$^{(6)}$ I.N.F.N., Sezione di Napoli, Complesso Universitario Monte S. Angelo, Via Cintia, I-80126 Napoli, Italy \\
$^{(7)}$ Unit\'e  Mixte de Physique, CNRS, Thales, Universit\'e Paris-Saclay, 91767, Palaiseau, France\\
$^{(8)}$ Department of Energy Conversion and Storage, Technical University of Denmark, 2800 Kongens Lyngby, Denmark
}

\date{\today}

\begin{abstract}

We compute  the spin-Hall conductance in a multiband model describing the two-dimensional electron gas formed at a 
LaAlO$_3$/SrTiO$_3$ interface 
in the presence of a finite concentration of impurities. Combining linear response theory with a systematic calculation of 
the impurity contributions to the self-energy, 
as well as to the vertex corrections of the relevant diagrams, we  recover the full spin-Hall vs. sheet conductance dependence of LaAlO$_3$/SrTiO$_3$ as reported  in Trier {\it et al.}  [Nano Lett. {\bf 20}, 395 (2020)], 
finding a  very good agreement with the experimental data below and above the Lifshitz transition. In particular, 
we demonstrate that the multiband electronic structure leads to only a partial, instead of a complete, 
screening of the spin-Hall conductance, which decreases with increasing the carrier density.
Our method can be generalized to other  two-dimensional systems characterized by a broken inversion
 symmetry and multiband physics. 

\end{abstract}

\maketitle

\section{Introduction}
\label{intro}

Recently, spintronics, a branch of electronics based on the manipulation of electron spin, rather than the charge, is emerging as a promising
 technology for information storing and processing, and for sensing  \cite{Wolf2001,Vaz2019,Hirohata2020}. Spin-injection in spintronic devices
  can be done by using ferromagnetic leads, however the efficiency of this process is usually low 
 \cite{Chauleau2016,Qi2017,Telesio2018}.  An alternative approach is to use semiconducting materials characterized by large charge to spin
  conversion efficiency. Here, a charge current can be converted into a spin current due to the Edelstein  
  \cite{Edelstein1990} and/or  
 the spin-Hall (SH) effects \cite{Dyakonov1971,Sinova2015}. This is a particularly efficient process, for example, in some two-dimensional systems 
 \cite{Qian2014,Nie2015,Safeer2019,Benitez2020,Safeer2020}, and  topological insulators   (see, for instance,
Ref. \cite{Farzaneh2020} and references therein),
 characterized by breaking of the inversion symmetry due to a  Rashba-type spin orbit interaction (SOI), 
 even in the absence of ferromagnetic leads \cite{Trier2020}.   

The Edelstein effect consists of  a spin-accumulation induced by an injected electrical current due to the presence of 
Rashba-split Fermi surfaces. The excess spin density diffuses across the system, thus giving rise to a 
net spin-polarized spin current perpendicular to the charge current, as the spin and the momentum of the carrier get `` locked''. 
At variance, the SH effect can either be extrinsic, i.e., related to the impurity scattering in the material, or intrinsic, due to the SOI related 
to  the direct coupling (in the response function) between the electric and the spin currents  \cite{Sinova2004}.

An ideal platform to realize a versatile and tunable Rashba SOI in realistic devices is provided by the two-dimensional electron gas (2DEG) that emerges at the surface of SrTiO$_3$ (STO) and at the interface between STO and large gap band insulating oxides, like LaAlO$_3$ (LAO)\cite{King2014,Ohtomo2004,Chen2011,Rodel2016}. Among other remarkable properties, the 2DEG possesses a strong tunability of the carrier density by gate voltages, which allows for a tuning of  a Rashba-type SOI by electric field effect \cite{Caviglia2008,Caviglia2010,Trama1,Lesne2016}. 

\begin{figure}
 \center
\includegraphics*[width=.9 \linewidth]{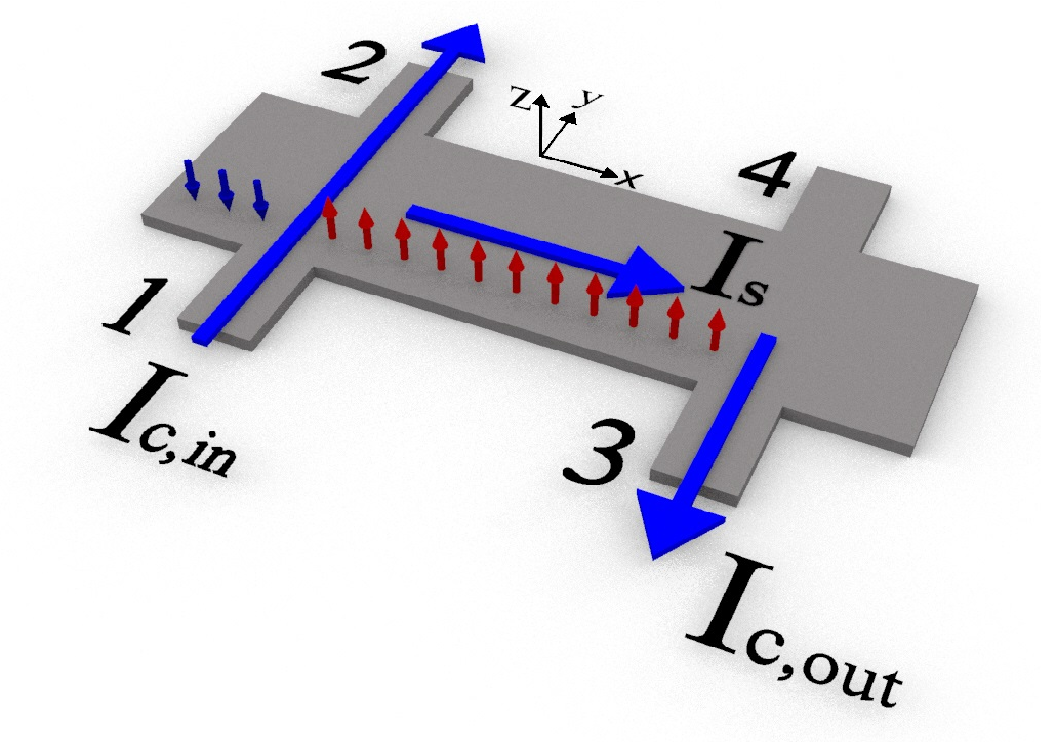}
\caption{Sketch of the experimental 
setup of \cite{Trier2020}. The  charge current $I_{\rm c,in}$   injected  at   contacts 1 and 2 
generates a net spin-current $I_{\rm s}$ diffusing in the bridge along the $x$-axis, which by a second spin-charge conversion at 
 contacts  3 and 4  turns again into the charge current $I_{\rm c,out}$,
} 
\label{fig_device}
\end{figure}

SrTiO$_3$-based nanodevices are very promising for spintronics, as they are characterized by one of the 
largest spin-charge conversion efficiencies among all materials, as shown in Refs. \cite{Varignon2018,Vaz2019}. 
   In Fig. \ref{fig_device} we sketch the experimental 
setup of Ref. \cite{Trier2020}. Here, the  charge current $I_{\rm c,in}$   injected (along the $y$ axis of the figure) at   contacts 1 and 2 
generates a net spin-current $I_{\rm s}$ diffusing in the bridge along the $x$ axis.  A second spin-charge conversion at 
nearby contacts (3 and 4) turns $I_{\rm s}$ back  into the charge current $I_{\rm c,out}$, thus giving rise to a nonlocal voltage drop and 
to a nonlocal resistance $R_{\rm NL}$. 

 $R_{\rm NL}$ is related to the spin diffusion length and to the spin-Hall angle $\gamma$ (note that $R_{\rm NL}$ is expected to be $\propto \gamma^2$ \: \cite{Abanin2009}), which, through a fitting, directly allows a measurements of the efficiency of the charge-to-spin conversion. In Ref. \cite{Trier2020} a spin-diffusion length up to 900 nm was estimated in the LAO/STO 2DEG. Moreover,  a strong, 
non-monotonic, dependence of $\gamma$ as a function of the gate voltage $V_g$ was reported. 

The spin-polarization direction, determined by measuring $R_{\rm NL}$ as a function of the magnetic field intensity and direction, is mainly 
along the out-of-plane $z$ direction (Fig. \ref{fig_device}), 
consistent with the SH effect as the main mechanism. The resulting spin-Hall vs. longitudinal conductance was then compared to calculations assuming a multiband tight-binding model in the absence of disorder. While the overall non-monotonic dependence is captured by a clean multiband mode
 \cite{Trier2020}, discrepancies can be noticed at low values of the chemical potential and at the Lifshitz transition, calling for further studies, in 
 particular considering the role of realistic impurity scattering induced by nonmagnetic disorder.

\noindent

In this paper we develop a general approach to the spin-Hall conductance (SHC) 
in a multiband model with a SOI in the form characteristic of the 2DEG at a LAO/STO interface. In particular, 
we employ linear response theory (LRT) to compute the  SHC  by accounting for 
elastic scattering effects of  nonmagnetic impurities in the 2DEG (which seems   appropriate at low density of carriers \cite{Trier2020}).
In addition,  we compute the sheet conductance and the SHC  in the multiband system with the band structure and the various parameters
 as in Ref. \cite{Trier2020}.

Along our derivation, we systematically take into account  the complex pattern of SOI in LAO/STO, both of atomic nature, 
as well as Rashba-type, due to the combined effect of the splitting of the Ti-$3d$ bands   (which support the conduction in the  
2DEG), determined by   the lateral confinement  \cite{Vaz2019}, and 
to the  inversion-symmetry breaking.
 
Throughout our analysis of the SH effect we show that, while any finite concentration of elastic impurity scatterers would fully screen to 
zero the SHC in a single-band Rashba 2DEG
 \cite{Mishchenko2004,Inoue2004,Khaetskii2006,Raimondi2004,Dimitrova2005},  the multiband structure of the model, 
combined with the intraband and the inter-band SOI, leads to only a partial screening of the SHC, which shows a non-monotonic carrier density dependence consistent with the experimental results of  Ref. \cite{Trier2020}. 
  
Our results for the sheet conductance show an excellent agreement with the experimental data, both from the qualitative,
 as well as from the quantitative point of view. When low-lying bands only are involved, our results for the SHC exhibit again an excellent quantitative (and qualitative) agreement with the experimental data.   The model captures also the presence of a peak in the SHC when 
 higher-energy bands come into play through the Lifshitz transition in the 2DEG, however, in this region the agreement is apparently less good, 
 which calls for a deeper discussion of  how  the SHC should behave across the Lifshitz transition. Eventually, the agreement is recovered at a good   level for larger values of $V_g$. 
   
 Our method allows us to   derive in detail   the impurity-induced vertex corrections to the SHC 
 in a multiband system such as the 2DEG at the LAO/STO: it 
is a substantial extension of single-band  models available to calculate the SHC in two dimensional systems, and can be easily generalized to any multi-band system.
 Here we focus on the  ``eight-band'' model to recover the results of Ref. \cite{Trier2020},   
but we show how it can be   generalized to other systems, such as a Rashba, single-band 2DEG, which we discuss in the Appendixes of 
our paper, or the ``six-band'' model, a simplified version of the eight-band one discussed, for instance, in Ref. \cite{Perroni2019}.  

In presenting our method and its applications, we organize  our paper as follows: 

\begin{itemize} 

\item In Sec. \ref{mbmodel} we present our method and how to apply it to a generic multiband,
tight-binding model in the presence of a finite density of impurities.

\item In Sec. \ref{sh8} we introduce and discuss in detail 
the 8-band Hamiltonian describing the LAO/STO interface.

\item In Sec. \ref{shsc} we compute the sheet and the spin-Hall conductance of 
the 8-band model in the presence of a finite density of impurities. Eventually, we 
discuss our results in relation to the experimental data of Ref. \cite{Trier2020}.

\item In Sec. \ref{concl}  we provide our conclusions and present some possible 
further developments of our work.

In the various appendices we provide several technical details of our derivation. In particular:

\item In Appendix \ref{lrt} we review the Kubo formulas for the response functions that we 
use to describe charge and/or spin transport.

\item In Appendix \ref{impures} we present our approach to describing the 
effects of a finite density of impurity scattering centers on the various response functions.

\item In Appendix \ref{lat2Rashba}
 we review the paradigmatic calculation of the  spin-Hall conductance in a lattice
model for the two-dimensional Rashba Hamiltonian.

\item In Appendix \ref{efra} we derive the effective, low-energy description of the 8-band model Hamiltonian 
in terms of a two-subband Rashba-like Hamiltonian.

\item Finally, in the last appendix \ref{anshsh} we provide the mathematical details of 
 our analytical derivation of the sheet conductance 
and of the spin-Hall conductance in the 8-band model.

\end{itemize}

\section{Multi-band model Hamiltonian with spin-orbit interaction}
\label{mbmodel}

In this paper we add the impurity Hamiltonian on top of a specific model Hamiltonian describing a multiband system in the presence of 
atomic spin-orbit coupling \cite{Bistritzer2011}, as well as inter-band inversion-symmetry breaking terms, providing an emergent Rashba interaction \cite{Caviglia2010,Karen2012}.  Given the grouping of the energy bands into quasi-degenerate doublets \cite{Khalsa2013}, the SOI generically has matrix elements both intradoublet (the same), as well as interdoublets (different)  \cite{Shanavas2014}.  Assuming lattice translational invariance, we employ a lattice Hamiltonian of the form 
 
\beq
H = \sum_{\vec{k}} \: \sum_{ \sigma , \sigma'} c_{\vec{k} , \sigma}^\dagger {\cal H}_{\sigma , \sigma'}  ( \vec{k} ) c_{\vec{k} , \sigma' } 
\:\:\:\: ,
\label{model.1}
\eneq
\noindent
with $\vec{k}$ being a generic point in the Brillouin zone and  $\sigma$ being the ``bare'' band index (not to be confused with the spin index, 
with which it can nevertheless coincide in 
some cases, such as the single-band two-dimensional Rashba Hamiltonian). The (``undressed'')
 single-fermion operators in Eq.(\ref{model.1}) obey 
the canonical anticommutation relations $\{c_{\vec{k} , \sigma},c_{\vec{k}' , \sigma'}^\dagger\} = \delta_{\vec{k} , \vec{k}'} \delta_{\sigma , \sigma'}$.
 We also use $\Gamma_{\vec{k} , \lambda}$ to denote 
the (``dressed'') eigenoperators of $H$, satisfying the anticommutation relations $\{\Gamma_{\vec{k},\lambda},\Gamma_{\vec{k}',\lambda'}\}
= \delta_{\vec{k} , \vec{k}'} \delta_{\lambda , \lambda'}$, as well as the canonical commutation relations 

\beq
[ \Gamma_{\vec{k} , \lambda} , H ] = \epsilon_{\vec{k} , \lambda} \Gamma_{\vec{k} , \lambda}
\:\:\:\: , 
\label{model.2}
\eneq
\noindent
so that $\lambda$ is the ``dressed band'' label.  The operators $\Gamma_{\vec{k}, \lambda}$ and $c_{\vec{k},\sigma}$ are related
to each other by an unitary transformation, according to

\beq
c_{\vec{k} , \sigma } = \sum_\lambda w_{\sigma , \lambda}^{\vec{k}} \Gamma_{\vec{k} , \lambda}
\:\:\:\: , 
\label{model.3}
\eneq
\noindent
with the transformation matrix elements satisfying the relations

\begin{eqnarray}
&& \sum_\lambda w^{\vec{k}}_{\sigma , \lambda} [w^{\vec{k}}_{\sigma' , \lambda}]^* = \delta_{\sigma , \sigma'} \;\;,
\nonumber \\
&& \sum_\sigma [w^{\vec{k}}_{\sigma , \lambda}]^*  w^{\vec{k}}_{\sigma , \lambda' } = \delta_{\lambda , \lambda'} 
\:\:\:\: . 
\label{model.4}
\end{eqnarray}
\noindent
According to Eq. (\ref{model.3}), it is possible to express any observable ${\cal O}$ that 
is bilinear in the undressed fermionic operators in the rotated basis, according to 
 
\begin{eqnarray}
{\cal O} &=& \sum_{\vec{k} , \vec{k}' } \sum_{\sigma , \sigma'} [{\cal O}]_{(\vec{k} , \sigma) ; (\vec{k}' , \sigma' ) } 
c_{\vec{k} , \sigma}^\dagger c_{\vec{k}' , \sigma'} \nonumber \\
&=&  \sum_{\vec{k} , \vec{k}' } \sum_{\lambda , \lambda'} 
[{\cal O}]_{(\vec{k} , \lambda) ; (\vec{k}' , \lambda' ) } 
\Gamma_{\vec{k} , \lambda}^\dagger \Gamma_{\vec{k}' , \lambda'} 
\:\:\:\: , 
\label{model.5}
\end{eqnarray}
\noindent
with 

\beq
[{\cal O}]_{ ( \vec{k} , \lambda ) ; (\vec{k}' , \lambda' ) } = 
\sum_{\sigma , \sigma'} [w_{\sigma , \lambda}^{\vec{k}} ]^* w_{\sigma' , \lambda'}^{\vec{k}'} 
[{\cal O}]_{ ( \vec{k} , \sigma ) ; (\vec{k}' , \sigma' ) } 
\:\:\:\: . 
\label{model.6}
\eneq
\noindent
On top of the ``clean'' Hamiltonian in Eq. (\ref{model.1}), we add disorder to our 
system by introducing an impurity scattering potential and a finite density of 
impurities. We do so by introducing, in real space, 
 the white-noise impurity potential $V_{\rm Imp} ( \vec{r} )$ given by  \cite{Borge2014} 
 
 \beq
 V_{\rm Imp} (\vec{r}) = \sum_{ \vec{R}_i } U ( \vec{r} - \vec{R}_i )
 \;\;\;\; , 
 \label{model.7}
 \eneq
 \noindent
 with $U (\vec{r})$ being the single-impurity scattering potential and the impurity scattering 
 centers $\vec{R}_j$ randomly distributed over the system lattice. In Appendix \ref{impures} we 
 perform a systematic derivation of the self-energy, as well as of the vertex corrections due to 
 the finite density of impurities in our system.

 In a single-band 2DEG with a Rashba-type SOI, it is well known that any 
 finite amount of impurities provides a vertex correction that fully screens to zero the spin-Hall conductance 
  \cite{Rashba2004,Mishchenko2004,Inoue2004,Khaetskii2006,Raimondi2004,Dimitrova2005}, 
 while the general conditions at which the cancellation 
  does, or does not, take place  for impurity scattering with an arbitrary angular dependence, and for an 
  arbitrary angular dependence of the spin-orbit field around the Fermi surface are discussed in 
  \cite{Shytov2006} within the Boltzmann equation approach. Also, in  Appendix \ref{lat2Rashba} 
 we apply our method to computing the 
SHC in a lattice model for a two-dimensional electron gas with Rashba 
SOI, finding  a perfect, impurity induced, screening.   At variance, as 
 we show in the following in the    multiband model  
  describing the electronic states in ${\rm SrTiO}_3$-nanodevices \cite{Vivek2017,Perroni2019,Trier2020},
  the multiband structure itself  determines a variable screening of the SHC. Moreover,   
 the amount of screening, at a given density of impurities, depends on the position of the Fermi energy in the system and, 
 therefore, it can be continuously tuned from being almost perfect  to being 
 negligible. 
 
 \section{The  8-band model Hamiltonian} 
 \label{sh8}

 We now focus onto the eight-band, tight-binding 
  Hamiltonian  describing    the 2DEG that forms at a 
 ${\rm LaAlO}_3$-${\rm SrTiO}_3$ (LAO/STO)-interface \cite{Huijben2017,Vaz2019,Trier2020}.  Basically, the perovskite structure of the 
 background lattice induces an isotropic dispersion relation in the $t_{2g}$-bands of the Ti, which are responsible for 
 the conduction in the 2DEG. In addition, the lateral confinement that determines the 2DEG splits the $d_{xy}$-subbands 
 from the $d_{yz}$- and the $d_{zx}$-ones \cite{Vaz2019}.  A ``minimal'' (six-band) model accounting for such effects 
is based on retaining one band of each kind, for a total of six different  subbands, taking into account the spin degree of 
freedom, as well \cite{Perroni2019}. Following Ref. \cite{Trier2020}, in this paper we include an additional pair of $d_{xy}$-subbands
($d_{xy;B}$), splitted above in energy with respect to the  lower ($d_{xy;A}$) subbands with the same orbital character, but still
below, with respect to the $d_{yz}$- and to the $d_{zx}$-subbands.   
 
Defining the various labels as outlined above, we write the eight-band model Hamiltonian in 
the lattice momentum representation as 

\beq
H_{\rm 8-band} = \sum_{\vec{k}} \: 
\sum_{\sigma , \sigma' } c_{\vec{k},\sigma}^\dagger [ {\cal H}_{\rm 8-band} ( \vec{k} )  ]_{\sigma , \sigma' } 
c_{\vec{k},\sigma'} 
\;\;\;\; , 
\label{tr.x1}
\eneq
\noindent
with the sum over $\vec{k}$ taken over the full Brillouin zone, and  $\sigma$ used to 
label single-particle operators $c_{\vec{k} , \alpha , s}$, with $\alpha \in \{ yz,zx,(xy;A) ,(xy;B) \}$ and 
$s \in \{\uparrow,\downarrow\}$. Using a block-notation with respect to the spin degree of freedom, we 
write the matrix ${\cal H}_{\rm 8-band} ( \vec{k} )$ in Eq.(\ref{tr.x1}) as

  \beq
{\cal H}_{\rm 8-band} (\vec{k})   ={\cal H}_0 (\vec{k}) + {\cal H}_{\rm SO} + {\cal H}_Z  ( {\vec{k}})  
\:\:\:\:,
\label{tr.2}
\eneq
\noindent 
with the various matrices at the right-hand side of Eq.(\ref{tr.2}) defined as it follows:

\begin{itemize}

{\item The band dispersion relation}: 
 \beq
{\cal H}_0(\vec{k}) = \left[ \begin{array}{cccc} \epsilon_{yz}  (\vec{k})  {\bf I} & {\bf 0} & {\bf 0} & {\bf 0} \\
{\bf 0} & \epsilon_{zx} (\vec{k}) {\bf I} & {\bf 0} & {\bf 0} \\
{\bf 0} & {\bf 0} & \epsilon_{xy;A} (\vec{k}) {\bf I} & {\bf 0} \\
{\bf 0} & {\bf 0} & {\bf 0} & \epsilon_{xy;B} ( \vec{k} ) {\bf I}  \end{array} \right] 
\:\:\:\: , 
\label{tr.3}
\eneq
\noindent
with 

\beq
{\bf 0} = \left[ \begin{array}{cc} 0 & 0 \\ 0 & 0 \end{array} \right] 
\;\; , \; 
{\bf I} = \left[ \begin{array}{cc} 1 & 0 \\ 0 & 1 \end{array} \right] 
\:\:\:\: , 
\label{tr.3x}
\eneq
\noindent
and

\begin{eqnarray}
\epsilon_{yz} (\vec{k})    &=&2 t_2 [1 - \cos ( k_x )] + 2t_1 [1 - \cos ( k_y )] \;, \nonumber \\
\epsilon_{zx}  (\vec{k}) &=&  2t_1 [1 - \cos ( k_x )] + 2t_2 [1 - \cos ( k_y )]\;, \nonumber \\
\epsilon_{xy;(A,B)}  (\vec{k}) &=&2 t_1 [2 - \cos (k_x) - \cos (k_y)] -  \Delta_{(A,B)} 
\:\:\:\: .
\label{tr.4}
\end{eqnarray}
\noindent
{\item The atomic spin-orbit Hamiltonian:} 
 
\beq
{\cal H}_{\rm SO} = \lambda_{\rm SO} \: \left[ \begin{array}{cccc} {\bf 0} &  i {\bf \sigma}^z & -i {\bf \sigma}^y & -i {\bf \sigma}^y \\
- i {\bf  \sigma}^z & {\bf 0} & i {\bf \sigma}^x & i {\bf \sigma}^x \\
i {\bf \sigma}^y & - i {\bf \sigma}^x& {\bf 0} & {\bf 0} \\
i {\bf \sigma}^y & -i {\bf \sigma}^x  & {\bf 0} & {\bf 0} 
\end{array} \right]
\;\;\;\;.
\label{tr.5}
\eneq
\noindent
{\item The inter-band inversion-symmetry breaking interaction:} 

\beq
{\cal H}_Z (\vec{k}) = \left[ \begin{array}{cccc} {\bf 0} & {\bf 0} & - i f_1^X (\vec{k}) {\bf I}  & - i f_2^X (\vec{k}){\bf I} \\
{\bf 0} & {\bf 0} & - i f_1^Y  (\vec{k}) {\bf I}  & - i f_2^Y (\vec{k}){\bf I} \\
i f_1^X (\vec{k}){\bf I} & i f_1^Y (\vec{k}) {\bf I} & {\bf 0} & {\bf 0} \\
if_2^X (\vec{k}) {\bf I} & i f_2^Y (\vec{k}) {\bf I}  & {\bf 0} & {\bf 0} 
\end{array} \right] 
\;\;\;\; ,
\label{tr.6}
\eneq
\noindent
with 

\begin{eqnarray}
f_{1,2}^X (\vec{k}) &=& 2 \gamma_{1 , 2} \sin (k_x )\;, \nonumber \\
f_{1,2}^Y (\vec{k}) &=& 2 \gamma_{1,2} \sin (k_y ) 
\:\:\:\: . 
\label{tr.6bis}
\end{eqnarray}
\noindent
\end{itemize}
 In addition to the various contributions at the right-hand side of Eq.(\ref{tr.2}), 
in a nonzero applied magnetic field, an additional spin-Zeeman interaction term $H_M$ term appears,
 $H_M=-\vec{M}\cdot \vec{S}$, with   $\vec{M}$ being the Zeeman field and $\vec{S}$ being the total spin. 
 As, throughout our derivation, we always set $\vec{M}=0$, we 
do not include $H_M$ in Eq.(\ref{tr.2}).   

  In doing our calculation, we took the numerical estimate of the various parameters as 
presented in Ref. \cite{Trier2020}, which we summarize in Table \ref{8bandT}. 

  \begin{table} [h!]
    \centering
    \begin{tabular}{ || c | c   || }
    \hline
      Parameter  & Numerical value (meV)     \\
    \hline   
                $t_1$ & 388  \nonumber \\  \hline
                                  $t_2$ &31  \nonumber \\ \hline  
                                    $\Delta_A$ & 150  \nonumber \\ \hline   
                                      $\Delta_B$ & 30   \nonumber \\ \hline
        $\gamma_1$ & 20   \nonumber \\ \hline
              $\gamma_2$ & 5  \nonumber \\  \hline
                     $\lambda_{\rm SO}$ &8.3  \nonumber \\   
   \hline       \end{tabular}
    \caption{Numerical values of the parameters of the eight-band model
    (measured in meV) as from \cite{Trier2020}. }
    \label {8bandT}
    \end{table}

\begin{figure}
 \center
\includegraphics*[width=0.85 \linewidth]{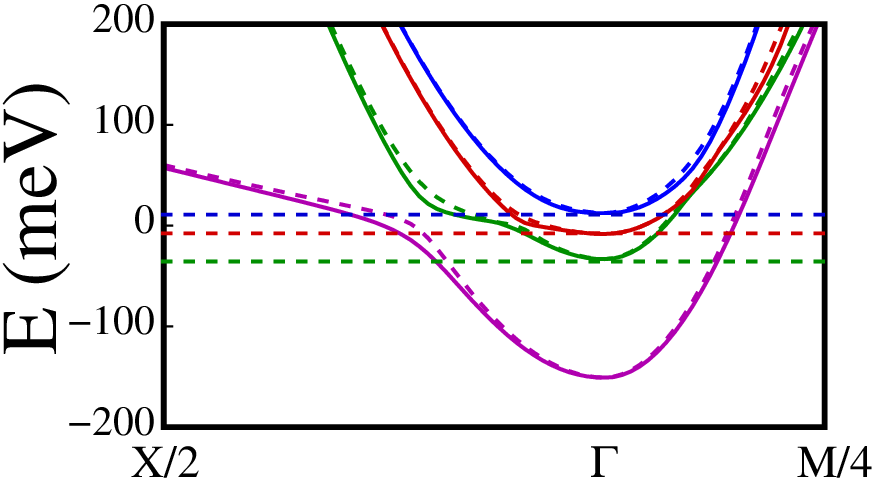}
\caption{Energy levels of the full Hamiltonian in Eq.(\ref{tr.2}) 
computed by choosing the other system parameters as in Table \ref{8bandT}. 
The colored, horizontal dashed lines mark the opening (on increasing the energy) of higher-energy 
doublets.
} 
\label{8_band}
\end{figure}

In Figure \ref{8_band}  we plot the energy levels of the full Hamiltonian in Eq.(\ref{tr.2})  along high-symmetry lines of the Brillouin zone, 
by choosing the other system parameters as in table \ref{8bandT}. 
The combined effect of the atomic  SOI  and of the inversion-symmetry breaking terms also splits each doublet,
as evidenced in the figure, even at zero Zeeman field. The colored horizontal dashed lines 
correspond to the ``opening'' of the higher-energy doublets. As we discuss in the following, 
as long as the chemical potential is much below the opening of the green doublet (dashed horizontal 
green line -- about 45 meV), 
the 2DEG behaves as a ``standard'' spinful, single-band Rashba 2DEG (see Appendix \ref{efra} for 
a detailed model calculation), although the multi-band structure of the system strongly affects
the impurity-induced screening of the spin-Hall conductance.

\noindent
Above the green dashed line the second doublet becomes available: as we 
discuss in the following, for what concerns spin transport, the green doublet 
is pretty similar to the magenta one, as they both share a high overlap with,
respectively, the $d_{xy;A}$  and the $d_{xy;B}$ subbands, very similar 
in their properties. 

Moving across the dashed horizontal red line, higher-energy subbands start to be
populated. As it may be readily checked by direct calculations, these subbands 
have a high overlap with the $d_{yz}$ and the $d_{zx}$ subbands, which have 
different symmetries, with respect to the lower-energy $d_{xy;A/B}$ subbands. 
At the onset of the higher-energy subbands, a Lifshitz transition (LT)  is expected to 
take place \cite{Joshua2012}, with a massive and sudden increase in the density of states due to 
the opening of the higher subbands, 
as evidenced in  Appendix \ref{anshsh}. As we evidence in 
the following, the LT carries along remarkable changes in the (spin) transport 
properties of the system, with relevant consequences on the experimental 
results. 

    \begin{figure}
 \center
\includegraphics*[width=1.05 \linewidth]{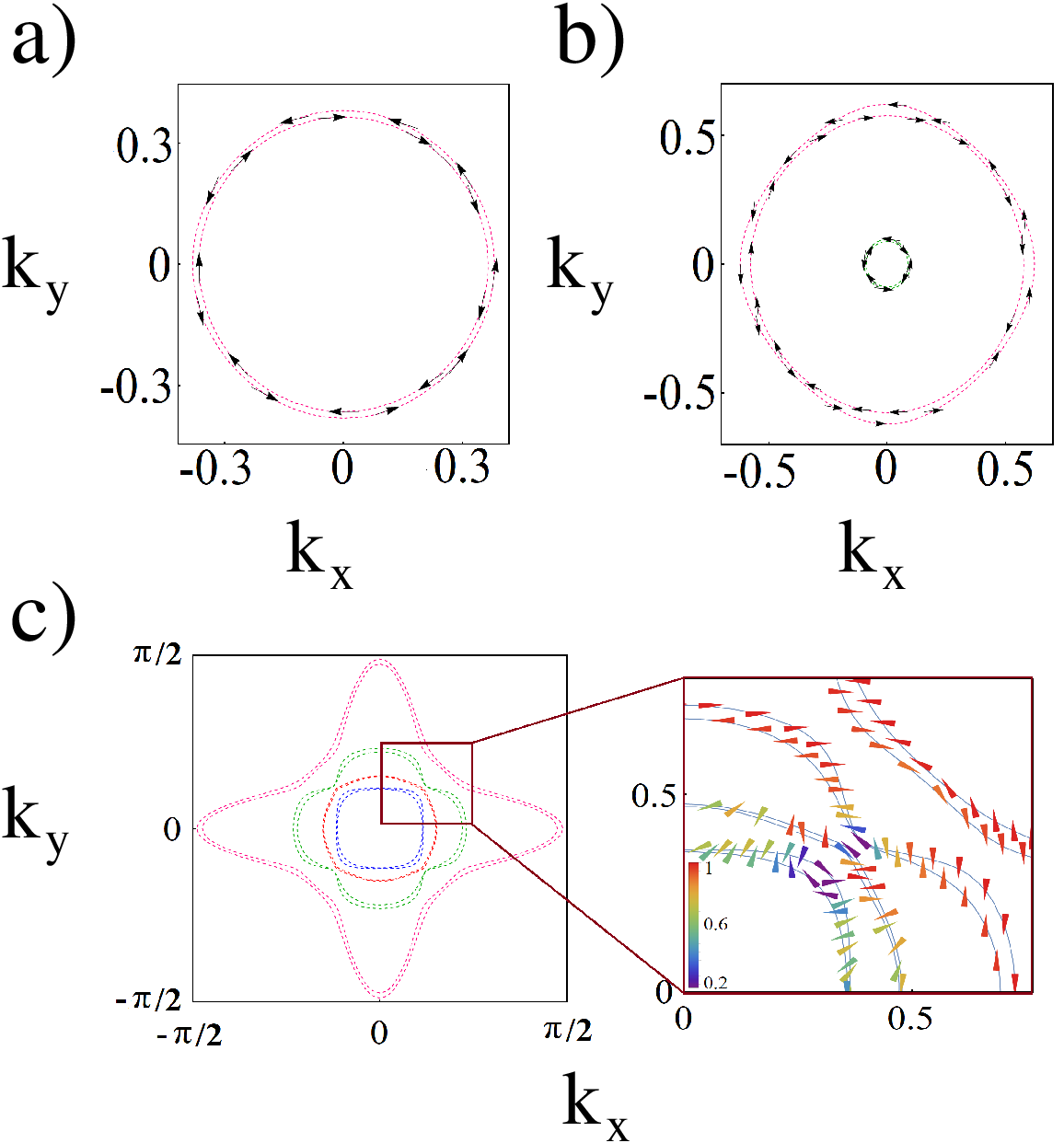}
\caption{
{\bf (a):}   Spin pattern 
 within the $xy$-plane in spin space as a function of the momentum 
 $\vec{k}$ computed using the eight-band model Hamiltonian in Eq. (\ref{tr.2}) with 
 $\vec{M} = 0$ and at energy $E=-100$ meV. 
 {\bf (b):} Same as in  {\bf (a)}, but with $E=-30$ meV. 
 {\bf (c):} Same as in  {\bf (a)} and  {\bf (b)},  but with $E=+55$ meV.  The 
 spin pattern is evidenced in the zoom in the right-hand subpanel. In all the cases
 the out-of-plane spin component is zero, due to the absence of an applied magnetic field.}   
\label{spinpol}
\end{figure}

Specifically,    on increasing the chemical potential, we have computed, at fixed energy $E$, the spin pattern 
 within the $xy$-plane in spin space as a function of the momentum 
 $\vec{k}$ within the Brillouin zone. In Fig. \ref{spinpol} we report the 
 results of our derivation at energy $E=-100$ meV, $E=-30$ meV, and $E=+55$ meV. 
 In Fig. \ref{spinpol}{\bf (a)} we show our results at $E=-100$ meV. We 
 clearly recognize the typical spin pattern of a Rashba 2DEG \cite{Sinova2004}.
 As Fig. \ref{spinpol}{\bf (a)}  evidences, the opposite spin orientation in the 
 two subbands corresponds to an opposite value of the Rashba effective 
 magnetic field acting over the electron spin and, accordingly, to 
  opposite, over-all contributions to the total spin current generated by 
  an applied electric field ${\cal E}^y$. 
In fact, that result is consistent with our derivation of Appendix \ref{efra}, where we show   
that, for $- \Delta_A < \mu < - \Delta_B$,  with the band offsets $\Delta_{A,B}$ 
defined in the last one of Eqs. (\ref{tr.4}),
the eight-band model is effectively 
described by the Rashba Hamiltonian $H_{xy}$ in Eq. (\ref{efra.8}). 
We therefore expect, in the absence of disorder, a quantized jump (by $e^2/h$) in 
 the spin-Hall conductance $\sigma_{xy}^z $ at the onset of the first doublet and 
 a featureless, constant   value $\sigma_{xy}^z$ on further increasing $\mu$
 (see Appendix  \ref{lat2Rashba}  for details). As we discuss in the 
 following, this is exactly what happens in the absence of disorder. 

In Fig.\ref{spinpol}{\bf (b)} we draw states at energy $E=-30$ meV. These belong to 
the first two pairs of subbands. As it appears from the figure, the sign of the Rashba 
SOI is the same in both pairs of subbands: this suggests that, at the onset of the 
second pair of subbands, an additional jump, similar to the one at the start of the first
doublet, should  appear, with a corresponding doubling in the quantized value of $\sigma_{xy}^z$. 

A drastic  change happens at $E=+55$ meV, when all four of the doublets are involved
[see Fig. \ref{spinpol}{\bf (c)}].  Indeed, from the zoom of the highlighted region, we clearly 
see how the mixing with higher effective mass bands implies that different pairs of subbands
show different signs in the net SOI (in fact, our result mimics the one of 
Ref. \cite{Johansson2021}, despite the different values of the model parameters
used in that paper). At the onset 
of the (mostly $yz$- and $zx$-like) higher-energy subbands, we therefore  
expect a breakdown of the monotonical increase of $\sigma_{xy}^z$ with $\mu$ and, 
due to the large difference in the effective mass between the lower and the upper doublets, 
even a possible change of sign in the spin-Hall conductance, at a certain value of $\mu$. As 
we will see in the following, this is, in fact, what happens when going through a full 
calculation of $\sigma_{xy}^z$ in the eight-band model, even after impurities are added
to the system.    In addition, while  our sketch of the spin orientation in 
 Fig. \ref{spinpol}{\bf (a)}-  \ref{spinpol}{\bf (c)} is intended just to evidence the spin pattern at 
 equal energy curves,  close to the Lifshitz transition, it is important to 
 consider the modulus of the average spin polarization, in particular close to the avoided 
 crossing directions (in our case the diagonals of the Brillouin zone), due to its   relation 
 with alternative possible sources of a nonlocal resistance signal, such as the Edelstein effect 
 \cite{Johansson2021}. For this reason, in the zoom of  Fig. \ref{spinpol}{\bf (c)}, we 
  provide the modulus of the average spin polarization in the color code specified in
 the figure itself, thus evidencing a   reduction in the modulus of the average spin
 polarization over the avoided crossing directions  similar to the 
 one found in Ref. \cite{Johansson2021}. 
 
 We now provide the results of our calculations of the sheet conductance, as well as of the 
 spin-Hall conductance, in the eight-band model, in the clean limit, as well as in the 
 presence of a finite impurity concentration. 
 
 \section{The sheet conductance and the  spin-Hall conductance  in the 8-band model}
\label{shsc}

To compute  the sheet conductance of the   8-band model of 
 section \ref{sh8}, $\sigma_s$, we use  the  analytical expression of  Appendix \ref{anshsh}, 
 which we derive by means of a systematic implementation of  the formalism of Appendixes \ref{lrt} and 
 \ref{impures}.  In particular, having numerically checked that the vertex corrections 
 in Eq.(\ref{lc.5}) provide a negligible contribution to $\sigma_s$, we perform our calculation by using 
 Eq. (\ref{recr.12}) of Appendix \ref{lrt}, with the single-particle Green's functions including the finite,
 impurity-induced, imaginary part of the self-energy. 
 Specifically, we set 
  
 \begin{eqnarray} 
  && \sigma_{s} =     \frac{1}{  V   }   
 \sum_{ \vec{k}  } \: \sum_{\lambda  , \lambda^{'} }  [   j_{{\rm ch},x} ]_{(\vec{k} , \lambda ) ; (
  \vec{k}, \lambda^{'} )}    [ j_{{\rm ch},x} ]_{(\vec{k} ,\lambda^{'}  ) ; (
  \vec{k}  , \lambda  )} \nonumber \\
  && \times  \int \frac{ d \bar{\omega}}{2 \pi}  \: \Biggl\{ 
 g_{ (  \vec{k}  , \lambda )   }^< ( \bar{\omega} ) \partial_\omega g_{ (  \vec{k},\lambda^{'}   ) }^R (
   \bar{\omega } ) +   g_{  (  \vec{k}  , \lambda )  }^A ( \bar{\omega} ) 
   \partial_\omega g_{(  \vec{k} , \lambda^{'}   )  }^< ( \bar{\omega}   )  \Biggr\} \:, \nonumber \\
   &&  
   \label{recr.12x}
\end{eqnarray}
\noindent
with $V$ being the system volume, $[j_{{\rm ch},a}]_{(\vec{k},\lambda);(\vec{k},\lambda')}$ being 
the charge current matrix elements in the basis of the single-particle eigenstates of ${\cal H}_{\rm 8-band}
(\vec{k})$, and the single particle Green-Keldysh functions $g_{(\vec{k},\lambda)}^{A,R,<} (\omega)$ being 
defined in Appendix \ref{selfeim}. The impurity concentration determines, according to Eq.(\ref{ims.10c}) of appendix 
\ref{impures}, a finite lifetime $\tau_{\rm Imp}^\lambda$ for 
the quasiparticle excitations in band $\lambda$, which enters the corresponding Green's functions according 
to Eq.(\ref{ims.13}). Due to the relation between the impurity concentration and the finite quasiparticle  
lifetime, as we discuss in Appendix \ref{impures}, 
we take into account the former through the latter, by using for $\tau_{\rm Imp}^\lambda$ the 
$\lambda$-independent value $\tau_{\rm Imp}=3\: {\rm ps}$, as from the measures of Ref.\cite{Trier2020}. 
 
\begin{figure}
 \center
\includegraphics*[width=1. \linewidth]{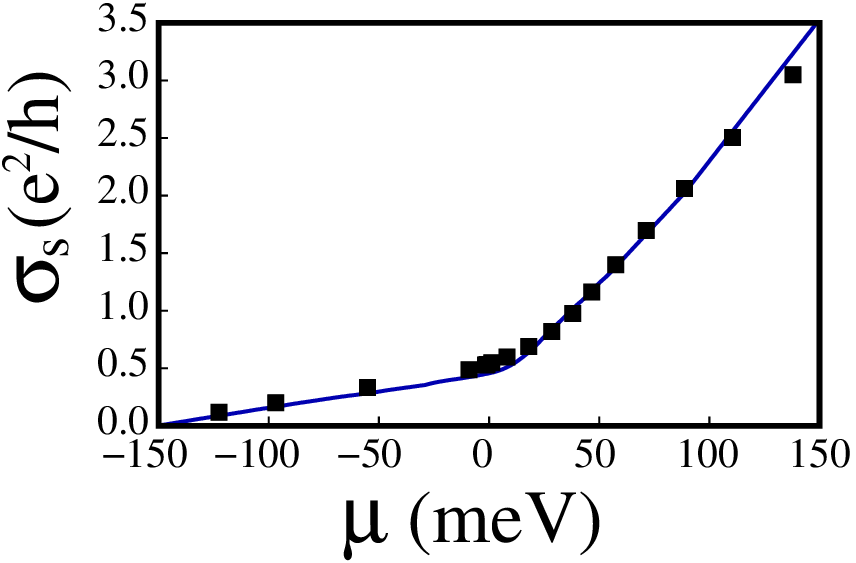}
\caption{$\sigma_s$ as a function of $\mu$ (in units of $e^2/h$), computed according to  Eq.(\ref{lc.4bis})  (blue curve);
$\sigma_s$ experimentally measured, as reported in Ref. \cite{Trier2020} (black full squares).   As discussed in
the main text, we have accounted for the finite impurity concentration by setting the quasiparticle lifetime 
$\tau_{\rm Imp} = 3\:{\rm ps}$, in agreement with the experimental results of  Ref. \cite{Trier2020} . 
} 
\label{sheets}
\end{figure}

In Figure \ref{sheets} we plot our result for  $\sigma_s$ (in units of $e^2/h$) as a function of $\mu$, computed using Eq.(\ref{recr.12})
 as discussed before (blue curve). For comparison, we also plot  the experimental data  corresponding to  the measures of
 Ref. \cite{Trier2020}  (black full squares). Apparently,  there is a pretty good agreement between the
two sets of data: in particular, both in the theoretical curve, as well as in the experimental data, 
we note the sudden change in the slope of $\sigma_s$  at the onset 
of the higher-energy bands, at $\mu $ slightly higher than 0, which evidences the sudden change in the density of state
at the LT. 

Close to the upper bound of the interval of values of $\mu$ that we consider ($\mu \in [-150 \: {\rm meV} , 150 \; {\rm meV} ]$)
the theoretical curve seems to become slightly  higher  than the last experimental point. This might be a signal of 
an (expected) breakdown of our approximation at large values of $\mu$. Yet, the agreement between the analytical and 
the experimental results is pretty good throughout almost all the interval of values covered by the experiment of Ref.
\cite{Trier2020}.

We now consider the spin-Hall conductance $\sigma_{xy}^z$.  
Along the derivation of Appendix \ref{lrt} and 
 \ref{impures}, we find that the corresponding mathematical expression 
can be derived from Eq.(\ref{recr.12x}) by substituting the first charge-current vertex, 
$[j_{{\rm ch},x}]_{(\vec{k},\lambda);(\vec{k},\lambda')}$,
with the fully dressed spin-current vertex $[j_{{\rm sp}, x}^z]_{(\vec{k},\lambda);(\vec{k},\lambda')}$ and the 
second one, $[j_{{\rm ch},x}]_{(\vec{k},\lambda');(\vec{k},\lambda)}$, with  $[j_{{\rm ch},y}]_{(\vec{k},\lambda');(\vec{k},\lambda)}$. 
  Within the approach of  Appendix \ref{lrt}, we  perform our calculation using the   formulas for $\sigma_{xy}^z$ for a homogeneous system. 
In fact,  as it is typical for the spin-Hall effect, the  spin current is established 
 over typical length scales of the order   of the electron mean free path \cite{Abanin2009},  which
we assume to be much smaller than the width  of the sample we consider, as well 
as of the distance between the contacts \cite{Trier2020}.  Eventually, this enables us 
not to consider possible inhomogeneities in the spin current in the sample due to finite-size effects
and allows us to resort to the formulas we employ in   Appendix \ref{lrt}.    
 
 For the sake of the discussion and 
also to interpret the results in terms of an effective Rashba-type SOI, we begin by computing $\sigma_{xy}^z$ as 
a function of $\mu$ in the ``clean limit'', that is, without adding impurities to the sample. 
To do so, we use Eq.(\ref{lcx.1}) of Appendix \ref{anshsh}, by replacing $g_{(\vec{k} , \lambda)}^{R/A} ( \omega )$
with their counterparts in the clean limit, $g_{(\vec{k} , \lambda)}^{R/A;(0)} ( \omega )$ in Eq.(\ref{ems.x2})

    \begin{figure}
 \center
\includegraphics*[width=1 \linewidth]{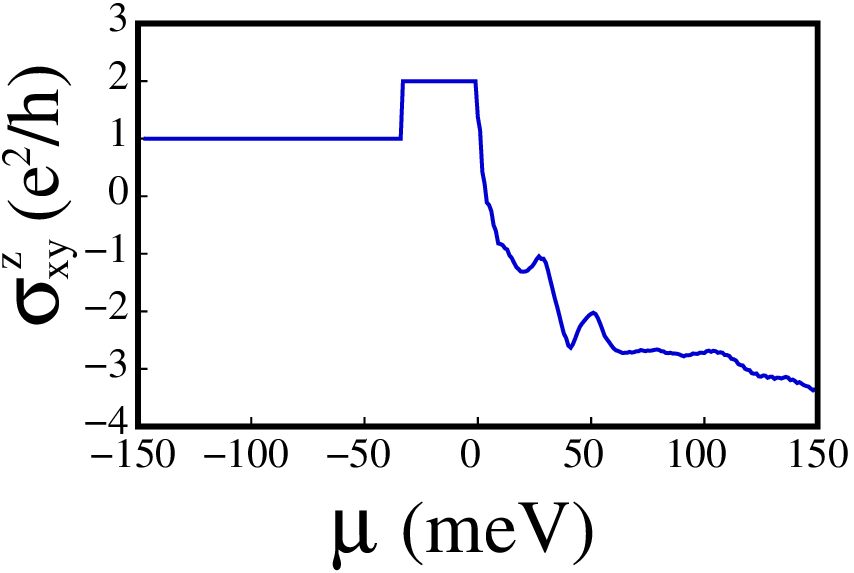}
\caption{Spin-Hall conductance $\sigma_{xy}^z$ computed in the eight-band model using the 
 full Hamiltonian in Eq.(\ref{tr.2}), by neglecting the effects of the impurities (``the clean limit''), as a 
function of the chemical potential $\mu$ for $-150 {\rm meV} \leq \mu \leq 150 {\rm meV}$. The 
quantized first two plateaus ($\mu \leq 0$) evidence the two-dimensional Rashba-type
nature of the effective spin-orbit interaction in the first two energy doublets of Fig. \ref{8_band}. 
} 
\label{sigmaH_clean}
\end{figure}
 
 In Fig. \ref{sigmaH_clean} we show the results of our 
calculation.  We note that, 
as soon as $\mu$ ``hits'' the bottom of the first $d_{xy;A}$-like subband, $\sigma_{xy}^z$ suddenly jumps from 
0 to $e^2/h$. A similar sudden, quantized jump takes place as $\mu$ crosses the bottom of the  second doublet of energy eigenstates. 
 As we review in detail in Appendix \ref{lat2Rashba}, this is the typical behavior of the SHC 
 in a lattice model of a Rashba 2DEG, which is consistent with the effective description of the 
eight-band model in terms of two, coupled Rashba-type Hamiltonians, at energies below the onset of the $yz$- and of the 
$zx$-like subbands (see Appendix \ref{efra} for details).  

Moving toward higher value of $\mu$, we see that the regular behavior of $\sigma_{xy}^z$, made out of 
sharp jumps between quantized plateaus, suddenly changes when the higher-energy  doublets 
set in. In fact, this comes along a drastical change in the effective Rashba SOI within each doublet. 
From Fig. \ref{spinpol} we see that, while  the first two doublets  
  both host an effective Rashba SOI with coupling strengths of the same sign (which is consistent 
with the two plateaus in the left part of the plot in Fig. \ref{sigmaH_clean}), as soon as the $yz$- and the $zx$-like subbands 
get involved, the sign of the Rashba SOI  from different  doublets of subbands can be different and, more 
importantly, the sign of the Rashba coupling strength in subbands with higher effective mass is opposite to 
the one of the $d_{xy;A}$-like doublet. Thus, two different effects are expected to arise at the onset of 
the higher subbands: a strong increase in $\sigma_{xy}^z$, due to the larger carrier density, accompanied by 
a sudden change in the sign of the SHC itself, for the reason discussed above and 
evidenced in Fig. \ref{spinpol}. Apparently, this is exactly the feature that shows up in Fig. \ref{sigmaH_clean}
for $\mu$ slightly greater than 0, the following oscillations being likely due to the sequential changes in sign 
of the effective Rashba SOI in the intermediate-energy doublets, as $\mu$ gets larger \cite{Trier2020}. 
We now discuss how Fig. \ref{sigmaH_clean} is modified by adding impurities to the system.

 The behavior of the SHC   in the presence of a finite density of   impurities is strictly related to the mechanism 
behind the nonlocal resistance in SrTiO$_3$-based devices. In Ref. \cite{Trier2020} a systematic analysis of all the possible 
different mechanisms that could potentially lead to a nonlocal resistance in the system has been gone through, similar to 
what has been done in Refs. \cite{Nachawaty2018,Tagliacozzo2019} for graphene close to the charge-neutrality point.
 Having ruled out all the possible alternatives, the only left over 
reasonable explanations rely on either the Edelstein effect \cite{Edelstein1990}, or over the spin-Hall effect \cite{Sinova2004} 
(or a combination of the two of  them).  Nevertheless,   the direct measurement of the spin polarization via the Hanle precession
revealed that the electron spins are mostly polarized orthogonal to the electronic 2DEG, which rules out the Edelstein 
effect, as well, leaving SH effect as the only possible mechanism responsible for the measured nonlocal resistance
\cite{Trier2020}. 

In the two-band Rashba model the presence of a finite impurity concentration is known to fully cancel
 the SHC  by ``compensating'' the terms $\sigma_{xy;A}^z$, computed as in 
 Eq. (\ref{lcx.1}), with the vertex correction $\sigma_{xy;B}^z$,  in Eq. (\ref{lcx.3})  
 \cite{Mishchenko2004,Inoue2004,Khaetskii2006,Raimondi2004,Dimitrova2005} (see also 
Appendix \ref{lat2Rashba} for details). At variance, as we are going to show by direct calculation, due to 
the complex inter-band mixing, the SHC  in the eight-band model is affected only partially by 
the presence of disorder: the cancellation strongly depends on $\mu$ and, more importantly, 
it is never complete. This leaves room for a finite  SHC in the  eight-band model,
even in the presence of a finite density of impurities in the system, which is apparently  consistent with the 
  experimental results of Ref. \cite{Trier2020}. 

\begin{figure}
 \center
\includegraphics*[width=1 \linewidth]{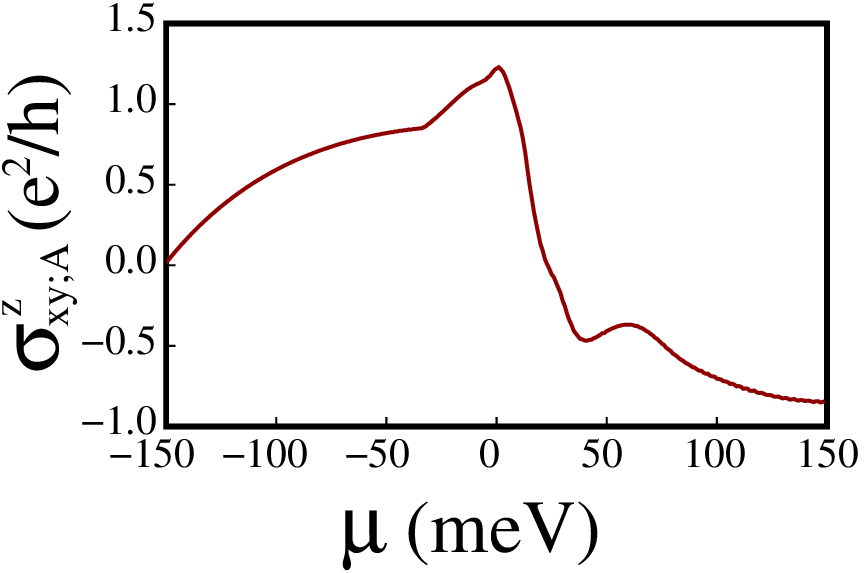}
\caption{Spin-Hall conductance $\sigma_{xy;A}^z$ (without vertex corrections) computed, as a function of $\mu$, 
 within the eight-band model,  by using the parameters in Table \ref{8bandT} and by setting $\tau_{\rm e} = 3$ ps in
 the  dressed single-particle Green's functions. 
} 
\label{sh_nov}
\end{figure}

To spell out in detail the role of the two contributions to $\sigma_{xy}^z$, we now separately discuss the
two of them. In Fig.\ref{sh_nov}, we plot our results for $\sigma_{xy;A}^z$ as a function of $\mu$ 
for $-150 \: {\rm meV} \leq \mu \leq 150  \: {\rm meV}$, computed according to  Eq. (\ref{lcx.1}) 
by using the parameters in Table \ref{8bandT} and by setting $\tau_{\rm e} = 3$ ps. Remarkably, we can already see how  adding the 
finite quasiparticle lifetime already strongly alters the behavior of  $\sigma_{xy}^z$, compared
to the clean limit of Fig.\ref{sigmaH_clean}. Yet, differently from what happens for the sheet conductance, 
we now show how, in this case,   the vertex corrections have an important weight in the final result.

To compute $\sigma_{xy;B}^z$, we use Eq.(\ref{lcx.3}) of Appendix \ref{anshsh}.  
In Fig.\ref{sh_yv} we plot the corresponding result for the ``full'' spin-Hall conductance 
$\sigma_{xy}^z = \sigma_{xy;A}^z + \sigma_{xy;B}^z$, including the 
effects of the vertex corrections. Looking at the plot, we note that there is a large window 
of values of $\mu$, from $\mu = -150$ meV to the onset of the high-energy subbands, where the 
vertex correction screens $\sigma_{xy}^z$ to a ``quasilinear'' dependence on $\mu$ 
at its onset. A minor, though clearly visible, change in the slope takes place 
at the onset of the second doublet: the change in the slope is the remnant, after the screening, of 
the second jump between the quantized values of $\sigma_{xy}^z$ in Fig.\ref{sigmaH_clean}. A relatively sharp
decrease in the spin-Hall conductance takes place as soon as $\mu$ crosses the bottom of the two 
high-energy subbands. For $\mu$ slightly larger than 0,  $\sigma_{xy}^z$ first increases and eventually crosses 0 and becomes 
negative. This is strictly connected with the onset of the higher-energy doublets which, as we highlighted before, 
provide contributions to $\sigma_{xy}^z$ opposite in sign with respect to the ones arising from the lower doublet. 
Higher-energy doublets  are indeed characterized by 
an effective Rashba SOI of opposite sign, with respect to the lower-energy ones, and by a much higher 
density of states (at the band onset).

    \begin{figure}
 \center
\includegraphics*[width=1 \linewidth]{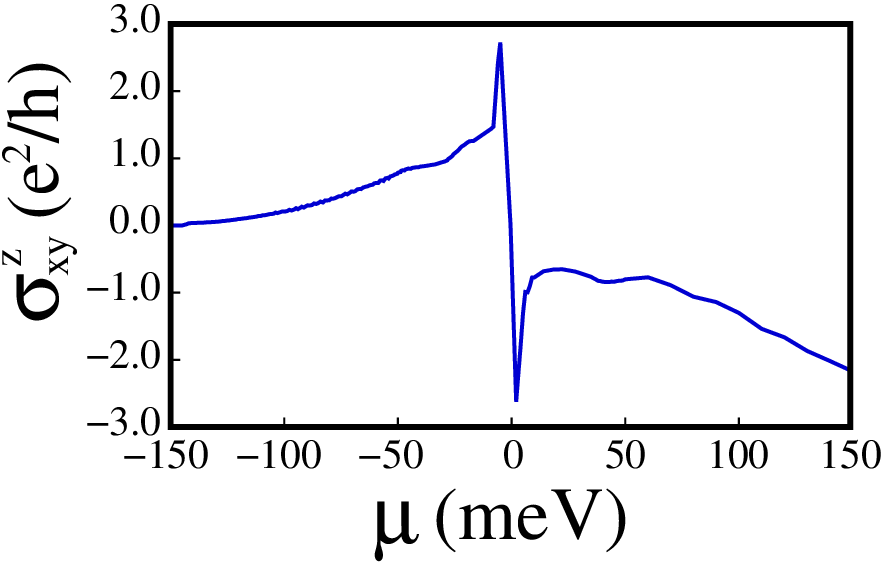}
\caption{Spin-Hall conductance $\sigma_{xy}^z = \sigma_{xy;A}^z+ \sigma_{xy;B}^z$(including vertex corrections) computed, as a function of $\mu$, 
 within the eight-band model,  by using the parameters in Table \ref{8bandT} and by setting $\tau_{\rm e} = 3$ ps in
 the  dressed single-particle Green's functions. 
} 
\label{sh_yv}
\end{figure}

\begin{figure}
 \center
\includegraphics*[width=1  \linewidth]{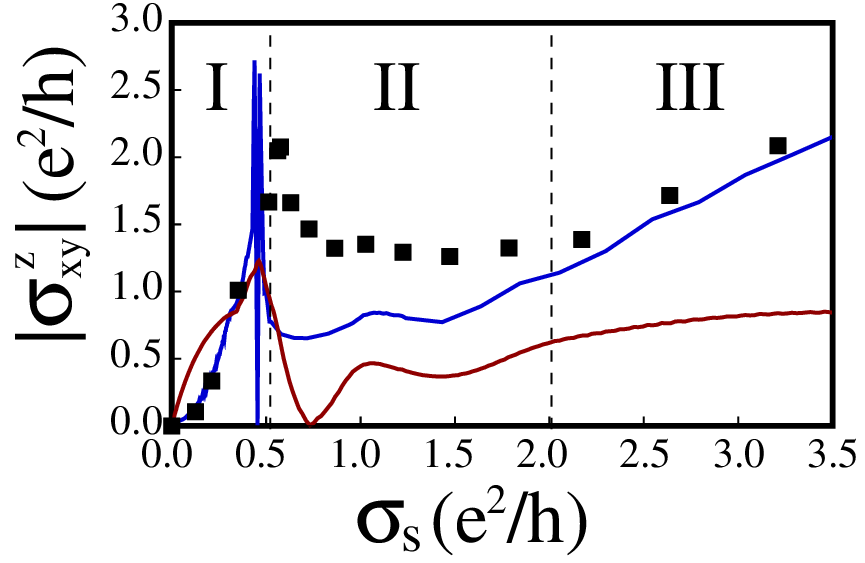}
\caption{Solid blue line: (absolute value of the) spin-Hall conductance $| \sigma_{xy}^z|$ as in Fig. \ref{sh_yv} drawn   
as a function of the sheet conductance $\sigma_s$ in Fig. \ref{sheets}; Solid red line: (absolute value of the) spin-Hall conductance 
{\it without vertex corrections}, $| \sigma_{xy;A}^z|$,  as in Fig.\ref{sh_nov} drawn   
as a function of the sheet conductance $\sigma_s$ in Fig.\ref{sheets}; Black full squares: experimental results
for $\sigma_{xy}^z$ as a function of $\sigma_s$ as presented in Ref. \cite{Trier2020}.  
The dashed vertical lines mark the splitting of the graph into 
the regions I,II and III, as discussed in the text. 
} 
\label{compar_exp}
\end{figure}

We now discuss our results in comparison with the experimental data of 
Ref. \cite{Trier2020}. To do so,  we mimic  the presentation 
of that paper by combining the results of Figs. \ref{sheets} and  \ref{sh_yv}  to display 
our results for $| \sigma_{xy}^z |$ as a function of our theoretically computed sheet conductance 
$\sigma_s$ (note that we consider the absolute value of $ | \sigma_{xy}^z|$ to make a rigorous comparison
with the experimental results of Ref. \cite{Trier2020}, where they measure the spin-Hall conductance from 
the square root of the squared spin-Hall angle, which appears in the formula for the nonlocal resistance, 
[Eq.(1) of the paper.   To evidence the importance of accounting for impurity-induced vertex corrections, 
in Fig. \ref{compar_exp}  we synoptically show the corresponding plots of both $| \sigma_{xy}^z|$ (computed by taking 
the absolute of $\sigma_{xy;A}^z + \sigma_{xy;B}^z$ -- blue curve) and of $ | \sigma_{xy;A}^z|$ (red curve), together with the experimental 
points corresponding to Fig. 4(c) of Ref. \cite{Trier2020} (black full squares).

\noindent
In analyzing Fig.\ref{compar_exp} we split the graph   
into three regions: region I, for $0 \leq \sigma_s \leq e^2/(2h)$, region II,  for $e^2/(2h) \leq \sigma_s \leq 2 e^2/h$, 
region III, for $2 e^2/h \leq \sigma_s$.

Region I roughly corresponds to the part of the diagram before the 
onset of the high-energy, $yz$-like and $zx$-like, subbands in which, as it emerges from our analysis of 
Appendix \ref{efra} and from Fig.\ref{spinpol}, the two active $xy$-like subbands 
exhibit an effective Rashba SOI of the same sign in both bands. 
A synoptical numerical analysis of the data shown in Figs.\ref{sheets},\ref{compar_exp},\ref{taus} also evidences how 
 region I corresponds to the interval of values of $\mu$ where only the lowest-energy doublets contributes the vertex 
 corrections,  $\sigma_{xy;B}^z$, 
which we reliably compute within our approach, that is expected to work well in the low-density region of 
the system, by using  Eq. (\ref{ic.a2}) of Appendix \ref{anshsh}. In fact, we see that there is a perfect collapse of the experimental data onto the analytical blue curve: 
on one hand this shows the reliability of our method in this region, on the other hand, it evidences the importance of 
pertinently accounting for the vertex corrections. Indeed, the corresponding (red) curve, derived without taking 
into account the term $\sigma_{xy;B}^z$, clearly does not fit the experimental points.

Leaving aside region II, for the time being, let us focus onto region III of Fig. \ref{compar_exp}.
Region III is characterized by a much higher carrier density than region I. Moreover, as we readily 
infer from Fig. \ref{taus}, there is also, in this region,  a finite contribution to 
Eq. (\ref{ash.1}) from all the four doublets of subband although, just as in region I, there is a single doublet (the lowest-energy one)
that takes over with respect to the other ones. This  makes us employ the approximate 
method outlined in Appendix \ref{anshsh} to pertinently weight the four involved doublets. Yet, despite the possibly 
oversimplified approach we use here, from the plot of Fig. \ref{compar_exp} we see again quite a good agreement between
the experimental points and the analytical curve, with just a minor undershooting of the latter which could possibly be
fixed by, for instance, refining our approach by 
going through a systematic fitting procedure of the normalization factor in front of the vertex correction. 
Importantly, by synoptically looking at the red and at the blue curve of Fig. \ref{compar_exp} we see how also in region III, 
it is fundamental to add vertex corrections to the final expression for 
$\sigma_{xy}^z$, in order to recover a good agreement between experiment and theory. 
 Apparently, our results imply that contributions 
to the spin-Hall conductance arising from vertex corrections
are always relevant, although their precise numerical estimate deserves a more refined modulation 
than the one behind Eq. (\ref{ic.a2}). In addition, as evidenced by Fig.S4 of the Supplemental 
Material of Ref. \cite{Trier2020}, the contribution to $\tau_{\rm Imp}$ arising from inelastic  scattering 
processes should be taken into account, as well. All these  further extensions of our derivation, although 
being definitely relevant to improve the numerical match between theory and experiment, go
beyond the scope of this work, where we focus on the role of impurities to catch the main features of the 
experiments of Ref. \cite{Trier2020}. We will possibly consider them in further developments 
of our investigation. Yet, it is worth stressing once more the good agreement of our theoretical calculations with the 
experimental data, despite the possibly oversimplified model we are employing here. 

We finally discuss region II. At a first glance, in this region the agreement between the experimental data of Ref. \cite{Trier2020} and 
our analytical results is not as good as in the other two regions. Indeed, first of all we note that the 
experimental data   seem  to exhibit  a peak in $ | \sigma_{xy}^z |$ for $\sigma_s^* \approx 0.6 e^2/h$,
while, as it is clearly evidenced in Figs. \ref{sh_nov} and \ref{sh_yv}, our analytical results, instead, show 
a zero around that value of $\sigma_s$, whether the vertex corrections are included, or not. 
As we discuss above, due to the different sign of the effective SOI interaction arising within the various energy doublet, 
it is somehow expected that  $\sigma_{xy}^z$   crosses zero at some finite value of $\sigma_s$, roughly corresponding 
to the onset of the higher-energy doublets. However, due to the rather small hybridization between those last subbands and 
the lower-lying ones, we also expect that the crossing takes place sharply, within a small interval of variation of 
$\mu$, as also apparently shown in Fig. \ref{sh_yv}. Once switching to $|\sigma_{xy}^z|$, as we do in 
drawing Fig. \ref{compar_exp}, the zero-crossing feature trades for a sharp double peak which, 
due to the limited number of experimental points in  Ref. \cite{Trier2020} around $\sigma_s^*$, might very well be
experimentally seen as a single peak. An additional observation concerns the apparent offset, present throughout 
all region II, between the experimental data and the analytical curve, which apparently underestimates 
$\sigma_{xy}^z$, particularly close to the left-hand endpoint of the region, although the agreement between 
the experiment and the theory becomes better on moving towards the right-hand end point of region II.  
While we discuss more about this point below, here we stress how the agreement could be in principle improved by 
a pertinent refining of our method for estimating the prefactor in front of the vertex corrections and that, more importantly, the 
main trend of the theoretical curve as a function of $\sigma_s$ is basically consistent with the experimental data 
throughout Region II, as well. In any case, at least from the conclusions one may recover from Fig.\ref{spinpol} about the 
Rashba SOI in each doublet, a zero-crossing of $ | \sigma_{xy}^z |$ should be expected, right after the onset 
of the higher-energy doublets. In addition to the previous arguments, we note a remarkable, experimentally 
measured,   decrease and increase of the spin-Hall angle $\gamma$ (although not all the way down to 0), which 
is evidenced in Fig.3{\bf (c)} of Ref. \cite{Trier2020}, at the value of the gate voltage that corresponds to $\sigma_s^*$.
In view of above arguments and given the good agreement in regions I and  III between our results and the 
experimental data, a refinement of the  $\sigma_{xy}^z$ measurements around $\sigma_s^*$  would be highly desirable. 

 On the theory side, one could improve the modeling of the system by introducing scattering from finite-range impurities.
Indeed, the scattering potential could present a correlation length consistent
with experiments on negative magnetoresistance driven by spin-orbit coupling and scattering by dislocations \cite{caviglia_tras}. 
This could give a contribution in the region II, since the anisotropic scattering could suppress
backscattering processes within the outer Fermi surface
with large average Fermi momentum, while still allowing for
interband scattering. This should be accompanied by a quasiparticle
lifetime which can be smaller for the inner bands with smaller average Fermi momenta. The induced imbalance could 
enhance the values of  spin-Hall conductance in the region II, while leaving the sheet conductance poorly affected.

A possible extension of our work could possibly involve carrying out the analysis of the spin-, as well as of the 
orbital-, Edelstein effect. Indeed, while in Ref. \cite{Trier2020} a contribution to the nonlocal resistivity from the Edelstein effect 
was ruled out   from the dependence of the Hanle signal on the applied magnetic field, it might also be possible that, 
due to a strong anisotropy in the spin lifetime and  spin-diffusion length with the spin direction, for shorter channels, with lengths 
comparable to the spin diffusion length for spins in the plane, one gets a significant contribution from the 
(spin) Edelstein effect,  in addition to the spin-Hall effect (SHE(.  Moreover, as evidenced in Refs. \cite{Johansson2021,mattia_new}, 
there might be also a
sizeable contribution from the orbital Edelstein effect, especially in region II, where the agreement between the experimental data
and our analytical results is apparently less striking than in regions I and III .

We are   developing an extension of our work that encompasses the analysis of the Edelstein effect, which 
should appear in a forthcoming work, presently in progress.

 \section{Conclusions}
 \label{concl}  
 
By means of a systematic implementation of linear response theory applied to  charge- and spin-transport 
in disordered systems, we have   computed the sheet conductance, as well as  the spin-Hall conductance, in a multiband model of 
 the 2DEG at a LAO/STO interface,
 in the presence of a finite density of impurities.  Systematically computing the impurity-induced single-particle 
 self-energy corrections as well as the vertex corrections to the relevant diagrams, we have   
derived  the sheet conductance and the spin-Hall conductance as a function of the chemical 
potential in the 2DEG (which ultimately determines the  density of carriers supporting conduction). 
The sheet conductance shows an 
 excellent qualitative and quantitative agreement with the measurements presented in Ref. \cite{Trier2020}. 
 The spin-Hall conductance shows a similar excellent  qualitative and quantitative agreement with the 
experimental data as long as only carriers from ${xy}$ subbands contribute the corresponding response 
function. In particular, we recover the monotonic increase of the spin-Hall conductance, as a function of 
the chemical potential, until the Lifshitz transition is reached, as well as its main behavior 
after the transition.

Our results highlight how the charge-to-spin conversion in LAO/STO is mainly due to the spin-Hall effect, 
how the spin-Hall conductance (and, therefore, the efficiency of the charge-to-spin conversion mechanism
itself) depends on the external control parameters and, finally, how the inpurity-induced screening of the spin-Hall
conductance in a Rashba 2DEG is strongly suppressed in multiband systems. 

While the model captures the spin-Hall vs longitudinal conductance trend as seen in the experiment, the quantitative
 discrepancies around the Lifshitz transition definitely call  for a refinement of our method in that region, 
 not only by introducing a finite range in the impurity scattering, but also 
by  improving our estimate of the vertex corrections. For instance, one could include at the right-hand side of 
Eq.(\ref{ic.a2})  contributions from higher-energy bands. Moreover, one could solve numerically from   scratch the Kubo response
 function for the calculation of spin-Hall conductance including this way all the single-particle scatterings even in high density regimes.
  We expect that these additional effects could only slightly improve the evaluation of spin-Hall conductance around the Lifshitz transition, 
  since the densities close to the transition are not high (a few $10^{13}$ ${\rm cm}^{-2}$).     

An improvement of the  quantitative agreement beyond 
the Lifshitz transition, instead, might be possibly recovered by taking into account inelastic scattering processes, 
as well, which would be consistent with the estimate for the corresponding characteristic time 
as a function of the applied gate voltage in Fig. S4 of the Supplemental Material of Ref. \cite{Trier2020}.
  
Aside for the necessary improvements of our method, we apparently catch a number of 
features of the 2DEG in LAO/STO that can be hardly  recovered by means of approaches 
to the spin-Hall effect different from our ``fully quantum'' one.   Also, 
the magnitude and nonmonotonic tunability of the spin-Hall conductance in the LAO/STO 2DEG stems 
from  the multi-orbital nature of STO where the individual and unequal contributions from different subbands 
($d_{zx}$, $d_{xy}$, $d_{yz}$) result in a large carrier density (Fermi energy) dependence of the system.
 This subband character can be present in some topological insulators or  dichalcogenides, but it 
 is naturally not shared by all materials or other 
 two-dimensional electronic systems, like, e.g., graphene, therefore, even though it is possible
  to demonstrate spin-Hall conductance in such systems, the non-monotonic
   tunability of spin-Hall conductance will most likely remain a unique feature of STO-based 2DEGs \cite{Trier2022}.

Given 
the effectiveness of our technique, as a further extension of our work we plan to extend it to compute the nonlocal 
resistivity  of the LAO/STO in the presence  of an applied magnetic field, as well as to study charge-to-spin
conversion in alternative devices with similar properties, such as the recently discovered ones in which 
SrTiO$_3$ is replaced with the KTaO$_3$.

 \vspace{0.3cm} 
 
{\bf Acknowledgements:}   D.G., A.N., C.A.P., and M.S. 
  acknowledge   financial support  from Italy's MIUR  PRIN project  TOP-SPIN  (Grant No. PRIN 20177SL7HC). 
  A. N. acknowledges funding by the Deutsche Forschungsgemeinschaft (DFG, German Research Foundation) 
  under Germany's Excellence Strategy - Cluster of Excellence Matter and Light for Quantum Computing 
  (ML4Q) EXC 2004/1 - 390534769.  M.S. acknowledges    financial support from PNRR MUR project PE0000023-NQSTI.  
  F.T. acknowledges support by research grant 37338 (SANSIT) from Villum Fonden. 
       
\appendix 

\section{Review of linear response theory}
\label{lrt}

In this appendix we review the Kubo formulas  for  the response 
functions relevant to describing  charge and/or spin  transport, derived within linear response theory. 
 As specific applications,  in the following we consider the (Ohmic) sheet conductance, as well 
as the spin-Hall conductance. To   describe their behavior in realistic samples, 
in both cases we have to take into account the effects of a scattering off the 
impurities. In Appendix \ref{impures} we will therefore  complement the derivation of this appendix 
 with the analysis of the effects of a finite density of impurities.

Over-all, we compute   the average value of an observable ${\cal O}_1 ( \vec{r} , t )$,
$\langle {\cal O}_1 ( \vec{r} , t ) \rangle$,  in response 
to an applied electric field ${\cal E}(\vec{r},t)$ along
the $a$ direction in a two-dimensional sample, which couples to the 
charge current in the $a$ direction,  $j_{{\rm ch},a} ( \vec{r})$. Therefore, 
letting ${\cal A} ( \vec{r} , t )$ be the corresponding vector potential, 
such that $- \frac{ \partial {\cal A} ( \vec{r} , t ) }{ \partial t} = {\cal E} ( \vec{r} , t )$, a nonzero  ${\cal E} (\vec{r} , t)$ 
generates a ``source'' term in the   system Hamiltonian, $H_{\rm Source} ( t )$, given by

\beq
H_{\rm Source} ( t ) = - \int \: d \vec{r}' \: {\cal A} ( \vec{r}' , t ) j_{{\rm ch},a} ( \vec{r}' )
\;\;\;\;. 
\label{revr.1}
\eneq
\noindent
Computing $\langle {\cal O}_1 ( \vec{r} , t ) \rangle$ within first-order, time-dependent perturbation theory in the 
applied field, we obtain

\beq
\langle {\cal O}_1 ( \vec{r} , t ) \rangle =
 i \int \: d t' \: \int d \vec{r}' \: {\cal G}^R ( \vec{r} , t ; \vec{r}' , t' ) \:  {\cal A} ( \vec{r}' , t' ) 
\:\:\:\: , 
\label{revr.2}
\eneq
\noindent
with the retarded Green's function 

\beq
 {\cal G}^R ( \vec{r} , t ; \vec{r}' , t' ) = \theta ( t - t' )  \: \langle [{\cal O}_1 (\vec{r},t) , 
j_{{\rm ch},a} (\vec{r}',t')]\rangle
\;\;\;\; , 
\label{revr.3}
\eneq
\noindent
and $\langle \ldots \rangle$ denoting the equilibrium averages computed with respect to the ``unperturbed'' Hamiltonian 
(which we generically refer to as  $H_0$). 

To compute the response to a time-independent electric field, we first apply a  modulated   field at frequency $\omega_0$, 
${\cal E} ( \vec{r} , t ) = {\cal E} ( \vec{r} , \omega_0 ) \cos ( \omega_0 t )$. Defining $\langle {\cal O}_1 ( \vec{r} , \omega_0 ) \rangle$
to be the corresponding average value of ${\cal O}_1$, we obtain 
 
\beq
\langle {\cal O}_1 (\vec{r},\omega_0) \rangle  = \sum_{a = \pm 1} \: 
\frac{1}{ 2 \omega_0} \left\{i  \int d \vec{r}' \: {\cal G}^R (\vec{r},\vec{r}', a \omega_0) \:  {\cal E} (\vec{r}' , a \omega_0 ) \right\} 
\:\:\:\: , 
\label{revr.4}
\eneq
\noindent
with $G^R (\vec{r},\vec{r}',\omega) = \int \: d t \:  e^{ i \omega t} \: G^R ( \vec{r} , t ; \vec{r}' , 0 )$.

As a next step, we now consider the decomposition of ${\cal O}_1 (\vec{r},t)$ in terms of single-fermion 
operators. To do so, we denote with $\psi_{\vec{k},\lambda } e^{ i \vec{k} \cdot \vec{r} }$ a generic set of lattice momentum 
eigenfunctions determining a basis in our state space (note that the $\psi_{\vec{k} , \lambda}$ may not necessarily
be a set of eigenfunctions of $H_0$ and  if they are, then $\lambda$ can be regarded as a ``dressed'' band index), and with 
$\Gamma_{ \vec{k} , \lambda}$ the corresponding eigenmodes. Resorting to Heisenberg representation,
we assume that, in terms of the $\Gamma_{\vec{k} , \lambda}$,
 the operators ${\cal O}_1 (\vec{r},t)$  can be written as 
 
 \beq
 {\cal O}_1 ( \vec{r} , t ) = \frac{1}{V } \: \sum_{ \vec{k} , \vec{k}' } \: \sum_{\lambda , \lambda'} \: 
  e^{ - i \vec{r} \cdot [ \vec{k} - \vec{k}']} \: [ {\cal O}_1 ]_{(\vec{k} , \lambda ) ; ( \vec{k}' , \lambda' ) } \: 
  \Gamma_{\vec{k} , \lambda}^\dagger ( t ) \Gamma_{\vec{k}' , \lambda'} ( t ) 
  \:\:\:\: . 
  \label{recr.5}
  \eneq
  
  For an applied field at a fixed momentum $\vec{Q}$, that is, if ${\cal E} (\vec{r},\omega_0)   
  = {\cal E} (\vec{Q} , \omega_0) e^{ i \vec{r} \cdot \vec{Q}}$,  resorting to the 
  Keldysh-Green function approach, we obtain that $\langle {\cal O}_1 ( \vec{r} , \omega_0 ) \rangle
 = {\cal O}_1 ( \vec{Q} ; \omega_0 ) e^{ i \vec{r} \cdot \vec{Q}}$, with 
 
\begin{eqnarray}
&& {\cal O}_1 ( \vec{Q} ; \omega_0 ) =  \frac{e^{ - i \omega_0 t  - i \vec{Q} \cdot \vec{r} }}{2 V \omega_0 } \: \sum_{a = \pm 1} \: a 
 \sum_{ \vec{k} } \: \sum_{\lambda_1 , \lambda_1^{'}} \: \sum_{\lambda_2 , \lambda_2^{'}} \nonumber \\
&& \times  [ {\cal O}_1]_{(\vec{k} , \lambda_1 ) ; (
  \vec{k} + \vec{Q} , \lambda_1^{'} )} \:  [ j_{{\rm ch},a} ]_{(\vec{k} + \vec{Q} , \lambda_2 ) ; (
  \vec{k}  , \lambda_2^{'} )} \nonumber \\
  && \times  \int \frac{ d \bar{\omega}}{2 \pi}  \: \Biggl\{ 
 g_{ [  \vec{k} ; ( \lambda_2^{'} , \lambda_1 ) ]  }^< ( \bar{\omega} ) g_{ [ \vec{k}  + \vec{Q} ; (  \lambda_1^{'} , \lambda_2 )] }^R (
   \bar{\omega } +  a \omega_0 )   \nonumber \\
  && +  g_{ [  \vec{k} ; ( \lambda_2^{'} , \lambda_1 ) ]  }^A ( \bar{\omega} ) 
   g_{ [ \vec{k}  + \vec{Q} ; (  \lambda_1^{'} , \lambda_2 )] }^< ( \bar{\omega} + a \omega_0 )  \Biggr\} {\cal E} ( \vec{Q} , \omega_0 )
    \: , \nonumber \\
&& \label{recr.7}
\end{eqnarray}
\noindent
with $g_{[\vec{k} ; ( \lambda , \lambda' ) ]}^{(\eta , \eta' ) } ( \omega )$ being the Fourier transform of 
the Keldysh-Green function 

\beq
g_{[\vec{k} ; ( \lambda , \lambda' ) ]}^{(\eta , \eta' ) } (t ) = - i \langle {\bf T}_K \Gamma_{ \vec{k} , \lambda} ( t ; \eta ) 
\Gamma_{\vec{k} , \lambda'}^\dagger ( 0 ; \eta' ) \rangle 
\;\;\;\; , 
\label{recr.8}
\eneq
\noindent
and ${\bf T}_K$ being the Keldysh path-ordering product.  The dc response to a time-independent applied field
is eventually recovered by taking the $\omega_0 \to 0$ limit of the right-hand side of Eq.(\ref{recr.7}).

A remarkable simplification occurs if the subscript $\lambda$ labels the dressed energy eigenstates of 
$H_0$. In this case, the single-particle Green's functions are   diagonal in that index and 
Eq.(\ref{recr.7}) reduces to

\begin{eqnarray}
&& {\cal O}_1 (\vec{Q};\omega_0) = \frac{e^{ - i \omega_0 t  - i \vec{Q} \cdot \vec{r} }}{2 V \omega_0 } \: \sum_{a = \pm 1} \: a 
 \sum_{ \vec{k} } \: \sum_{\lambda  , \lambda^{'} } \times \nonumber \\
 &&  [ {\cal O}_1]_{(\vec{k} , \lambda ) ; (
  \vec{k} + \vec{Q} , \lambda^{'} )} \:  [ j_{{\rm ch},a} ]_{(\vec{k} + \vec{Q} , \lambda^{'}  ) ; (
  \vec{k}  , \lambda  )}  \times \nonumber \\
  && \int \frac{ d \bar{\omega}}{2 \pi}  \: \Biggl\{ 
 g_{ (  \vec{k}  , \lambda )   }^< ( \bar{\omega} ) g_{ (  \vec{k}  + \vec{Q} , \lambda^{'}   ) }^R (
   \bar{\omega } + a \omega_0 ) + \nonumber \\
   &&   g_{  (  \vec{k}  , \lambda )  }^A ( \bar{\omega} ) 
   g_{(  \vec{k}  + \vec{Q} , \lambda^{'}   )  }^< ( \bar{\omega} + a \omega_0 )  \Biggr\} {\cal E} ( \vec{Q} , \omega_0 )
    \:\:\:\: ,
\label{recr.9}
\end{eqnarray}
\noindent
In the case of a uniform applied field,  ${\cal E} ( \vec{Q} ; \omega_0 ) \propto \delta^{(2)} ( \vec{Q} )$, 
from Eq.(\ref{recr.9}) we define the DC   response function $\Sigma_{1,2}$ according to

 \begin{eqnarray} 
  \Sigma_{1,2} = && \lim_{\omega_0 \to 0 } \Biggl\{  \frac{1}{2 V   \omega_0 } \: 
 \sum_{a = \pm 1} \: a 
 \sum_{ \vec{k}  } \: \sum_{\lambda  , \lambda^{'} } \times 
 \nonumber \\ && [ {\cal O}_1]_{(\vec{k} , \lambda ) ; (
  \vec{k}   , \lambda^{'} )}    [ j_{{\rm ch},a} ]_{(\vec{k}  , \lambda^{'}  ) ; (
  \vec{k}  , \lambda  )}  \times \nonumber \\
  && \int \frac{ d \bar{\omega}}{2 \pi}  \: \Biggl\{ 
 g_{ (  \vec{k}  , \lambda )   }^< ( \bar{\omega} ) g_{ (  \vec{k}   , \lambda^{'}   ) }^R (
   \bar{\omega } +  a \omega_0 ) + \nonumber \\
   &&  g_{  (  \vec{k}  , \lambda )  }^A ( \bar{\omega} ) 
   g_{(  \vec{k}    , \lambda^{'}   )  }^< ( \bar{\omega} + a \omega_0 )   \Biggr\}    \:\:\:\: .
   \label{recr.12}
\end{eqnarray}
\noindent
To further simplify Eq.(\ref{recr.12}), we set

\begin{eqnarray}
&& [ {\cal S}_{1,2} ]_{(\vec{k} , \lambda )  ; ( \vec{k}' , \lambda' ) } = \Re e \{ [{\cal O}_1]_{(\vec{k}  \lambda ) ; (
  \vec{k}^{'}   , \lambda^{'} )} \:  [j_{{\rm ch},a} ]_{(\vec{k}^{'}   , \lambda^{'}  ) ; (
  \vec{k}  , \lambda )}  \} \; ,
  \nonumber \\
&& [ {\cal I}_{1,2} ]_{(\vec{k} , \lambda )  ; ( \vec{k}' , \lambda' ) } =\Im m  \{[{\cal O}_1]_{(\vec{k}  \lambda ) ; (
  \vec{k}^{'}   , \lambda^{'} )} \:  [j_{{\rm ch},a} ]_{(\vec{k}^{'}   , \lambda^{'}  ) ; (
  \vec{k}  , \lambda )}  \}  \:\:\:\: .  \nonumber \\ 
 \label{recr.11}
 \end{eqnarray}
 \noindent
 [Note that, if ${\cal O}_1 = j_{{\rm ch},a}$, then $  [ {\cal I}_{1,2} ]_{(\vec{k} , \lambda )  ; ( \vec{k}' , \lambda' ) } = 0 $.]
  Taking into account the splitting  in Eqs.(\ref{recr.11}), we recast Eq.(\ref{recr.12}) in the form 
 
 \beq
 \Sigma_{1,2}  = \Sigma_{1,2}^{\cal S}  + \Sigma_{1,2}^{\cal I}  
 \;\;\;\; , 
 \label{recr.13}
 \eneq
 \noindent
 with

 \begin{eqnarray}
 && \Sigma_{1,2}^{\cal S}  =- \frac{1}{2 V  } \: 
 \sum_{ \vec{k}  } \: \sum_{\lambda  , \lambda^{'} } \:[ {\cal S}_{1,2} ]_{(\vec{k} , \lambda )  ; ( \vec{k}  , \lambda' ) } 
 \int \frac{ d \bar{\omega}}{2 \pi}   \times 
\label{recr.16} \\
  &&  [   g_{ (  \vec{k}  , \lambda )   }^A ( \bar{\omega} ) - g_{ (  \vec{k}   , \lambda  ) }^R (  \bar{\omega } ) ]
   [   g_{ (  \vec{k}  , \lambda' )   }^A ( \bar{\omega}  ) - g_{ (  \vec{k}   , \lambda'  ) }^R (  \bar{\omega }  )) ]
\frac{ \partial f ( \bar{\omega} ) }{ \partial \bar{\omega} } 
     \:\:\:\: , \nonumber 
\end{eqnarray}
\noindent
  and 
  
\begin{eqnarray}
 && \Sigma_{1,2}^{\cal I} =- \frac{i}{ 2 V  } \: 
 \sum_{ \vec{k}  } \: \sum_{\lambda  , \lambda^{'} } \:[ {\cal I}_{1,2} ]_{(\vec{k} , \lambda )  ; ( \vec{k}  , \lambda' ) } 
 \int \frac{ d \bar{\omega}}{2 \pi}   \times \label{recr.17} \\
  &&  [   g_{ (  \vec{k}  , \lambda )   }^A ( \bar{\omega} ) - g_{ (  \vec{k}   , \lambda  ) }^R (  \bar{\omega } ) ]
 \partial_{\bar{\omega}}   [   g_{ (  \vec{k}  , \lambda' )   }^A ( \bar{\omega}  ) + g_{ (  \vec{k}   , \lambda'  ) }^R (  \bar{\omega }  )) ]
  f ( \bar{\omega} )    
     \:\:\:\: .
     \nonumber 
\end{eqnarray}
\noindent
Equations (\ref{recr.16}) and (\ref{recr.17}), pertinently improved by introducing the effects of a finite impurity density, are
the main formulas we used throughout our paper to derive the response functions of our system. 

 \section{The effects of a finite density of impurity scattering centers}
\label{impures}

In this appendix we   review  the impurity-related corrections to the
Kubo-conductance formulas derived in the clean limit. 

Following \cite{Mishchenko2004,Inoue2004,Khaetskii2006,Raimondi2004,Dimitrova2005}, in the following we employ a simple
model of short-range, uncorrelated impurity scatterers.  Accordingly, we encode the effects 
of the disorder on our system with the   impurity potential 

\beq
V_{\rm Imp}  ( \vec{r} ) = \sum_{\vec{R}_i } \: U ( \vec{r} - \vec{R}_i ) 
\:\:\:\: , 
\label{ims.3}
\eneq
\noindent
with  $ \{ \vec{R}_i \}$ being the (randomly distributed over the plane) impurity centers. 
 Denoting with an overbar
the ensemble average with respect to the position of the impurity centers 
and with $ V_{\rm Imp}  ( \vec{Q} )$ the Fourier transform of $V_{\rm Imp} ( \vec{r} )$, defined as 

\beq
V_{\rm Imp}  (\vec{Q}) = \int \: d^2 \vec{r} \: e^{ i \vec{Q} \cdot \vec{r}} \: V_{\rm Imp} ( \vec{r} ) 
\;\;\;\; , 
\label{addi.1}
\eneq
\noindent
we set 
 
 \begin{eqnarray}
  \overline{ V_{\rm Imp} ( \vec{Q} ) } &=& \int \: d^d \vec{r} \: e^{ i \vec{r} \cdot \vec{Q}} \: U ( \vec{r} )  
\: \sum_{ \vec{R}_i } \overline{ e^{ i \vec{R}_i \cdot \vec{Q} } }    \nonumber \\
&=&  N_{\rm Imp} \: \int \: d^d \vec{r} \: e^{ i \vec{r} \cdot \vec{Q}} \: U ( \vec{r} )  
\: \delta_{ \vec{Q}} \label{ims4aa}  \;\;\;\; ,
\end{eqnarray}
\noindent
and 

\begin{eqnarray}
&& \overline{ V_{\rm Imp} ( \vec{Q} ) V_{\rm Imp} ( \vec{Q}' ) } =\nonumber \\
&&  \int\: d^d \vec{r} \: \int \: d^d \: \vec{r}^{'} \: 
e^{ i \vec{r} \cdot \vec{Q} } e^{ i \vec{r}' \cdot \vec{Q}'} \: U ( \vec{r} ) U ( \vec{r}' ) \: \sum_{\vec{R}_i , \vec{R}_i^{'}}  \overline{ 
e^{ i \vec{R}_i \cdot \vec{Q} } e^{ i \vec{R}_i^{'} \cdot \vec{Q}^{'}} } = \nonumber \\
&& N_{\rm Imp} ( N_{\rm Imp} - 1 ) \:  \int\: d^d \vec{r} \: \int \: d^d \: \vec{r}^{'}  
  \: U ( \vec{r} ) U ( \vec{r}' ) \: \delta_{\vec{Q}}\:  \delta_{\vec{Q}'} +  \nonumber \\
&& N_{\rm Imp} \:  \int\: d^d \vec{r} \: \int \: d^d \: \vec{r}^{'} \: 
e^{ i \vec{r} \cdot \vec{Q} } e^{ i \vec{r}' \cdot \vec{Q}'} \: U ( \vec{r} ) U ( \vec{r}' ) \: \delta_{\vec{Q}+ \vec{Q}'} 
\: ,
\label{ims.4a}
 \end{eqnarray}
\noindent
with $N_{\rm Imp}$ being the number of impurity scattering centers in the system. 

A minimal model describing the coupling between the impurities and the band 
electrons is provided by  the impurity Hamiltonian $H_{\rm Imp}$ given by   
 
\begin{eqnarray}
H_{\rm Imp} &=& \int \: d^2 r \: V_{\rm Imp}  ( \vec{r} ) \: \sum_\sigma c_\sigma^\dagger ( \vec{r} ) c_\sigma ( \vec{r} ) \nonumber \\
 &=& 
\frac{1}{V} \: \sum_{\vec{k },\vec{k}'} \: V_{\rm Imp} ( - \vec{k} + \vec{k}' ) \: \sum_{\sigma } c_{\vec{k} , \sigma}^\dagger c_{\vec{k}' , \sigma}
\:\:\:\: , 
\label{adimp.2}
\end{eqnarray}
\noindent
with $\sigma$ being a generic band index, including (but not necessarily coinciding with) the actual spin index. 
For the following applications, it is also useful to express the right-hand side of Eq.(\ref{adimp.2}) in terms 
of the eigenmodes $\Gamma_{\vec{k} , \lambda}$ of the (clean) system Hamiltonian, with $\lambda$ being 
the (dressed) band index. The correspondence between the operators $c_{\vec{k} , \sigma}$ and the 
$\Gamma_{\vec{k} , \lambda}$ is given by 
 
\beq
c_{\vec{k} , \sigma}  =  \sum_\lambda \: w_{\sigma , \lambda}^{\vec{k}} \: \Gamma_{\vec{k},\lambda} 
\:\:\:\: , 
\label{adimp.3}
\eneq
\noindent
with the matrix $\{ w_{\sigma , \lambda}^{\vec{k}} \}$ being unitary, at any given $\vec{k}$. As a result, 
  Eq.(\ref{adimp.2}) can be rewritten as

 \begin{eqnarray}
&& H_{\rm Imp} =
\label{adimp.4} \\
&& \frac{1}{V} \sum_{\vec{k } , \vec{k}' } V_{\rm Imp} ( - \vec{k} + \vec{k}' )  \sum_{\lambda , \lambda'}   
  \sum_{\sigma } [ w_{\sigma , \lambda}^{\vec{k} } ]^* w_{\sigma , \lambda'}^{\vec{k}'} ]  
\Gamma_{\vec{k} , \lambda}^\dagger \Gamma_{\vec{k}' , \lambda'} \nonumber \\
&& \equiv \frac{1}{V}   \sum_{\vec{k } , \vec{k}' }  V_{\rm Imp} ( - \vec{k} + \vec{k}' ) \: \sum_{\lambda , \lambda'}   {\cal A}_{(\vec{k} , \lambda) ; (\vec{k}' , \lambda')} 
\Gamma_{\vec{k} , \lambda}^\dagger \Gamma_{\vec{k}' , \lambda'} \nonumber 
 \:\:\:\: . 
\end{eqnarray}
\noindent
In the following, unless explicitly stated otherwise, we   will  resort to the approximation of short-range impurity scattering potential, that is, 
we will assume $U ( \vec{r} ) = \bar{u}\delta ( \vec{r} )$.

We now discuss the main effects of having a finite impurity density: the impurity induced  single-fermion self-energy imaginary part
and the interaction vertex renormalization. 

\subsection{Impurities and self-energy imaginary part}
\label{selfeim}

To discuss the impurity-induced self-energy imaginary part, 
here we consider a generic  disordered Hamiltonian  given by 
 
\beq
H = H_0 + H_{\rm Imp} 
\;\;\;\; , 
\label{ims.1}
\eneq
\noindent
with $H_0 = \sum_{\vec{k}} \: \sum_\lambda \: \epsilon_{ \vec{k} , \lambda } \: \Gamma_{\vec{k} , \lambda}^\dagger \Gamma_{\vec{k} , \lambda}$,
$\epsilon_{\vec{k} , \lambda}$ being the (dressed) energy dispersion relations 
of the system, and $H_{\rm Imp}$ given in Eq.(\ref{adimp.4}). In the  
absence of impurities,  the retarded and advanced Green's functions for the eigenmode belonging to the energy 
eigenvalue $E_{\vec{k} , \lambda}$, $g^{R/A ; (0) }_{(\vec{k} , \lambda )} ( t )$, are respectively 
given by

\begin{eqnarray}
g^{R ; (0)}_{(\vec{k} , \lambda )} ( t ) &=& - i \theta ( t ) \: \langle \{ \Gamma_{\vec{k} , \lambda} ( t ) , 
\Gamma_{\vec{k} , \lambda}^\dagger ( 0 ) \}\rangle \; , \nonumber \\
g^{A ; (0)}_{(\vec{k} , \lambda )} ( t ) &=& i \theta ( - t ) \: \langle \{ \Gamma_{\vec{k} , \lambda} ( t ) , 
\Gamma_{\vec{k} , \lambda}^\dagger ( 0 ) \}\rangle 
\;\;\;\; ,
\label{ems.x1}
\end{eqnarray}
\noindent
in the time domain, and by   
 
 \beq
 g^{ R / A ; (0)}_{(\vec{k} , \lambda)} ( \omega ) = \frac{1}{ \omega - \xi_{\vec{k} , \lambda} \pm i \eta} 
 \;\;\;\; , 
 \label{ems.x2}
 \eneq
 \noindent
 in Fourier space, with $\xi_{\vec{k} , \lambda} = \epsilon_{\vec{k} , \lambda} - \mu$, $\mu$ being the chemical potential,
 and $\eta = 0^+$. At a fixed realization of the disorder (that is, at a given $V_{\rm Imp}$), computing the self-energy correction
 to the retarded and to the advanced Green's functions,   $\Sigma^{R/A}_{(\vec{k} , \lambda) ; (\vec{k}' , \lambda')} ( \omega)$, yields, 
 to first order in $V_{\rm Imp}$, the result

\beq
\hat{\Sigma}^{R/A ; (I)}_{(\vec{k} , \lambda) ; (\vec{k}' , \lambda')} ( \omega) = 
\frac{\delta_{\lambda , \lambda'}  }{ V } V_{\rm Imp} ( - \vec{k} + \vec{k}' ) \; {\cal A}_{(\vec{k} , \lambda) ; (\vec{k}' , \lambda' ) } 
\;\;\;\; ,
\label{ims.6a}
\eneq
\noindent
with $V$ being the total volume of the system and    ${\cal A}_{(\vec{k} , \lambda ) ; (\vec{k} , \lambda')} = \sum_\sigma
 \: [ w_{\sigma , \lambda}^{\vec{k} } ]^* w_{\sigma , \lambda'}^{\vec{k}}$ . 
On  ensemble-averaging the result in Eq. (\ref{ims.6a}) over the impurity distribution  according to Eqs. (\ref{ims.4a}),  one 
obtains 

\beq
\Sigma^{R/A ; (I)}_{(\vec{k} , \lambda) ; (\vec{k}' , \lambda')} ( \omega) = \overline{ \hat{\Sigma}^{R/A ; (I)}_{(\vec{k} , \lambda) ; (\vec{k}' , \lambda')} ( \omega) }  = 
N_{\rm Imp} \bar{u}_\lambda  \; \delta_{-\vec{k} + \vec{k}' } \: \delta_{\lambda , \lambda' } 
\:\:\:\: , 
\label{ims.7a}
\eneq
\noindent
with $\bar{u}_\lambda = U ( \vec{Q} = 0 )$ [note that, in going through the last step of Eq. (\ref{ims.7a}), we 
employed the identity  ${\cal A}_{(\vec{k} , \lambda ) ; (\vec{k} , \lambda')} =  \delta_{\lambda  , \lambda'}  $, which is 
a trivial consequence of the orthogonality of the eigenvectors of the Hamiltonian at fixed $\vec{k}$]. 
The (first-order in the impurity interaction) correction in Eq. (\ref{ims.7a}) just implies a uniform shift in the offset of the dressed band 
$\lambda$. 

In considering  the second-order contribution to the self-energy we note that  the former term at the right-hand side 
of Eq.(\ref{ims.4a}) can be  lumped into an additional 
constant correction to the real part of the self-energy. At variance, the latter contribution  
 yields a second-order contribution to the self-energy that, at fixed realization of the disorder, 
 is given by 

\begin{eqnarray}
  && \hat{\Sigma}^{R/A ; (II)}_{(\vec{k} , \lambda) ; (\vec{k}' , \lambda')} ( \omega)  =  \frac{1}{V} \: 
\sum_{\vec{q}  } \: \sum_\mu  \{  
\: V_{\rm Imp}  ( - \vec{k} + \vec{q} ) V_{\rm Imp}  ( - \vec{q} + \vec{k}' )  \nonumber \\
&&  \times {\cal A}_{(\vec{k} , \lambda ) ; (\vec{q} , \mu )} 
{\cal A}_{(\vec{q} , \mu ) ; (\vec{k}' , \lambda' ) } \: g_{( \vec{q} , \mu) }^{R/A ; (0)} ( \omega ) \}
\:\:\:\: .
\label{ims.9a}
\end{eqnarray}
\noindent
Averaging over the impurities and leaving aside terms that just provide a further renormalization to 
the uniform part of the self-energy, we obtain 

\begin{eqnarray}
&& \Sigma^{R/A ; (II)}_{(\vec{k} , \lambda) ; (\vec{k}' , \lambda')} ( \omega) = \overline{ 
\hat{\Sigma}^{R/A ; (II)}_{(\vec{k} , \lambda) ; (\vec{k}' , \lambda')} ( \omega) } =\frac{N_{\rm Imp} }{V} \: 
\sum_{\vec{q}  } \: \sum_\mu \times \label{ims.10a} \\
&&  \{  
\: |  U  ( - \vec{k} + \vec{q} ) |^2 \:  g_{( \vec{q} , \mu )}^{R/A ; (0)} ( \omega ) \: {\cal A}_{(\vec{k} , \lambda ) ; (\vec{q} , \mu )}  
\: {\cal A}_{(\vec{q} , \mu ) ; (\vec{k} , \lambda')} \}  \:  \delta_{- \vec{k} + \vec{k}' } 
\:\:\:\: . 
\nonumber 
\end{eqnarray}
\noindent
Assuming   $U ( \vec{r} ) = u \delta ( \vec{r} ) $, Eq.(\ref{ims.10a}) simplifies to 

\begin{eqnarray}
&& \Sigma^{R/A ; (II)}_{(\vec{k} , \lambda) ; (\vec{k}' , \lambda')} ( \omega) \approx 
\frac{N_{\rm Imp} \: \bar{u}^2  }{V} \: 
\sum_{\vec{q}  } \: \sum_\mu \times \nonumber \\
&& \{   {\cal A}_{(\vec{k} , \lambda ) ; (\vec{q} , \mu )}  \: 
{\cal A}_{(\vec{q} , \mu ) ; (\vec{k} , \lambda')} 
 g_{( \vec{q} , \mu )}^{R/A ; (0)} ( \omega ) \} \:  \delta_{- \vec{k} + \vec{k}' } 
\: . 
\label{ims.10b}
\end{eqnarray}
\noindent
Focusing on the right-hand side of Eq.(\ref{ims.10b}) we note that impurity-triggered 
inter-band scattering processes can, in principle, take place both as {\it virtual}  processes 
(that is, $\mu \neq \lambda , \lambda'$), as well as {\it real} processes, yielding, at 
a given $\vec{k}$, an inter-band finite transition amplitude, corresponding to having 
$\lambda' \neq \lambda$. However, as discussed in detail in Ref. \cite{Schwab2002}, 
when computing the self-energy corrections in the weak impurity scattering limit, one may safely 
neglect inter-band real transition. Based on this observation, in the following we will assume
$\lambda = \lambda'$ at the right-hand side of Eq.(\ref{ims.10b}). Now, 
by definition, ($\frac{1}{2}$ times) the inverse particle lifetime due to a finite impurity concentration, 
$\tau_{\rm Imp}$, is given by the imaginary part of the self-energy. In our further derivation we 
will neglect the dependence on
$\omega$ [which is equivalent to only taking into account elastic scattering processes  at
the impurity sites (see the main text for our motivation of such an approximation), given the 
experimental results of Ref. \cite{Trier2020}]. Accordingly, reabsorbing the real part of the self-energy in a ``trivial'' shift of 
$\mu$, we assume for the dressed (by the interaction with the impurities) Green's function 
the form 

\beq
g_{(\vec{k} , \lambda)}^{R/A} ( \omega ) = \frac{1}{ \omega - \xi_{\vec{k} , \lambda} \pm \frac{i}{ 2 \tau^\lambda_{\rm Imp}}} 
\:\:\:\: .
\label{ims.13}
\eneq
\noindent

    \begin{figure}
 \center
\includegraphics*[width=0.8 \linewidth]{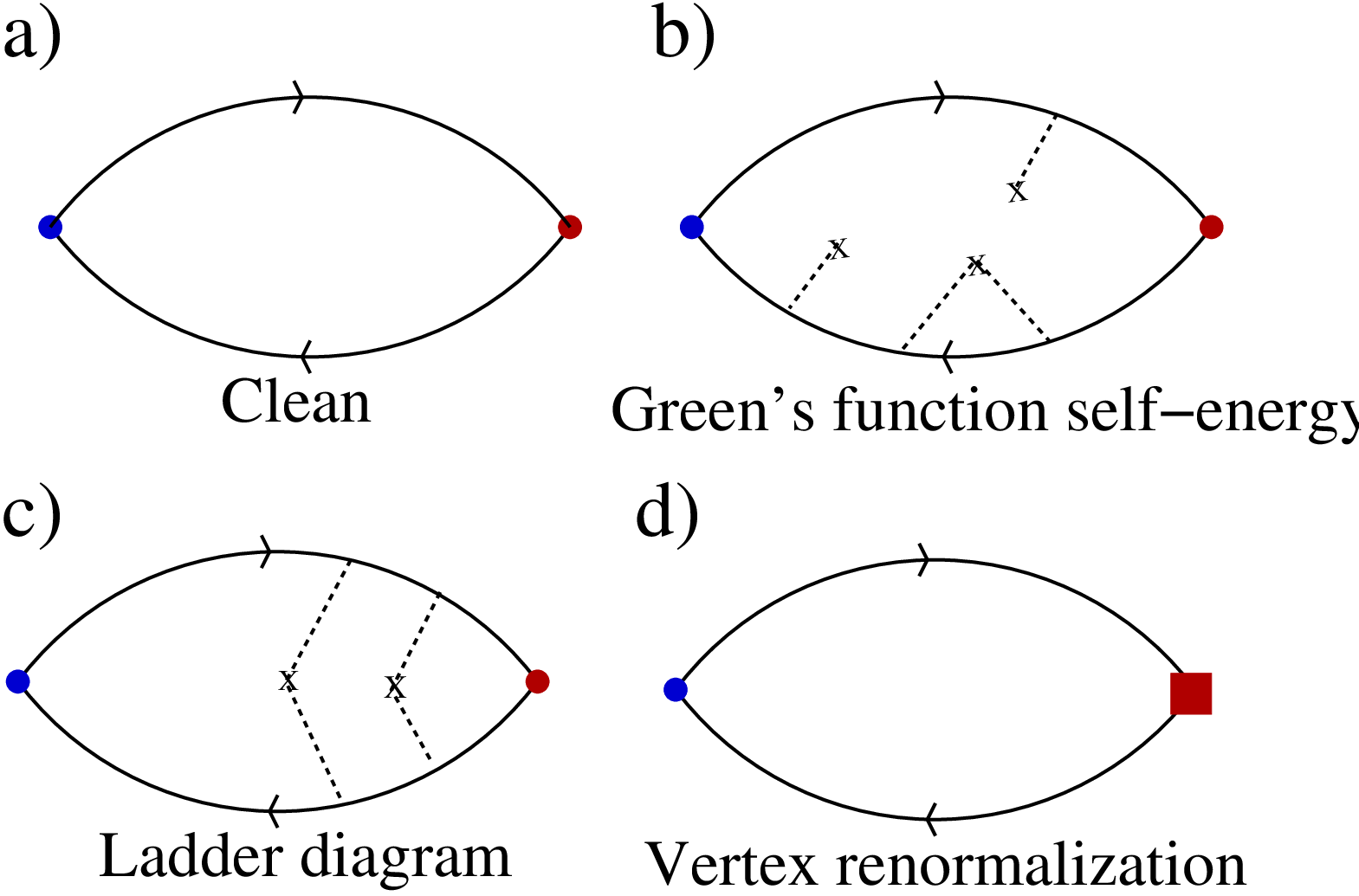}
\caption{ {\bf (a)}: Feynman diagram (``bubble'') representing the 
 conductance tensor computed n the absence  of impurity effects.  
 {\bf (b)}: Sketch of the typical diagrams contributing to the self-energy 
 renormalization (finite $\tau_{\rm Imp}^{-1}$).  
 {\bf (c)}: Sketch of typical diagrams (``ladder'') contributing to the vertex renormalization. 
 {\bf (d)}: Pictorial representation of the renormalization of the spin-hall current vertex 
 (red square).  
} 
\label{bubble.1}
\end{figure}

In principle the inverse lifetime $\tau_{\rm Imp}^\lambda$ (that is, the imaginary part of the single-electron 
self-energy) should be computed separately for each band. To second order in the impurity interaction strength, 
it is 
determined by the condition 
 
\begin{eqnarray}
&& \frac{1}{2 \tau_{\rm Imp}^\lambda} = - \frac{N_{\rm Imp} \: \bar{u}^2  }{V} \: 
\sum_{\vec{q}  } \: \sum_\mu \times \nonumber \\
&&   \{   {\cal A}_{(\vec{k} , \lambda ) ; (\vec{q} , \mu )}  \: 
{\cal A}_{(\vec{q} , \mu ) ; (\vec{k} , \lambda)}  \: \Im m \{ 
 g_{( \vec{q} , \mu )}^{R  ; (0)} ( \omega ) \}  \} 
\:\:\:\: . 
\label{ims.10c}
\end{eqnarray}
\noindent
Substituting, in the right-hand side of Eq.(\ref{ims.10c}), $ g_{( \vec{q} , \mu )}^{R  ; (0)} ( \omega )$
with  $g_{( \vec{q} , \mu )}^{R  } ( \omega )$ yields a set of self-consistent equations for the 
$\tau_{\rm Imp}^\lambda$. In general, solving, even numerically, the full set of equations is quite 
a formidable task to achieve, especially because, due to the dependence on the momenta of 
${\cal A}_{(\vec{k} , \lambda) ; (\vec{q} , \mu )}$, one cannot exclude {\it a priori} an explicit dependence 
of $\tau_{\rm Imp}^\lambda$ on the momentum $\vec{k}$, as well. This  eventually reflects into 
the formulas for the response functions, with a corresponding substantial increase in the computational 
complexity of the problem. For this reason it is compelling to find pertinent approximations that might 
possibly simplify the calculation of the imaginary part of the single-particle self-energy. In some cases,
such as in the two-band Rashba model, the calculation simplifies with no further approximations, due to the fact 
that summing over $\vec{q}$ allows for $\tau_{\rm Imp}$ fully getting rid of any dependence on both 
$\vec{k}$ and $\lambda$ (see Appendix \ref{lat2Rashba} for details). In the eight-band model the situation is not so 
simple and, as we discuss in the following, we have to resort to some ``educated''  approximations to 
make our problem tractable. 
   
\subsection{Impurity induced vertex renormalization}
\label{impver}

To derive the impurity induced vertex renormalization, we consider
 the diagrams contributing a specific conductance tensor 
  {\it in the absence} of impurity effects, such as 
 the ``bubble'' diagram reported in Fig. \ref{bubble.1}{\bf (a)}, with the 
colored dots representing the vertex insertions respectively corresponding to the 
operator coupled to the external source field (blue dot, in our formalism this
corresponds to $[ j_{{\rm ch},a} ]_{( \vec{k} , \lambda ) ; ( \vec{k}' , \lambda' ) }$), 
and the vertex insertion represented by the red  dot corresponding to the specific 
operator that we measure (in our case this corresponds to $[ {\cal O}_1]_{( \vec{k} , \lambda ) ; 
( \vec{k}' , \lambda' )}$). Next, we ``dress'' the retarded and the advanced single-particle Green's 
functions with  the self-energy corrections described in the previous subsection. 
Diagrammatically, this amounts to adding to the ``bare'' bubble the contributions of 
diagrams such as the ones sketched in Fig. \ref{bubble.1}{\bf (b)}.  Finally, we introduce diagrams such as the ones in 
Fig. \ref{bubble.1}{\bf (c)}, which renormalize the vertices, as well.  Carefully doing the calculation 
and singling out only the total contributions arising from diagrams such as those in Fig. \ref{bubble.1}{\bf (c)}, 
it is possible to recover the corresponding contribution to the fermion bubble that yields the 
required vertex renormalization. For a generic pair of bilinear operators in the fermionic fields, 
${\cal O}_1 ( \vec{r} , t )$ and ${\cal O}_2 ( \vec{r}' , t' )$, we have therefore to recover 
the vertex renormalization from the retarded Green's function 
$\theta ( t - t' ) \: \langle [ {\cal O}_1 ( \vec{r} , t ) , {\cal O}_2 ( \vec{r}' , t' ) ] \rangle$.  

We now consider the impurity-induced renormalization of the equal-momentum vertex  corresponding to 
an insertion of a generic operator  ${\cal O}_1 ( \vec{r} , t )$. Taking into account 
the impurity-induced imaginary part of the self-energy,  we obtain the modified version of 
Eq.(\ref{recr.12}) in the presence of a finite density of impurities, that is 

\begin{widetext}

\begin{eqnarray}
&& \Sigma_{1,2}= \lim_{\omega_0 \to 0} \sum_{a = \pm 1} \frac{a}{2 \omega_0}
 \Biggl\{ \frac{1}{V} \: \sum_{\vec{k}} \: \sum_{ \lambda , \lambda'} \: \int \: \frac{ d \bar{\omega}}{2 \pi} 
\: [{\cal O}_1 ]_{(\vec{k} , \lambda) ; ( \vec{k} , \lambda' ) } [{\cal O}_2]_{( \vec{k} , \lambda' ) ; ( \vec{k } , \lambda) } \times \nonumber \\
&& \left\{ f ( \bar{\omega} ) [ g_{(\vec{k} , \lambda )}^R ( \bar{\omega) } -  g_{(\vec{k} , \lambda )}^A ( \bar{\omega) }]
g_{(\vec{k} , \lambda' )}^R ( \bar{\omega} +a  \omega_0 ) + 
f ( \bar{\omega}+a  \omega_0 ) g_{(\vec{k} , \lambda)}^A ( \bar{\omega}) [ g_{( \vec{k} , \lambda' )}^R ( \bar{\omega} + 
a \omega_0 ) -  g_{( \vec{k} , \lambda' )}^A ( \bar{\omega} + 
a \omega_0 ) ] \right\} \nonumber \\
&& + \frac{1}{V} \: \sum_{\vec{k}} \: \sum_{ \lambda , \lambda'} \: \int \: \frac{ d \bar{\omega}}{2 \pi} 
\: [\delta {\cal O}_1 ]_{( \vec{k} ,  \lambda ) ; (\vec{k} ,  \lambda' ) }^{( \bar{\omega} , \bar{\omega} + a \omega_0) ; [R,R]}
 [{\cal O}_2]_{( \vec{k} , \lambda' ) ; ( \vec{k } , \lambda) }  \:  f ( \bar{\omega} )  g_{(\vec{k} , \lambda )}^R ( \bar{\omega) }  
g_{(\vec{k} , \lambda' )}^R ( \bar{\omega} +a \omega_0 ) 
\nonumber \\ && - 
 \frac{1}{V} \: \sum_{\vec{k}} \: \sum_{ \lambda , \lambda'} \: \int \: \frac{ d \bar{\omega}}{2 \pi} 
\: [\delta {\cal O}_1 ]_{(  \vec{k} , \lambda ) ; ( \vec{k} ,  \lambda' ) }^{( \bar{\omega} , \bar{\omega} + a \omega_0) ; [A,A] } [{\cal O}_2]_{( \vec{k} , \lambda' ) ; ( \vec{k } , \lambda) } \: 
 f ( \bar{\omega}  + \omega_0)  g_{(\vec{k} , \lambda )}^A( \bar{\omega) }  
g_{(\vec{k} , \lambda' )}^A( \bar{\omega} +a  \omega_0 ) \nonumber \\
&& + \frac{1}{V} \: \sum_{\vec{k}} \: \sum_{ \lambda , \lambda'} \: \int \: \frac{ d \bar{\omega}}{2 \pi} 
\: [ \delta {\cal O}_1 ]_{( \vec{k} , \lambda ) ; (\vec{k} ,  \lambda' ) }^{( \bar{\omega} , \bar{\omega} +a  \omega_0) ; [A,R]} [{\cal O}_2]_{( \vec{k} , \lambda' ) ; ( \vec{k } , \lambda) }\: 
 [ f ( \bar{\omega}  + a \omega_0)  - f ( \bar{\omega} ) ] g_{(\vec{k} , \lambda )}^A( \bar{\omega) }  
g_{(\vec{k} , \lambda' )}^R ( \bar{\omega} + a \omega_0 ) \Biggr\} 
\;\;\;\; ,
\label{resp.x2}
\end{eqnarray}
\noindent
\end{widetext}
with the vertex correction $ [\delta {\cal O}_1 ]_{( \vec{k} , \lambda  ) ; (\vec{k} ,  \lambda' ) }^{( \bar{\omega} , \bar{\omega} + \omega_0) ; [X,Y] } $ 
($X,Y=A,R$) satisfying the equation 

\beq
 [\delta {\cal O}_1 ]_{( \vec{k} , \lambda  ) ; (\vec{k} ,  \lambda' ) }^{( \bar{\omega} , \bar{\omega} + \omega_0) ; [X,Y] }  = 
 \sum_{ n = 1}^\infty  
\{   [\delta {\cal O}_1 ]_{( \vec{k} , \lambda ); (\vec{k}   , \lambda' ) }^{( \bar{\omega} , \bar{\omega} + \omega_0) ; [X,Y] } \}_{(n)}
\;\;\;\; ,
\label{resp.x3}
\eneq
\noindent
with 
 
\begin{eqnarray}
&& \{   [\delta {\cal O}_1 ]_{( \vec{k} , \lambda ); (\vec{k}   , \lambda' ) }^{( \bar{\omega} , \bar{\omega} + \omega_0) ; [X,Y] } \}_{(1)} = 
 \frac{ \bar{u}^2 N_{\rm Imp} }{V} \: \sum_{\vec{q} } \; \sum_{\mu , \mu' }  {\cal A}_{(\vec{k} , \lambda) ; (\vec{q} , \mu ) } 
\times  \nonumber \\
&&g_{(\vec{q} , \mu)}^X ( \bar{\omega} )   [{\cal O}_1]_{(\vec{q} , \mu ) ; (\vec{q} , \mu' ) } 
g_{(\vec{q} , \mu' ) }^Y ( \bar{\omega} + \omega_0 ) \: {\cal A}_{(\vec{q} , \mu' ) ; ( \vec{k} , \lambda' ) } 
\label{resp.x4a}
\; , 
\end{eqnarray}
\noindent

and 

\begin{eqnarray}
&& \{   [\delta {\cal O}_1 ]_{( \vec{k} , \lambda ); (\vec{k}   , \lambda' ) }^{( \bar{\omega} , \bar{\omega} + \omega_0) ; [X,Y] } \}_{(n+1)} =
\label{resp.x4b} \\
&& \frac{ \bar{u}^2 N_{\rm Imp} }{V} \: \sum_{\vec{q} } \; \sum_{\mu , \mu' } \: {\cal A}_{(\vec{k} , \lambda) ; (\vec{q} , \mu ) } 
 g_{(\vec{q} , \mu)}^X ( \bar{\omega} ) \times \nonumber \\
 &&  \{  [\delta {\cal O}_1]_{(\vec{q} , \mu ) ; (\vec{q} , \mu' ) }^{(\bar{\omega} , \bar{\omega} + \omega_0 ) ; [X,Y]} \}_{(n)}  
g_{(\vec{q} , \mu' ) }^Y ( \bar{\omega} + \omega_0 ) \: {\cal A}_{(\vec{q} , \mu' ) ; ( \vec{k} , \lambda' ) }  \nonumber 
\;\;\;\; . 
\end{eqnarray}
\noindent
To formally solve the iterative system in Eqs.(\ref{resp.x4a}) and (\ref{resp.x4b}), we set 

\begin{eqnarray}
&& \{   [\delta {\cal O}_1 ]_{( \sigma ); (\sigma' ) }^{( \bar{\omega} , \bar{\omega} + \omega_0) ; [X,Y] } \}_{(n)} = 
\label{resp.x5}  \\
&& \sum_{\vec{k} } \: 
\sum_{\lambda  , \lambda'} 
[ w^{\vec{k}}_{\sigma , \lambda} ]^* w_{\sigma' , \lambda'}^{\vec{k}}   \{  [\delta {\cal O}_1 ]_{ ( \sigma , \sigma' ) }^{( \bar{\omega} , \bar{\omega} + \omega_0) ; [X,Y] } \}_{(n)}
\; ,  \nonumber 
\end{eqnarray}
\noindent
which enables us to respectively rewrite Eqs.(\ref{resp.x4a}) and  Eqs.(\ref{resp.x4b}) in the form 

\begin{eqnarray}
&& \{   [\delta {\cal O}_1 ]_{( \sigma , \sigma' ) }^{( \bar{\omega} , \bar{\omega} + \omega_0) ; [X,Y] } \}_{(1)} = 
\frac{ \bar{u}^2 N_{\rm Imp} }{V} \: \sum_{\vec{q} } \; \sum_{\sigma_1 , \sigma_1^{'} } \times \nonumber \\
&&   g_{[ \vec{q} ; (\sigma , \sigma_1 ) ]}^X ( \bar{\omega} ) [{\cal O}_1]_{(\sigma_1 , \sigma_1^{'} ) } 
g_{[ \vec{q} ; (\sigma_1^{'} , \sigma') ]  }^Y ( \bar{\omega} + \omega_0  )
\;\;\;\; , 
\label{resp.x6a}
\end{eqnarray}
\noindent
and 

 \begin{eqnarray}
&& \{   [\delta {\cal O}_1 ]_{(  \sigma , \sigma' ) }^{( \bar{\omega} , \bar{\omega} + \omega_0) ; [X,Y] } \}_{(n+1)} = 
\frac{ \bar{u}^2 N_{\rm Imp} }{V} \: \sum_{\vec{q} } \; \sum_{ \sigma_1 , \sigma_1^{'} } \times 
\label{resp.x6b} \\ 
&& g_{[ \vec{q} ; (\sigma , \sigma_1 ) ]  }^X ( \bar{\omega} ) \{  [\delta {\cal O}_1]_{(\sigma_1 , \sigma_1^{'} ) }^{(\bar{\omega} , \bar{\omega} + \omega_0 ) ; [X,Y]} \}_{(n)}  
g_{[\vec{q} ; (\sigma_1^{'} , \sigma' ) ]  }^Y ( \bar{\omega} + \omega_0 ) \ 
\;\;, \nonumber 
\end{eqnarray}
\noindent
with 

\beq
g_{[ \vec{q} ; (\sigma , \sigma' ) ]  }^X ( \bar{\omega} )  = \sum_{\lambda} \: w^{\vec{q}}_{\sigma , \lambda} [ w^{\vec{q}}_{\sigma' , \lambda}]^* \: g^X_{(\vec{q} , \lambda) } ( \bar{\omega} ) 
\:\:\:\: . 
\label{resp.x7}
\eneq
\noindent
As a further formal simplification, we   only retain the vertex corrections at the right-hand side of 
Eq. (\ref{resp.x2}) containing the $[A,R]$-terms \cite{Dimitrova2005}. Accordingly we eventually
obtain the vertex-correction induced additional contribution to $\Sigma_{1,2}$, $\delta \Sigma_{1,2}$, given by 
(in the zero-temperature limit) 

\begin{eqnarray}
&& \delta \Sigma_{1,2}  =\frac{1}{V} \sum_{\vec{k}} \sum_{\lambda , \lambda'} \: \int \frac{ d \bar{\omega}}{2 \pi}  \times \label{resp.x8}   \\
&&
[ \delta {\cal O}_1 ]_{(\vec{k} , \lambda) ; ( \vec{k} , \lambda' ) } ^{(\bar{\omega} , \bar{\omega} ); [ A , R ] } [
{\cal O}_2 ]_{(\vec{k} , \lambda' ) ; ( \vec{k} , \lambda ) } g^A_{( \vec{k} , \lambda ) } ( \bar{\omega} ) 
g^R_{(\vec{k } , \lambda' ) } ( \bar{\omega} ) \partial_{\bar{\omega}} f ( \bar{\omega} )
 = 
\nonumber \\
&& - \frac{2 \pi}{V}   \sum_{\vec{k}} \sum_{\lambda , \lambda'} \:  
[ \delta {\cal O}_1 ]_{(\vec{k} , \lambda) ; ( \vec{k} , \lambda' ) }^{(0 , 0); [ A , R ] } [
{\cal O}_2 ]_{(\vec{k} , \lambda' ) ; ( \vec{k} , \lambda ) } g^A_{( \vec{k} , \lambda ) } ( 0) 
g^R_{(\vec{k } , \lambda' ) } (0 )
\:\:\:\: . \nonumber 
\end{eqnarray}
\noindent

\section{Spin-Hall conductance in a lattice model for the 
two-dimensional Rashba Hamiltonian }
\label{lat2Rashba}

As a test bed of our approach, in this appendix we compute the SHC in a lattice 
model for a two-dimensional electron gas with Rashba SOI, described by the  Hamiltonian $H_{\rm R}$ given
by 

\beq
H_{\rm R} = \frac{1}{V} \:  \sum_{\vec{k}} \:  \sum_{\sigma , \sigma' } 
c_{\vec{k} , \sigma}^\dagger [ {\cal H}_{\rm R} ( \vec{k} )  ]_{\sigma, \sigma'} c_{\vec{k}, \sigma'}  
\:\:\:\: , 
\label{rashba.3}
\eneq
\noindent
with $V$ being the (two-dimensional) lattice ``volume'', $\sigma$ and $\sigma'$ being 
actual spin labels, and 

\beq
{\cal H}_{\rm R} ( \vec{k} ) = \left[\begin{array}{cc} 
\epsilon_{\vec{k}} & \delta_{\vec{k}} \\
\delta^*_{\vec{k}} & \epsilon_{\vec{k}} \end{array} \right]
\;\;\;\; , 
\label{ras.x1}
\eneq
\noindent
with 

\begin{eqnarray}
\epsilon_{\vec{k}} &=& - 2 t \{ \cos ( k_x ) + \cos ( k_y ) \}\; ,  \nonumber \\
\delta_{\vec{k}} &=&2 \alpha \{ i \sin ( k_x ) + \sin ( k_y ) \} 
\: . 
\label{ras.x2}
\end{eqnarray}
\noindent
 The relevant parameters
of $H_{\rm R}$  are the single-fermion hopping amplitude $t$ (which we assume to be 
the same both along the $x$ and   $y$ directions over the lattice) and  the Rashba spin-orbit coupling 
$\alpha$.  Eventually, we use  the chemical potential $\mu$ as our tuning parameter.  

To compute the SHC we need the SH  current and the charge-current 
operators, respectively given by (in the $2\times2$ matrix notation)

\begin{eqnarray} 
 j_{{\rm sp} , x}^z   &=&  t \sin ( k_x ) \sigma^z \;\; ,  \nonumber \\
 j_{{\rm ch} , y } &=&  2 e t \sin ( k_y ) {\bf I} + 2 e \alpha \cos ( k_y ) \sigma^x 
 \;\;\;\; . 
 \label{ras.x5}
 \end{eqnarray}
 \noindent 
The energy eigenvalues, $\epsilon_{\vec{k} , \lambda}$, with $\lambda = \pm$, 
and the corresponding eigenmodes $\Gamma_{\vec{k} , \lambda}$ are 
given by 

\beq
\epsilon_{\vec{k} , \lambda }  = \epsilon_{\vec{k}}+\lambda | \delta_{\vec{k}} |   \;\;\;\; , 
\label{ras.x3}
\eneq
\noindent
and by 

\begin{eqnarray}
\Gamma_{\vec{k} , + } &=& \frac{1}{\sqrt{2}} \{ c_{\vec{k} , \uparrow} + e^{ i \varphi_{\vec{k}} } c_{\vec{k} , \downarrow}\} \; ,\nonumber \\
\Gamma_{\vec{k} , - } &=& \frac{1}{\sqrt{2}} \{ - e^{ - i \varphi_{\vec{k} }} c_{\vec{k} , \uparrow} + c_{\vec{k} , \downarrow}\} 
\; , 
\label{ras.x4}
\end{eqnarray}
\noindent
with $\varphi_{\vec{k}} = {\rm arg} ( \delta_{\vec{k}} )$.  Once rotated to the energy eigenmode basis, 
the operators in Eq.(\ref{ras.x5}) become

\begin{eqnarray}
j_{{\rm sp} , x}^z &\to& - t \sin (k_x ) \{ \cos ( \varphi_{\vec{k}} ) \sigma^x - \sin ( \varphi_{\vec{k}} ) \sigma^y\} \;,
\label{res.x5}\\
j_{{\rm ch} , y} &\to& 2 e t \sin (k_y ) {\bf I} + 2 e \alpha \cos ( k_y ) \cos ( \varphi_{\vec{k}} ) \sigma^z \nonumber \\
&-& 2 e \alpha i \sin ( \varphi_{\vec{k}} ) \cos ( k_y )
  \{ \cos ( \varphi_{\vec{k}} ) \sigma^x - \sin ( \varphi_{\vec{k}} ) \sigma^y \} \nonumber \; . 
\end{eqnarray}
\noindent
Summing over the possible directions 
of $\vec{k}$  (within the Brillouin zone) allows for rewriting Eq.(\ref{ims.10c})  as 

\beq
1 = \frac{ N_{\rm Imp} \bar{u}^2}{2  V } \: \sum_{\vec{k}} \sum_{\lambda = \pm 1} 
\left\{\frac{1}{ ( \epsilon_{\vec{k}} + \lambda | \delta_{\vec{k}} | - \mu )^2 + ( 2 \tau_{\rm Imp} )^{-2} }\right\}
\:\:\:\: . 
\label{resn.1}
\eneq
\noindent

From  Eq.(\ref{resn.1})  we see that 
any dependence on the momentum, as well as on the spin index, has disappeared at the right-hand side of 
the equation, leaving only an over-all factor of $\frac{1}{2}$. While, this 
is just a special feature of the two-band Rashba model,  it is in any case a useful guideline to 
choosing the appropriate approximations to implement in the more complicated multiband  calculation. 

We now compute the SHC  of the two-band Rashba model, $\sigma_{xy}^z$. 
To do so, we label with subscript  $1$ the vertex corresponding to the $z$-component of the spin 
current along $x$ and with subscript $2$ the vertex corresponding to the  charge current along $y$.
As in the main text, we denote with $\sigma_{xy;A}^z$ the contribution to the SHC
without  vertex corrections and with $\sigma_{xy;B}^z$ the total contribution arising from 
vertex corrections. As a next step, we consider  the splitting in Eqs.(\ref{recr.11}), which in 
this case yields 

\begin{eqnarray}
  [ {\cal S}_{1,2} ]_{(\vec{k} , \lambda); (\vec{k} , \lambda' )} &=& 0 \;\; , \; \forall \lambda , \lambda '  \; , \nonumber \\
  [ {\cal I}_{1,2} ]_{(\vec{k} , + ) ; ( \vec{k } , + )}&=& [ {\cal I}_{1,2} ]_{(\vec{k} , - ) ; ( \vec{k} , - ) } = 0 \; ,  \nonumber \\
 [ {\cal I}_{1,2} ]_{(\vec{k} , + ) ; ( \vec{k } , - )} &=& -  [ {\cal I}_{1,2} ]_{(\vec{k} , - ) ; ( \vec{k } , + )} = \nonumber \\
&& \frac{ 2 e t \alpha \sin^2 ( k_x ) \cos ( k_y ) }{ \sqrt{ \sin^2 ( k_x ) + \sin^2 ( k_y ) } } 
\:\:\:\: . 
\label{resn.2}
\end{eqnarray}
\noindent
Taking into account Eqs.(\ref{resn.2}), we obtain 

\beq
\sigma_{xy;A}^z  =  \frac{2 e t \alpha }{\pi V} \: \sum_{\vec{k}} \frac{ \sin^2 ( k_x ) \cos (k_y ) \Phi_{\vec{k}} ( \mu ) }{ \sqrt{
\sin^2 ( k_x ) + \sin^2 ( k_y ) } } 
\;\;\;\; , 
\label{resn.4}
\eneq
\noindent
with 

\begin{widetext}
\begin{eqnarray}
\Phi ( \mu ) &=&  - i \int_{- \infty}^0  \frac{ d \bar{\omega}}{ 2 \pi } \: \sum_{\lambda = \pm }  \lambda [ g_{(\vec{k} , \lambda )}^A ( \bar{ \omega } ) 
- g_{(\vec{k} , \lambda )}^R ( \bar{ \omega } ) ] \partial_{\bar{\omega}}  [ g_{(\vec{k} , - \lambda )}^A ( \bar{ \omega } ) 
+ g_{(\vec{k} , - \lambda )}^R ( \bar{ \omega } ) ]    \nonumber \\
&=&  \frac{ 2 \tau_{\rm Imp}  [  (  | \delta_{\vec{k}} |^2 -  (\epsilon_{\vec{k}} - \mu)^2 ) 
- ( 2 \tau_{\rm Imp} )^{-2}  ]}{ \pi | \delta_{\vec{k} } | \{( 2 \tau_{\rm Imp} )^{-2}  + 
2  [ ( \epsilon_{\vec{k}} - \mu)^2 + | \delta_{\vec{k}} |^2 + 2 \tau_{\rm Imp}^2 ( ( \epsilon_{\vec{k}} - \mu)^2 - |\delta_{\vec{k}} |^2 )^2 ]
\} } \nonumber \\
&+& \left\{  \frac{ {\rm arctan} [ 2 \tau_{\rm Imp} ( | \delta_{\vec{k}} | - \epsilon_{\vec{k}}  + \mu ) ] + 
{\rm arctan} [ 2 \tau_{\rm Imp} (  | \delta_{\vec{k}} | +  \epsilon_{\vec{k}} - \mu ) ] }{2  \pi | \delta_{\vec{k}} |^2} \right\} 
\:\:\:\: . 
\label{resn.3}
\end{eqnarray}
\noindent
\end{widetext}
The contribution determined by the vertex corrections, $\sigma_{xy;B}^z$, can be computed according 
to Eq.(\ref{resp.x8}). It is given by 

 \begin{eqnarray}
\sigma_{xy;B}^z &=& \frac{ 2}{\pi V} \sum_{\vec{k}} \sum_{\sigma , \sigma'} 
\sum_{\rho , \rho'} [ \delta j_{{\rm sp} , x}^z ]^{(0,0); [A,R]}_{(\vec{k} , \sigma ) ; (\vec{k} , \sigma' )} \times \nonumber \\ 
&& g_{\vec{k} ; (\rho  , \sigma )}^A ( 0 ) g_{\vec{k} ; (\sigma' , \rho' )}^R ( 0 )  [ j_{{\rm ch} , y} ]_{ (\vec{k} , \rho' ); 
(\vec{k} , \rho )} 
\: . 
\label{resn.5}
\end{eqnarray}
\noindent
(Note that, for computational purposes,  we wrote Eq.(\ref{resn.5}) in  the spin-eigenstate basis).
Finally, going through the systematic procedure discussed in appendix \ref{impures}, we   find for the 
renormalized vertex the expression 

\beq
 [ \delta j_{{\rm sp} , x}^z ]^{(0,0); [A,R]}_{(\vec{k} , \sigma ) ; (\vec{k} , \sigma' )} 
 = \frac{ N_{\rm Imp} \bar{u}^2 \Lambda}{ 1 - N_{\rm Imp} \bar{u}^2  \nu } [\sigma^x]_{\sigma , \sigma'} 
 \;\;\;\; , 
 \label{resn.6}
 \eneq
 \noindent
 that is independent of $\vec{k}$, with 
 
 \begin{widetext}
 \begin{eqnarray}
 \Lambda &=& \frac{\alpha t}{\tau_{\rm Imp} V} \sum_{\vec{q}} 
 \left\{\frac{ \sin^2 ( q_x ) }{ ( 2 \tau_{\rm Imp} )^{-4} + 2 ( 2 \tau_{\rm Imp} )^{-2} [ ( \epsilon_{\vec{q}} - \mu)^2 + 
 | \delta_{\vec{q}} |^2 ] + [ ( \epsilon_{\vec{q}} - \mu)^2  - | \delta_{\vec{q}} |^2 ]^2 }\right\} \; , \nonumber \\
 \nu &=& \frac{1}{V}  \sum_{\vec{q}} \left\{\frac{ 
  ( \epsilon_{\vec{q}} - \mu)^2 + ( 2 \tau_{\rm Imp} )^{-2} }{( 2 \tau_{\rm Imp} )^{-4} + 
 2 ( 2 \tau_{\rm Imp} )^{-2} [(  \epsilon_{\vec{q}} - \mu)^2 + 
 | \delta_{\vec{q}} |^2 ] + [ ( \epsilon_{\vec{q}} - \mu)^2  - | \delta_{\vec{q}} |^2 ]^2} \right\}
\:\:\:\: . 
\label{resn.7}
\end{eqnarray}
\noindent
\end{widetext}
 Finally, plugging Eq. (\ref{resn.1}) into (\ref{resn.6}), we eventually get 

\beq
  [ \delta j_{{\rm sp} , x}^z ]^{(0,0); [A,R]}_{(\vec{k} , \sigma ) ; (\vec{k} , \sigma' )} 
 =   \frac{t}{ 4 \alpha \tau_{\rm Imp}}  \sigma^x
 \:\:\:\: .
 \label{resn.9}
 \eneq
 \noindent 
Inserting  Eq. (\ref{resn.9}) into (\ref{resn.5}), we derive 
 $\sigma_{xy;B}^z$ as a function of   $\mu$. 
    \begin{figure}
 \center
\includegraphics*[width=0.9 \linewidth]{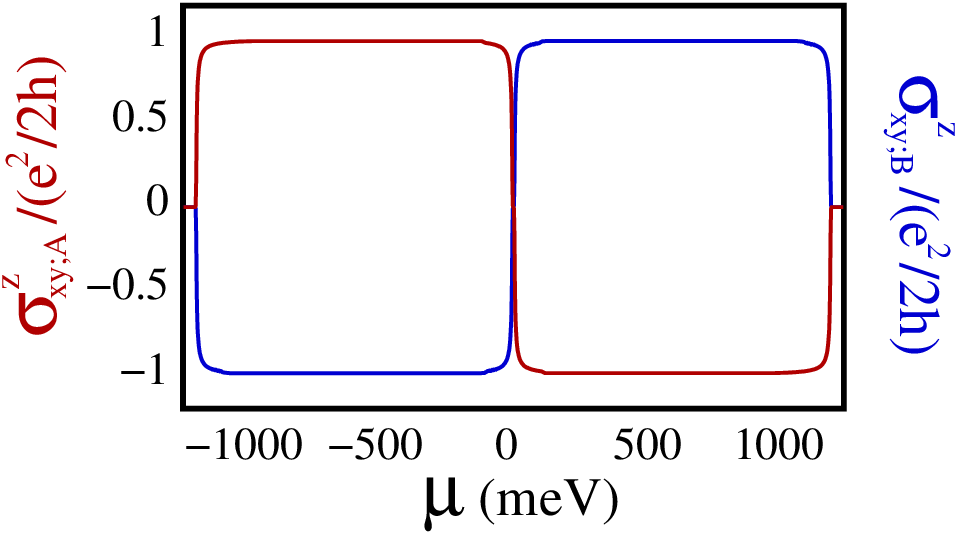}
\caption{Contributions to the spin-Hall conductance $\sigma_{xy;A}^z$ (red curve) and $\sigma_{xy;B}^z$
(blue curve) in the model described by the lattice Rashba Hamiltonian in Eq. (\ref{rashba.3}), in the presence 
of a finite impurity concentration,  with parameters 
chosen so that  $t=300 {\rm meV}$, $\alpha = 10 {\rm meV}$ and  $\tau_{\rm Imp}  = 3 {\rm ps}$.  
} 
\label{rashba_c}
\end{figure}

 In Fig. \ref{rashba_c} we 
 separately plot the results for the two terms, computed using the Hamiltonian $H_{\rm R}$
 with  t=300  meV, $\alpha = 10 \; {\rm meV}$ and with $\tau_{\rm Imp}  = 3 \; {\rm ps}$. 
From the plot in Fig. \ref{rashba_c}   we  note the main features of the SHC  in 
a Rashba-2DEG in the presence of impurity. First, we note that, in the absence of vertex corrections, 
$\sigma_{xy}^z$ is either zero, or it is quantized (in units of $e^2 / 2h$) \cite{Kane2005}. When including 
vertex corrections, these fully screen the  SHC, which results in the absence
of the effect, as widely discussed in the literature 
\cite{Rashba2004,Mishchenko2004,Inoue2004,Khaetskii2006,Raimondi2004,Dimitrova2005,Shytov2006}. 
In the main text we, instead, note how the 
behavior is completely different in the multiband model, due to the various, impurity induced,
 interband processes. This notwithstanding the fact that, as we show in the next appendix,
 at low enough values of $\mu$, the 
 8-band model can be effectively described as a Rashba-like Hamiltonian such as the one 
 we discuss here.

 \section{Effective two-subband Rashba-like Hamiltonian}
 \label{efra}
 
 In this appendix we show how the eight-band model of Sec. \ref{sh8}, when $\mu$ is within the first pair of 
 subbands but still below the bottom of higher-energy subbands,  
 can be effectively described by a  Rashba  Hamiltonian 
 $H_{\rm xy}$,  involving the effects of virtual transitions  from the low-lying subbands 
 to, and from, higher-energy bands. 
 
Employing the approach developed in Ref.\cite{Lepori2021} and 
referring to the  eight-band Hamiltonian that we discuss in Sec. \ref{sh8},
 we define the energy eigenmodes $\Gamma_{\vec{k} , \lambda}$ as 
 $\Gamma_{\vec{k} , \lambda} = \sum_{\alpha} \sum_s u_{\alpha , s}^* c_{\alpha , s}$,
 with $\alpha\in\{( xy;A ),(xy;B),zx,yz\}$ and  $s$ being the spin polarization. In the following, 
 we will simply denote with $u_\alpha$ the (bi)spinor $[ u_{\alpha , \uparrow} ,  u_{\alpha , \downarrow} ]_t$. 
 We now recover the reduced, effective Hamiltonian for the first two subbands in the bispinor representation, 
 by  systematically going  through a projection over the $xy;A$ and the $xy;B$ subbands. To do so, we 
 introduce the projection operators ${\cal P}$ and ${\cal Q}$ which, in the block notation of 
Sec. \ref{sh8}  and in the basis of the spinors such as the one in Eq.(\ref{tr.x1}) are written as 
 
 \beq
 {\cal P}  =  \left[ \begin{array}{cccc} {\bf 0} &  {\bf 0} & {\bf 0} &  {\bf 0} \\
 {\bf 0} &  {\bf 0} & {\bf 0} &  {\bf 0} \\  {\bf 0} &  {\bf 0} & {\bf I} &  {\bf 0} \\ {\bf 0} &  {\bf 0} & {\bf 0} &  {\bf I} \end{array} 
 \right]\;\;\; , \; 
  {\cal Q}  =  \left[ \begin{array}{cccc} {\bf I} &  {\bf 0} & {\bf 0} &  {\bf 0} \\
 {\bf 0} &  {\bf I} & {\bf 0} &  {\bf 0} \\  {\bf 0} &  {\bf 0} & {\bf 0} &  {\bf 0} \\ {\bf 0} &  {\bf 0} & {\bf 0} &  {\bf 0} \end{array} 
 \right]
 \:\:\:\: . 
 \label{efra.1}
 \eneq
 \noindent
 Projecting with ${\cal P}$ and ${\cal Q}$ the Schr\"odinger equation corresponding to the 
 Hamiltonian in Eq.(\ref{tr.2}),  we obtain 
 
 \begin{widetext}
 \begin{eqnarray}
 && \left[ \begin{array}{cc} \epsilon -  \epsilon_{xy;A} ( \vec{k} ) &  {\bf 0} \\
 {\bf 0 } & \epsilon - \epsilon_{xy;B} ( \vec{k} ) \end{array} \right] \left[
 \begin{array}{c} u_{xy;A} \\ u_{xy;B} \end{array} \right] = 
 i \left[ \begin{array}{cc} f_1^X ( \vec{k} ) + \lambda_{\rm SOI} \sigma^y & 
  f_1^Y ( \vec{k} ) - \lambda_{\rm SOI} \sigma^x \\   f_2^X ( \vec{k} ) + \lambda_{\rm SOI} \sigma^y & 
  f_2^Y ( \vec{k} ) - \lambda_{\rm SOI} \sigma^x  \end{array} \right] 
 \left[ \begin{array}{c} u_{yz} ( \vec{k} ) \\ u_{zx} ( \vec{k} ) \end{array} \right] \nonumber \\
 && \left[\begin{array}{cc} \epsilon - \epsilon_{yz} ( \vec{k} ) & - i \lambda_{\rm SOI} \sigma^z \\
 i \lambda_{\rm SOI} \sigma^z & \epsilon - \epsilon_{zx} ( \vec{k} ) \end{array} \right]
 \left[ \begin{array}{c} u_{yz} ( \vec{k} ) \\ u_{zx} ( \vec{k} ) \end{array} \right] 
 = - i   \left[ \begin{array}{cc} f_1^X ( \vec{k} ) + \lambda_{\rm SOI} \sigma^y & 
  f_2^X ( \vec{k} ) + \lambda_{\rm SOI} \sigma^y \\   f_1^Y ( \vec{k} ) -  \lambda_{\rm SOI} \sigma^x  & 
  f_2^Y ( \vec{k} ) - \lambda_{\rm SOI} \sigma^x  \end{array} \right] \left[
 \begin{array}{c} u_{xy;A} \\ u_{xy;B} \end{array} \right] 
 \:\:\:\: . 
 \label{efra.2}
 \end{eqnarray}
 \noindent
 Getting rid of the $u_{yz} ( \vec{k} )$ and of the $u_{zx} ( \vec{k})$, Eqs.(\ref{efra.2}) yields the 
 reduced time-independent Schr\"odinger equation, with energy eigenvalue $\epsilon$, projected over 
 the low-lying doublets in the form 
 
 \beq
\left[\begin{array}{cc}  \epsilon - \epsilon_{xy;A} ( \vec{k} ) - h_{AA} ( \vec{k} ) &-h_{AB} ( \vec{k} ) \\
- h_{BA} ( \vec{k} ) & \epsilon - \epsilon_{xy;B} ( \vec{k} ) - h_{BB} ( \vec{k} ) \end{array}
\right] \left[ \begin{array}{c} u_{xy;A} ( \vec{k} ) \\ u_{xy; B } ( \vec{k} ) \end{array} \right] = 0 
  \:\:\:\: , 
  \label{efra.3}
  \eneq
  \noindent
  with 
  
  \begin{eqnarray}
  && \left[ \begin{array}{cc} h_{AA} ( \vec{k} ) & h_{AB} ( \vec{ k } ) \\
  h_{BA} ( \vec{k} ) & h_{BB} ( \vec{k} ) \end{array} \right] = 
  \frac{1}{ [ \epsilon - \epsilon_{yz} ( \vec{k} ) ] [ \epsilon - 
  \epsilon_{zx} ( \vec{k} ) ] - \lambda_{\rm SOI}^2 } \: \left[ \begin{array}{cc} f_1^X ( \vec{k} ) + \lambda_{\rm SOI} \sigma^y & 
  f_1^Y ( \vec{k} ) - \lambda_{\rm SOI} \sigma^x \\   f_2^X ( \vec{k} ) + \lambda_{\rm SOI} \sigma^y & 
  f_2^Y ( \vec{k} ) - \lambda_{\rm SOI} \sigma^x  \end{array} \right] \times \nonumber \\
 && \left[ \begin{array}{cc} \epsilon - \epsilon_{zx} ( \vec{k} ) & i \lambda \sigma^z \\
 - i \lambda \sigma^z & \epsilon - \epsilon_{yz} ( \vec{k} ) \end{array} \right] 
  \left[ \begin{array}{cc} f_1^X ( \vec{k} ) + \lambda_{\rm SOI} \sigma^y & 
  f_2^X ( \vec{k} ) + \lambda_{\rm SOI} \sigma^y \\   f_1^Y ( \vec{k} ) -  \lambda_{\rm SOI} \sigma^x  & 
  f_2^Y ( \vec{k} ) - \lambda_{\rm SOI} \sigma^x  \end{array} \right] \left[
 \begin{array}{c} u_{xy;A} \\ u_{xy;B} \end{array} \right] 
 \:\:\:\: . 
 \label{efra.4}
 \end{eqnarray}
 \noindent
  Explicitly expanding the right-hand side of Eq.(\ref{efra.4}), we eventually obtain 
  
  \begin{eqnarray}
  && h_{AA} ( \vec{k} ) = {\cal D}^{-1} ( \vec{k} ) \Biggl\{ ( \epsilon - \epsilon_{zx} ( \vec{k} ) ) [ f_2^X ( \vec{k} ) ]^2 +  ( \epsilon - \epsilon_{yz} ( \vec{k} ) ) [ f_2^Y ( \vec{k} ) ]^2
  + \lambda_{\rm SOI}^2 \: [ 2 \epsilon - \epsilon_{zx} ( \vec{k} ) - \epsilon_{yz} ( \vec{k} ) ]   \label{efra.5} \\
  && + 2 \lambda_{\rm SOI} \: \{ - ( \epsilon - \epsilon_{yz} ( \vec{k} ) ) f_1^Y ( \vec{k} ) \sigma^x +  
     ( \epsilon - \epsilon_{zx} ( \vec{k} ) ) f_1^X ( \vec{k} ) \sigma^y \}   \Biggr\} \nonumber \\
     && h_{BB} ( \vec{k} ) = {\cal D}^{-1} ( \vec{k} ) \Biggl\{  ( \epsilon - \epsilon_{zx} ( \vec{k} ) ) [ f_1^X ( \vec{k} ) ]^2 +  ( \epsilon - \epsilon_{yz} ( \vec{k} ) ) [ f_1^Y ( \vec{k} ) ]^2
  + \lambda_{\rm SOI}^2 \: [ 2 \epsilon - \epsilon_{zx} ( \vec{k} ) - \epsilon_{yz} ( \vec{k} ) ] \nonumber \\
  && + 2 \lambda_{\rm SOI} \: \{ - ( \epsilon - \epsilon_{yz} ( \vec{k} ) ) f_2^Y ( \vec{k} ) \sigma^x +  
     ( \epsilon - \epsilon_{zx} ( \vec{k} ) ) f_2^X ( \vec{k} ) \sigma^y \}  \Biggr\}  \nonumber \\ 
     && h_{AB} ( \vec{k} ) = h_{BA}^* ( \vec{k} ) = \nonumber \\
     &&  {\cal D}^{-1} ( \vec{k} ) \Biggl\{   (\epsilon - \epsilon_{zx} ( \vec{k} ) ) ( f_1^X ( \vec{k} ) + \lambda_{\rm SOI} \sigma^y ) ( f_2^X ( \vec{k} ) 
     + \lambda_{\rm SOI} \sigma^y ) 
     +  (\epsilon - \epsilon_{yz} ( \vec{k} ) ) ( f_1^Y ( \vec{k} ) - \lambda_{\rm SOI} \sigma^x) ( f_2^Y ( \vec{k} ) -  \lambda_{\rm SOI} \sigma^x ) 
     \nonumber \\
     && + i \lambda_{\rm SOI} \{ - ( \epsilon - \epsilon_{zx} ( \vec{k} ) ) ( f_1^Y ( \vec{k} ) - \lambda_{\rm SOI} \sigma^x ) ( f_2^X ( \vec{k} ) +
      \lambda_{\rm SOI} \sigma^y ) + 
   ( \epsilon - \epsilon_{yz} ( \vec{k} ) ) ( f_1^X ( \vec{k} )+  \lambda_{\rm SOI} \sigma^y  ) ( f_2^Y ( \vec{k} ) - \lambda_{\rm SOI} \sigma^x )  
   \} \sigma^z \Biggr\}
   \:\:\:\: , 
\nonumber 
   \end{eqnarray}
   \noindent
   with 
   
   \beq
   {\cal D}^{-1} ( \vec{k} ) = \frac{1}{[ \epsilon - \epsilon_{yz} ( \vec{k} )] [\epsilon - 
  \epsilon_{zx} ( \vec{k} )] - \lambda_{\rm SOI}^2 } 
  \:\:\:\: . 
  \label{efra.6}
  \eneq
  \noindent
  At a given chemical potential $\mu$, for  $- \Delta_A < \mu < - \Delta_B$ we  approximate the matrix elements in Eqs. (\ref{efra.5}) by 
 simply substituting $\epsilon$ with $\epsilon_{xy;A} ( \vec{k})$.    For $| \mu | \sim \Delta_A $ we may further neglect the effective inter-band couplings in the right-hand side of Eqs. (\ref{efra.4}) and (\ref{efra.5}) 
 by therefore introducing a simple effective Rashba-type  Hamiltonian for the $xy;A$ doublet, $H_{AA} (\vec{k})$, given by 
 
 \begin{eqnarray}
 H_{\rm AA} ( \vec{k} ) &=& \epsilon_{xy;A} (\vec{k}) +\{ (\epsilon_{xy;A} (\vec{k}) - \epsilon_{yz} (\vec{k}))   (\epsilon_{xy;A} (\vec{k}) - \epsilon_{zx} (\vec{k})) - \lambda_{\rm SOI}^2 \}^{-2} 
 \: \times \nonumber \\
 && \biggl\{( \epsilon_{xy;A} (\vec{k}) -  \epsilon_{zx} (\vec{k})) [f_2^X (\vec{k}) ]^2 +  (\epsilon_{xy;A} (\vec{k}) -  \epsilon_{yz} (\vec{k})) 
 [f_2^Y (\vec{k})]^2
 +  \lambda_{\rm SOI}^2 \: [ 2 \epsilon_{xy;A} ( \vec{k} ) - \epsilon_{zx} ( \vec{k} ) -   \epsilon_{yz} ( \vec{k} ) ] \nonumber \\
 &+&  2 \lambda_{\rm SOI} \: \biggl[ -  ( \epsilon_{xy;A} ( \vec{k} ) -  \epsilon_{yz} ( \vec{k} ) ) f_1^Y ( \vec{k} ) \sigma^x +  
     ( \epsilon_{xy;A} ( \vec{k} ) -   \epsilon_{zx} ( \vec{k} ) ) f_1^X ( \vec{k} ) \sigma^y \biggr] \biggr\} 
     \:\:\:\: . 
     \label{efra.7}
     \end{eqnarray}
     \noindent
Expanding the right-hand side of Eq.(\ref{efra.7}) up to second-order in $\vec{k}$, we obtain 
that $H_{AA} ( \vec{k} ) \approx H_{xy } ( \vec{k} )$, with 

\beq
H_{xy} ( \vec{k} ) \approx - \hat{\Delta}_A + \hat{t}^2 \vec{k}^2 + \hat{\alpha } \{ - k_x \sigma^y + k_y \sigma^x \} 
\;\;\;\; , 
\label{efra.8}
\eneq
\noindent
and 

\begin{eqnarray}
\hat{\Delta}_A &=& \Delta_A + \frac{ 2 \Delta_A \lambda_{\rm SOI}^2 }{ \Delta_A^2 - \lambda_{\rm SOI}^2} \; ,  \nonumber \\
\hat{t} &=&t_1 + \frac{ \lambda_{\rm SOI}^2 ( t_1 - t_2 ) - 4 \Delta_A \gamma_2^2}{ \Delta_A^2 - \lambda_{\rm SOI}^2} - 
\frac{ 2 \Delta_A^2 \lambda_{\rm SOI}^2 ( t_1 - t_2 ) }{ [ \Delta_A^2 - \lambda_{\rm SOI}^2 ]^2} \; ,  \nonumber \\
\hat{\alpha} &=& \frac{ 4 \gamma_1 \lambda_{\rm SOI} \Delta_A}{ \Delta_A^2 - \lambda_{\rm SOI}^2 } 
\:\:\:\: . 
\label{efra.9}
\end{eqnarray}
\noindent
The emergence of an effective Hamiltonian, at low $\vec{k}$, with a linear  Rashba spin-orbit coupling 
is consistent with the apparent suppression of the spin-Hall conductance due to the vertex renormalization
in the low-$\mu$ part of the plot of Fig.\ref{sh_yv}. Of course, on increasing $\mu$, nonlinear contributions to the
Rashba coupling as well as inter-band terms in the system Hamiltonian are expected to spoil the 
perfect cancellation of the spin-Hall conductance from impurity-induced vertex corrections and, 
again, this is consistent with our result of Fig. \ref{sh_yv}.
 \end{widetext}
 
 \section{Analytical derivation of the sheet conductance and of the spin-Hall conductance in 
 the 8-band model}
 \label{anshsh}

 In this appendix we provide the explicit derivation of the formulas for the sheet conductance and for 
 the spin-Hall conductance in the eight-band model,  that we used to analytically derive the plots that we show in the main text
 of the paper.   All the formulas we derive in the following are grounded over a systematic implementation of the 
 formalism that we review in Appendixes \ref{lrt} and  \ref{impures}.   Before going through the details 
 of the derivation of the conductances, it is crucial to extensively discuss and motivate the way in which 
 we employ (the self-consistent version of) Eq.(\ref{ims.10c}) for the inverse single-particle 
 lifetime $\tau_{\rm Imp}^\lambda$.

 In general, $\tau^\lambda_{\rm Imp}$ should be self-consistently determined from Eq.(\ref{ims.10c}), with the 
 `` bare''  single-particle retarded Green's function $g_{(\vec{q};\mu)}^{R;(0)} ( \omega )$ at the right-hand side
 of the equation substituted with 
 the ``dressed'' one  $g_{(\vec{q};\mu)}^{R } ( \omega )$  in Eq.(\ref{ims.13}). However, in the specific case of 
 the LAO/STO interface,  as it is evidenced in the Supplemental Material 
of Ref. \cite{Trier2020}, within all the interval of values of  the gate voltage that they consider (which corresponds to our interval of 
values of $\mu$), the elastic 
contribution to $\tau_{\rm Imp}$, $\tau_{\rm e}$, keeps smaller than the inelastic one, $\tau_{\rm i}$ (by even two to three orders of 
magnitude  in the first part of the interval of values of $\mu$). 
 This evidences how we can safely neglect $\tau_{\rm i}$ and, accordingly, 
compute $\tau_{\rm e}$ by setting to 0 the frequency $\omega$  in (the self-consistent version of) Eq. (\ref{ims.10c}).
Still keeping, for the time being, the explicit dependence on $\lambda$ we therefore simplify 
Eq.(\ref{ims.10c}) to

\beq
\frac{1}{2 \tau_{\rm e}^\lambda} = \frac{ N_{\rm Imp} \bar{u}^2}{V} \sum_{\vec{q}} \sum_\mu \left[
\frac{ ( 2 \tau_{\rm e}^\mu )^{-1} {\cal A}_{(\vec{k} , \lambda) ; (\vec{q} , \mu)}  {\cal A}_{(\vec{q} , \mu) ; (\vec{k} , \lambda) 
 }}{ \xi_{\vec{q} , \mu}^2 + ( 2 \tau_{\rm e}^\mu )^{-2}} \right]
\:\: . 
\label{ash.1}
\eneq
\noindent
To recover a numerical estimate for the $\tau_{\rm e}^\lambda$, we rely on Ref. \cite{Trier2020}, where 
the non-$\lambda$-resolved 
$\tau_{\rm e}$ is obtained from   the measured magnetoconductance by means of 
a fit to the Maekawa-Fukuyama formula  \cite{Huijben2017}. Having only a band-independent 
estimate for $\tau_{\rm e}$ we dropped the dependence on $\lambda$ in the inverse lifetimes 
appearing at the right-hand side of   Eq.(\ref{ash.1}) and approximated it by means 
of the rough estimate  $\tau_{\rm e} \approx 3$ ps \cite{Trier2020}.  Accordingly 
dropping the dependence on $\lambda$ in $\tau_{\rm e}$ 
in Eq.(\ref{ash.1}), we further approximate it by averaging over $\lambda$, so that we eventually obtain

\beq
1 = \frac{N_{\rm Imp} \bar{u}^2}{N V} \sum_{\vec{q}} \sum_\mu \left[
\frac{1 }{ \xi_{\vec{q} , \mu}^2 + ( 2 \tau_{\rm e} )^{-2}} \right]
\;\;\;\; . 
\label{ash.2}
\eneq
\noindent
While slightly changing the value of 
$\tau_{\rm e}$ and even adding a slight dependence on $\mu$ and/or on the 
band index $\lambda$ does almost not affect at all the contributions to the 
conductances from the nonzero, impurity induced single-fermion self energy. 
At variance, especially when computing $\sigma_{xy}^z$ in the multiband model, 
the implementation of the above approximations in computing the vertex corrections 
is crucial in determining the weight of that latter contribution on the finite result. 

    \begin{figure}
 \center
\includegraphics*[width=0.9 \linewidth]{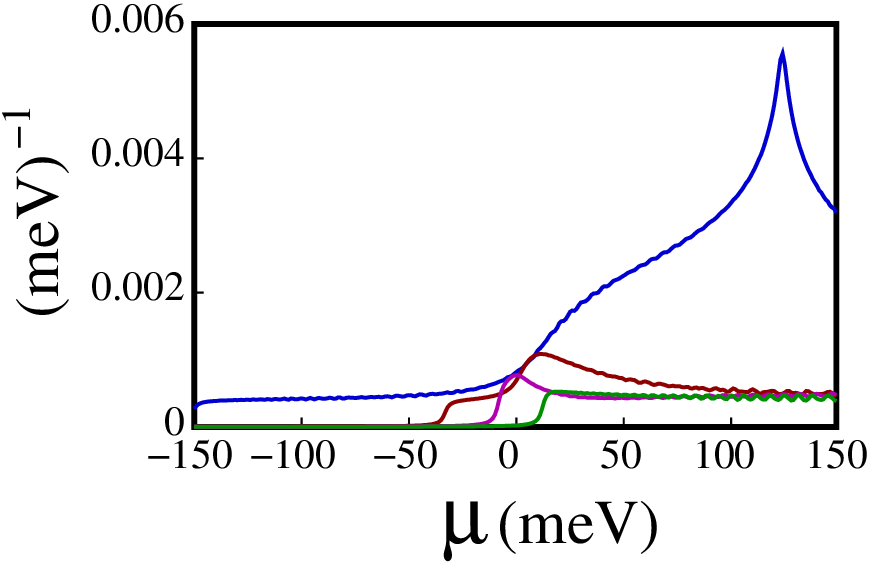}
\caption{Contributions to the factor multiplying $N_{\rm Imp} \bar{u}^2$ at the right-hand side of 
Eq. (\ref{ash.2}) determined by setting $\tau_{\rm e} = 3$ ps and averaged over band doublets, as a function
of the chemical potential $\mu$. 
In order of increasing energy, the contributions 
are drawn in blue,  red, magenta, and green.
} 
\label{taus}
\end{figure}
\noindent
In fact, looking at 
the right-hand side of Eq.(\ref{ash.2}),  we see that, on varying $\mu$, 
 terms with different index $\lambda$ contribute in a 
largely different way to the total sum. To evidence this, 
in Fig. \ref{taus} we plot (aside from the over-all factor $N_{\rm Imp} \bar{u}^2$) 
the contributions from the four different doublets,
averaged over each doublet, as a function of $\mu$. As in Fig.\ref{compar_exp}, 
it is useful to ideally split the plot in Fig. \ref{taus} into three regions.
The first region approximatively corresponds to $-150\;{\rm mev} \leq \mu \leq -35\:{\rm meV}$. In this 
region we   see that the contribution to the vertex correction just comes from 
the first doublet. This suggests us 
that the crude approximation resulting into Eq. (\ref{ash.2}) should not 
apply in this region: only two, out of eight, bands are contributing to
$\tau_{\rm e}$ and, therefore, the average encoded in the right-hand side 
of the equations has to   be performed over just 2, rather than $N=8$, bands. 
Consistently with this result, in computing the vertex correction 
we resort to a different approximation, by trading Eq. (\ref{ash.2}) for

\beq
1 = \frac{ N_{\rm Imp} \bar{u}^2 }{ 2 V} \sum_{\vec{k}} \sum_{\lambda = d_{xy;A}^\pm}  
\left[ \frac{ 1 }{ \xi_{\vec{k} , \lambda}^2  + ( 2 \tau_{\rm e} )^{-2} } \right]
\:\:\:\: , 
\label{ic.a2}
\eneq
\noindent
with $d_{xy;A}^\pm$ denoting the $d_{xy;A}$ doublet. Having numerically set $\tau_{\rm e}$ as discussed 
above, Eq. (\ref{ic.a2}) eventually trades for a numerical estimate of the (otherwise unknown) 
factor $N_{\rm Imp} \bar{u}^2$. To move ahead,  leaving aside, for the 
time being, the second region, we now focus onto the third one, roughly 
ranging over the interval $50\;{\rm meV} \leq \mu \leq 150\;{\rm meV}$. While, in 
this region, the first doublet still provides by a large amount the leading contribution 
to the right-hand side of Eq. (\ref{ash.2}), there is, now, a finite contributions from 
the other three doublets, which is approximatively the same over the various subbands. 
This suggests to weight the contributions of all four the doublets by first separately assuming 
that the right-hand side of Eq. (\ref{ash.2}) is contributed by a single doublet only, for all four the 
doublets, and by then attributing  to all the integrals a weight that is the same for the three
higher-energy doublet and, obviously, higher for the lowest-energy one. This eventually results into 
an effective normalization factor at the 
right-hand side of Eq. (\ref{ash.2}), $N_{\rm Eff}$, such that $2<N_{\rm Eff}<8$, with 
$N_{\rm Eff}$ numerically determined as stated above. This allows us to estimate $N_{\rm Imp} \bar{u}^2$
throughout the third region, as well. Finally, a straightforward extension of the method we used for 
the third region to the second region, as well, with different weights for different doublets, in general, allows for 
estimating $N_{\rm Imp} \bar{u}^2$ throughout this region, as well.

  We now proceed with  computing the sheet- and the spin-Hall conductances. Let us begin with 
the sheet conductance. 
  
Within the linear response theory of Appendix \ref{lrt}, for a system in a Hall 
bar arrangement  such as the one we draw in Fig. \ref{fig_device},  
the sheet conductance $\sigma_s$  is recovered from the current response 
along $x$ to an electric field applied in the same direction.  Adding also the effects
of the impurities we obtain,  according to the result 
of Appendixes \ref{lrt} and  \ref{impures},  $\sigma_s = \sigma_{s;A} + \sigma_{s;B}$, with 
$\sigma_{s;A}$ accounting for the nonzero, impurity-induced imaginary part of the single-electron 
self-energy and $\sigma_{s;B}$ determined by vertex corrections. 

From the derivation of Appendix \ref{impures} and taking into account our discussion above, we 
obtain 

\begin{eqnarray}
\sigma_{s;A} 
 &\approx&  \frac{   1  }{ 2 \pi w V} \: \sum_{\vec{k}} \: \sum_{ \lambda , \lambda'} \: [j_{{\rm ch},x}]_{(\vec{k} , \lambda) ; (\vec{k} , 
\lambda' )} \: [j_{{\rm ch},x}]_{(\vec{k} , \lambda' ) ; (\vec{k} , \lambda) }  \times \nonumber \\
 && \left\{ \frac{ ( 2 \tau_{\rm e} )^{-1} }{ \xi_{\vec{k} , \lambda}^2 + ( 2\tau_{\rm e})^{-2} } 
\:  \frac{ ( 2 \tau_{\rm e})^{-1} }{ \xi_{\vec{k} , \lambda'}^2 + ( 2 \tau_{\rm e})^{-2} } \right\}
\;\;\;\; ,
\label{lc.4bis}
\end{eqnarray}
\noindent
with  $w$ being the sample width along the $z$ direction, 
which we drop from our calculation, as we are only interested in the dependence of $\sigma_s$ on 
the chemical potential $\mu$.  The electric current operator matrix elements $ [j_{{\rm ch},x}]_{(\vec{k} , \lambda) ; (\vec{k} , 
\lambda' )}$ are derived by acting   with the transformation in Eq. (\ref{model.5}) over the operator 
$[j_{{\rm ch},x}]_{(\vec{k} , \sigma ) ; (\vec{k} , \sigma' ) } = e \left[ \frac{ \partial {\cal H}_{\rm 8-band} ( \vec{k} )}{ \partial k_x} 
\right]_{\sigma , \sigma'}=e \left[v^x \right]_{\sigma , \sigma'}$. 

Along the same guidelines leading to Eq. (\ref{lc.4bis}),  we  derive the 
contribution to the sheet conductance arising from the vertex corrections, $\sigma_{s;B}$.
This is given by

 \begin{eqnarray}
 \sigma_{s;B} 
 &=&  \frac{   e^2  }{ 2 \pi w V} \: \sum_{\vec{k}} \: \sum_{ \lambda , \lambda'} \: [v^x]_{(\vec{k} , \lambda) ; (\vec{k} , 
\lambda' )} \: [ \delta v^x ]_{(\vec{k} , \lambda' ) ; (\vec{k} , \lambda) } \times \nonumber \\
&&  \left\{ \frac{ 1 }{ \xi_{\vec{k} , \lambda} + i( 2 \tau_{\rm e})^{-1} } 
\:  \frac{ 1 }{ \xi_{\vec{k} , \lambda'} -i ( 2 \tau_{\rm e})^{-1} } \right\}
\;\;\;\; .
\label{lc.5}
\end{eqnarray}
\noindent
The quantity $\delta v^x$ in Eq.(\ref{lc.5}) is  the correction to the interaction vertex $ [v^x ]_{(\vec{k} , \lambda ) ; (\vec{k}' , \lambda')}$,
which we numerically compute  through an iterative process in 
the impurity interaction strength, as we outline in Appendix \ref{impures}.  Doing so, we explicitly verified   that 
$\sigma_{s;B}$, is several orders of magnitude lower than the contribution $\sigma_{s;A}$ 
in Eq. (\ref{lc.4bis}). We therefore used only this last contribution to estimate $\sigma_s$ and to accordingly
draw the plot  in Fig. \ref{sheets}.

To compute the  spin-Hall conductance $\sigma_{xy}^z$  we proceed in the same way. First of all, we 
split again $\sigma_{xy}^z$ as $\sigma_{xy}^z = \sigma_{xy;A}^z + \sigma_{xy;B}^z$, with 
$\sigma_{xy;A}^z$ taking into account the finite self-energy of the single-particle Green's functions 
and $\sigma_{xy;B}^z$ accounting for the impurity-induced vertex corrections.  To perform our
derivation, we need the retarded
Green's functions of the operators $j_{{\rm sp},x}^z = \frac{1}{2} \{\frac{\partial {\cal H}_{\rm 8-band} (\vec{k}) }{ \partial k_x},{\bf I} \otimes \sigma_z \}$ and 
$j_{{\rm ch},y} = e \frac{ \partial {\cal H}_{\rm 8-band} (\vec{k})  }{ \partial k_y}$. By direct inspection we verified that 
$ \Re e \{ [ j_{{\rm sp} , x}^z]_{(\vec{k} , \lambda ) ; ( \vec{k} , \lambda' ) } [ j_{{\rm ch},y} ]_{(\vec{k} , \lambda' ) ; ( \vec{k} , \lambda ) } \} = 0$.
Therefore,  $\sigma_{xy;A}^z$ is only contributed by the term in 
Eq.(\ref{recr.17}), while the contribution   from Eq.(\ref{recr.16}) is equal to 0. Specifically, we obtain

    \begin{eqnarray}
 && \sigma_{xy;A}^z = - \frac{i}{ 2 V  } \: 
 \sum_{ \vec{k}  } \: \sum_{\lambda  , \lambda^{'} } \times \nonumber \\ && \Im  m
  \{ [ j_{{\rm sp} , x}^z]_{(\vec{k} , \lambda ) ; ( \vec{k} , \lambda' ) } [ j_{{\rm ch},y} ]_{(\vec{k} , \lambda' ) ; ( \vec{k} , \lambda ) } \} 
  \times \nonumber \\
  && \int \frac{ d \bar{\omega}}{2 \pi}  \: \Biggl\{
   [   g_{ (  \vec{k}  , \lambda )   }^A ( \bar{\omega} ) - g_{ (  \vec{k}   , \lambda  ) }^R (  \bar{\omega } ) ]
  \times \nonumber \\
&& \partial_{\bar{\omega}}   [   g_{ (  \vec{k}  , \lambda' )   }^A ( \bar{\omega}  ) + g_{ (  \vec{k}   , \lambda'  ) }^R (  \bar{\omega }  )) ]
  f ( \bar{\omega} )   \Biggr\} 
     \: ,
     \nonumber \\
     \label{lcx.1}
\end{eqnarray}
\noindent
with (see Appendix \ref{lrt} for details) 

\beq
g_{( \vec{k} , \lambda )}^{R/A} ( \bar{\omega} ) = \frac{1}{ \bar{\omega} - \xi_{\vec{k} , \lambda} \pm  \frac{i}{ 2 \tau_{\rm e}} }
\:\:\:\: , 
\label{lcx.2}
\eneq
\noindent
and $f ( \bar{\omega} )$ being the Fermi distribution function.

The contribution from vertex corrections, $\sigma_{xy;B}^z$, takes a form similar to $\sigma_{s;B}$ in 
Eq. (\ref{lc.5}), provided the operators entering the corresponding equation are pertinently replaced. 
As a result, we obtain

 \begin{eqnarray}
 \sigma_{xy;B}^z 
 &=&  \frac{   1   }{ 2 \pi V} \: \sum_{\vec{k}} \: \sum_{ \lambda , \lambda'}   [\delta j_{{\rm sp} , x}^z]_{(\vec{k} , \lambda) ; (\vec{k} , 
\lambda' )} \: [ j_{{\rm ch},y} ]_{(\vec{k} , \lambda' ) ; (\vec{k} , \lambda) } \times \nonumber \\
&&  \left\{ \frac{ 1 }{ \xi_{\vec{k} , \lambda} + i( 2 \tau_{\rm e})^{-1} } 
\:  \frac{ 1 }{ \xi_{\vec{k} , \lambda'} -i ( 2 \tau_{\rm e})^{-1} } \right\}
\;\;\;\; ,
\label{lcx.3}
\end{eqnarray}
\noindent
with the vertex correction $\delta j_{{\rm sp} , x}^z$ computed according to Eqs. (\ref{resp.x5})-(\ref{resp.x6b}) 
and the self-consistent expression for $\tau_{\rm e}$ implemented as discussed above. 

Eqsations (\ref{lc.4bis}) and (\ref{lc.5}) and (\ref{lcx.1}) and (\ref{lcx.2}) provide  the expressions we used in the main 
text to draw the plots of the sheet conductance and of the spin-Hall conductance, in the clean limit, as well as 
in the presence of impurities.

\begin{figure}
 \center
\includegraphics*[width=0.9 \linewidth]{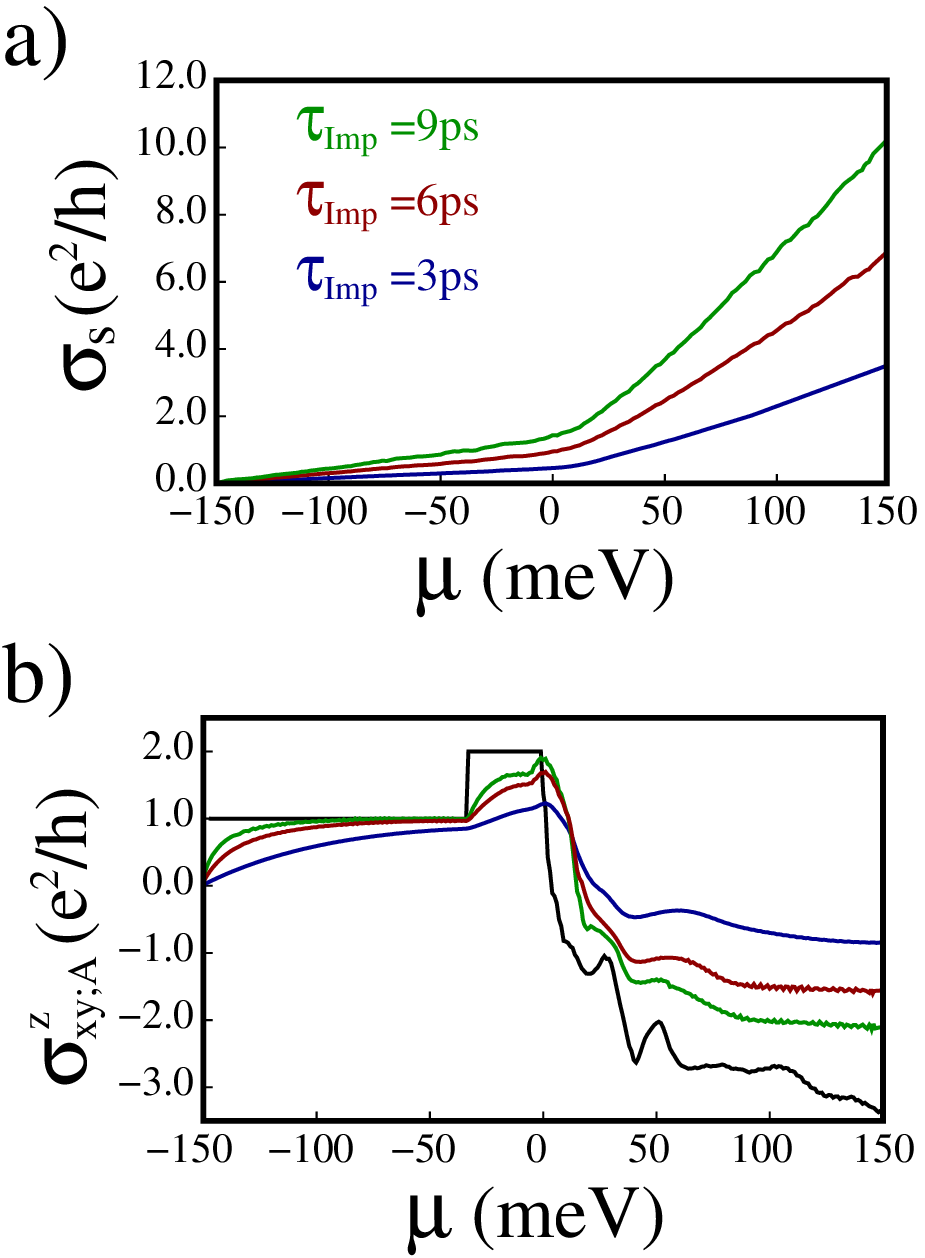}
\caption{{\bf (a):} $\sigma_s$ as a function of $\mu$ computed with the parameters listed in Table \ref{8bandT},
and with $\tau_{\rm Imp}=3\:{\rm ps}$, $\tau_{\rm Imp}=6\:{\rm ps}$ (red curve), and $\tau_{\rm Imp}=9\:{\rm ps}$ (green curve).
{\bf (b):}$\sigma_{xy;A}^{z}$  as a function of $\mu$ computed with the parameters listed in Table \ref{8bandT},
and with $\tau_{\rm Imp}^{-1}=0$ (black curve), 
$\tau_{\rm Imp}=3\:{\rm ps}$, $\tau_{\rm Imp}=6\;{\rm ps}$ (red curve), and $\tau_{\rm Imp}=9\;{\rm ps}$ (green curve).
} 
\label{various_taus}
\end{figure}

Before concluding this appendix, we briefly investigate how changing the density of impurities and/or
the impurity interaction potential affects $\sigma_s$ and $\sigma_{xy}^z$. To do so, 
  it is worth pointing out that, throughout our derivation, 
we lump all the details about the density of impurity and the impurity interaction into 
the parameter $\tau_{\rm Imp}$, which we regard as a phenomenological parameter, 
fitted from the data of Ref.\cite{Trier2020}. Therefore,  
we analyze the effects of changing the   density of impurities and/or
the impurity interaction potential by  simply computing the conductances as functions of 
$\mu$  for different
values of  $\tau_{\rm Imp}$. In Fig. \ref{various_taus}{\bf (a)} we plot the corresponding 
results for $\sigma_s$ as a function of $\mu$ computed using the parameters
listed in Table \ref{8bandT}, for $\tau_{\rm Imp}=3,\:6$ and 9 ps. As expected, at a given 
$\mu$, $\sigma_s$ increases monotonically, as a function of $\tau_{\rm Imp}$, with no 
particular additional features in the main behavior (for $\tau_{\rm Imp}=6$ and 9 ps), 
compared to the case $\tau_{\rm Imp}=3$ ps, corresponding to the value estimated from the data
of Ref. \cite{Trier2020}.   In Fig. \ref{various_taus}{\bf (b)} we plot 
 $\sigma_{xy;A}^{z}$ as a function of $\mu$ for the same values of $\tau_{\rm Imp}$ as
 we used for $\sigma_s$ and, in addition, for $\tau_{\rm Imp}^{-1}=0$.  The over-all main 
 trend shows how, varying $\tau_{\rm Imp}$, the plot in the clean limit
 evolves into the one at finite $\tau_{\rm Imp}$, with no particular features
 emerging at values in between $\tau_{\rm Imp}^{-1}=0$ (the clean limit) and
 the experimentally estimated value  $\tau_{\rm Imp}=3\:{\rm ps}$.

\noindent

\bibliography{transport}

\begin{thebibliography}{52}%
\makeatletter
\providecommand \@ifxundefined [1]{%
 \@ifx{#1\undefined}
}%
\providecommand \@ifnum [1]{%
 \ifnum #1\expandafter \@firstoftwo
 \else \expandafter \@secondoftwo
 \fi
}%
\providecommand \@ifx [1]{%
 \ifx #1\expandafter \@firstoftwo
 \else \expandafter \@secondoftwo
 \fi
}%
\providecommand \natexlab [1]{#1}%
\providecommand \enquote  [1]{``#1''}%
\providecommand \bibnamefont  [1]{#1}%
\providecommand \bibfnamefont [1]{#1}%
\providecommand \citenamefont [1]{#1}%
\providecommand \href@noop [0]{\@secondoftwo}%
\providecommand \href [0]{\begingroup \@sanitize@url \@href}%
\providecommand \@href[1]{\@@startlink{#1}\@@href}%
\providecommand \@@href[1]{\endgroup#1\@@endlink}%
\providecommand \@sanitize@url [0]{\catcode `\\12\catcode `\$12\catcode
  `\&12\catcode `\#12\catcode `\^12\catcode `\_12\catcode `\%12\relax}%
\providecommand \@@startlink[1]{}%
\providecommand \@@endlink[0]{}%
\providecommand \url  [0]{\begingroup\@sanitize@url \@url }%
\providecommand \@url [1]{\endgroup\@href {#1}{\urlprefix }}%
\providecommand \urlprefix  [0]{URL }%
\providecommand \Eprint [0]{\href }%
\providecommand \doibase [0]{http://dx.doi.org/}%
\providecommand \selectlanguage [0]{\@gobble}%
\providecommand \bibinfo  [0]{\@secondoftwo}%
\providecommand \bibfield  [0]{\@secondoftwo}%
\providecommand \translation [1]{[#1]}%
\providecommand \BibitemOpen [0]{}%
\providecommand \bibitemStop [0]{}%
\providecommand \bibitemNoStop [0]{.\EOS\space}%
\providecommand \EOS [0]{\spacefactor3000\relax}%
\providecommand \BibitemShut  [1]{\csname bibitem#1\endcsname}%
\let\auto@bib@innerbib\@empty
\bibitem [{\citenamefont {Wolf}\ \emph {et~al.}(2001)\citenamefont {Wolf},
  \citenamefont {Awschalom}, \citenamefont {Buhrman}, \citenamefont {Daughton},
  \citenamefont {von Molnár}, \citenamefont {Roukes}, \citenamefont
  {Chtchelkanova},\ and\ \citenamefont {Treger}}]{Wolf2001}%
  \BibitemOpen
  \bibfield  {author} {\bibinfo {author} {\bibfnamefont {S.~A.}\ \bibnamefont
  {Wolf}}, \bibinfo {author} {\bibfnamefont {D.~D.}\ \bibnamefont {Awschalom}},
  \bibinfo {author} {\bibfnamefont {R.~A.}\ \bibnamefont {Buhrman}}, \bibinfo
  {author} {\bibfnamefont {J.~M.}\ \bibnamefont {Daughton}}, \bibinfo {author}
  {\bibfnamefont {S.}~\bibnamefont {von Molnár}}, \bibinfo {author}
  {\bibfnamefont {M.~L.}\ \bibnamefont {Roukes}}, \bibinfo {author}
  {\bibfnamefont {A.~Y.}\ \bibnamefont {Chtchelkanova}}, \ and\ \bibinfo
  {author} {\bibfnamefont {D.~M.}\ \bibnamefont {Treger}},\ }\href {\doibase
  10.1126/science.1065389} {\bibfield  {journal} {\bibinfo  {journal}
  {Science}\ }\textbf {\bibinfo {volume} {294}},\ \bibinfo {pages} {1488}
  (\bibinfo {year} {2001})}\BibitemShut {NoStop}%
\bibitem [{\citenamefont {Vaz}\ \emph {et~al.}(2019)\citenamefont {Vaz},
  \citenamefont {No{\"e}l}, \citenamefont {Johansson}, \citenamefont
  {G{\"o}bel}, \citenamefont {Bruno}, \citenamefont {Singh}, \citenamefont
  {McKeown-Walker}, \citenamefont {Trier}, \citenamefont {Vicente-Arche},
  \citenamefont {Sander}, \citenamefont {Valencia}, \citenamefont {Bruneel},
  \citenamefont {Vivek}, \citenamefont {Gabay}, \citenamefont {Bergeal},
  \citenamefont {Baumberger}, \citenamefont {Okuno}, \citenamefont
  {Barth{\'e}l{\'e}my}, \citenamefont {Fert}, \citenamefont {Vila},
  \citenamefont {Mertig}, \citenamefont {Attan{\'e}},\ and\ \citenamefont
  {Bibes}}]{Vaz2019}%
  \BibitemOpen
  \bibfield  {author} {\bibinfo {author} {\bibfnamefont {D.~C.}\ \bibnamefont
  {Vaz}}, \bibinfo {author} {\bibfnamefont {P.}~\bibnamefont {No{\"e}l}},
  \bibinfo {author} {\bibfnamefont {A.}~\bibnamefont {Johansson}}, \bibinfo
  {author} {\bibfnamefont {B.}~\bibnamefont {G{\"o}bel}}, \bibinfo {author}
  {\bibfnamefont {F.~Y.}\ \bibnamefont {Bruno}}, \bibinfo {author}
  {\bibfnamefont {G.}~\bibnamefont {Singh}}, \bibinfo {author} {\bibfnamefont
  {S.}~\bibnamefont {McKeown-Walker}}, \bibinfo {author} {\bibfnamefont
  {F.}~\bibnamefont {Trier}}, \bibinfo {author} {\bibfnamefont {L.~M.}\
  \bibnamefont {Vicente-Arche}}, \bibinfo {author} {\bibfnamefont
  {A.}~\bibnamefont {Sander}}, \bibinfo {author} {\bibfnamefont
  {S.}~\bibnamefont {Valencia}}, \bibinfo {author} {\bibfnamefont
  {P.}~\bibnamefont {Bruneel}}, \bibinfo {author} {\bibfnamefont
  {M.}~\bibnamefont {Vivek}}, \bibinfo {author} {\bibfnamefont
  {M.}~\bibnamefont {Gabay}}, \bibinfo {author} {\bibfnamefont
  {N.}~\bibnamefont {Bergeal}}, \bibinfo {author} {\bibfnamefont
  {F.}~\bibnamefont {Baumberger}}, \bibinfo {author} {\bibfnamefont
  {H.}~\bibnamefont {Okuno}}, \bibinfo {author} {\bibfnamefont
  {A.}~\bibnamefont {Barth{\'e}l{\'e}my}}, \bibinfo {author} {\bibfnamefont
  {A.}~\bibnamefont {Fert}}, \bibinfo {author} {\bibfnamefont {L.}~\bibnamefont
  {Vila}}, \bibinfo {author} {\bibfnamefont {I.}~\bibnamefont {Mertig}},
  \bibinfo {author} {\bibfnamefont {J.-P.}\ \bibnamefont {Attan{\'e}}}, \ and\
  \bibinfo {author} {\bibfnamefont {M.}~\bibnamefont {Bibes}},\ }\href
  {\doibase 10.1038/s41563-019-0467-4} {\bibfield  {journal} {\bibinfo
  {journal} {Nature Materials}\ }\textbf {\bibinfo {volume} {18}},\ \bibinfo
  {pages} {1187} (\bibinfo {year} {2019})}\BibitemShut {NoStop}%
\bibitem [{\citenamefont {Hirohata}\ \emph {et~al.}(2020)\citenamefont
  {Hirohata}, \citenamefont {Yamada}, \citenamefont {Nakatani}, \citenamefont
  {Prejbeanu}, \citenamefont {Diény}, \citenamefont {Pirro},\ and\
  \citenamefont {Hillebrands}}]{Hirohata2020}%
  \BibitemOpen
  \bibfield  {author} {\bibinfo {author} {\bibfnamefont {A.}~\bibnamefont
  {Hirohata}}, \bibinfo {author} {\bibfnamefont {K.}~\bibnamefont {Yamada}},
  \bibinfo {author} {\bibfnamefont {Y.}~\bibnamefont {Nakatani}}, \bibinfo
  {author} {\bibfnamefont {I.-L.}\ \bibnamefont {Prejbeanu}}, \bibinfo {author}
  {\bibfnamefont {B.}~\bibnamefont {Diény}}, \bibinfo {author} {\bibfnamefont
  {P.}~\bibnamefont {Pirro}}, \ and\ \bibinfo {author} {\bibfnamefont
  {B.}~\bibnamefont {Hillebrands}},\ }\href {\doibase
  https://doi.org/10.1016/j.jmmm.2020.166711} {\bibfield  {journal} {\bibinfo
  {journal} {Journal of Magnetism and Magnetic Materials}\ }\textbf {\bibinfo
  {volume} {509}},\ \bibinfo {pages} {166711} (\bibinfo {year}
  {2020})}\BibitemShut {NoStop}%
\bibitem [{\citenamefont {Chauleau}\ \emph {et~al.}(2016)\citenamefont
  {Chauleau}, \citenamefont {Boselli}, \citenamefont {Gariglio}, \citenamefont
  {Weil}, \citenamefont {de~Loubens}, \citenamefont {Triscone},\ and\
  \citenamefont {Viret}}]{Chauleau2016}%
  \BibitemOpen
  \bibfield  {author} {\bibinfo {author} {\bibfnamefont {J.-Y.}\ \bibnamefont
  {Chauleau}}, \bibinfo {author} {\bibfnamefont {M.}~\bibnamefont {Boselli}},
  \bibinfo {author} {\bibfnamefont {S.}~\bibnamefont {Gariglio}}, \bibinfo
  {author} {\bibfnamefont {R.}~\bibnamefont {Weil}}, \bibinfo {author}
  {\bibfnamefont {G.}~\bibnamefont {de~Loubens}}, \bibinfo {author}
  {\bibfnamefont {J.-M.}\ \bibnamefont {Triscone}}, \ and\ \bibinfo {author}
  {\bibfnamefont {M.}~\bibnamefont {Viret}},\ }\href {\doibase
  10.1209/0295-5075/116/17006} {\bibfield  {journal} {\bibinfo  {journal}
  {Europhysics Letters}\ }\textbf {\bibinfo {volume} {116}},\ \bibinfo {pages}
  {17006} (\bibinfo {year} {2016})}\BibitemShut {NoStop}%
\bibitem [{\citenamefont {Song}\ \emph {et~al.}(2017)\citenamefont {Song},
  \citenamefont {Zhang}, \citenamefont {Su}, \citenamefont {Yuan},
  \citenamefont {Chen}, \citenamefont {Xing}, \citenamefont {Shi},
  \citenamefont {Sun},\ and\ \citenamefont {Han}}]{Qi2017}%
  \BibitemOpen
  \bibfield  {author} {\bibinfo {author} {\bibfnamefont {Q.}~\bibnamefont
  {Song}}, \bibinfo {author} {\bibfnamefont {H.}~\bibnamefont {Zhang}},
  \bibinfo {author} {\bibfnamefont {T.}~\bibnamefont {Su}}, \bibinfo {author}
  {\bibfnamefont {W.}~\bibnamefont {Yuan}}, \bibinfo {author} {\bibfnamefont
  {Y.}~\bibnamefont {Chen}}, \bibinfo {author} {\bibfnamefont {W.}~\bibnamefont
  {Xing}}, \bibinfo {author} {\bibfnamefont {J.}~\bibnamefont {Shi}}, \bibinfo
  {author} {\bibfnamefont {J.}~\bibnamefont {Sun}}, \ and\ \bibinfo {author}
  {\bibfnamefont {W.}~\bibnamefont {Han}},\ }\href {\doibase
  10.1126/sciadv.1602312} {\bibfield  {journal} {\bibinfo  {journal} {Science
  Advances}\ }\textbf {\bibinfo {volume} {3}},\ \bibinfo {pages} {e1602312}
  (\bibinfo {year} {2017})},\ \Eprint
  {http://arxiv.org/abs/https://www.science.org/doi/pdf/10.1126/sciadv.1602312}
  {https://www.science.org/doi/pdf/10.1126/sciadv.1602312} \BibitemShut
  {NoStop}%
\bibitem [{\citenamefont {Telesio}\ \emph {et~al.}(2018)\citenamefont
  {Telesio}, \citenamefont {Moroni}, \citenamefont {Pallecchi}, \citenamefont
  {Marré}, \citenamefont {Vinai}, \citenamefont {Panaccione}, \citenamefont
  {Torelli}, \citenamefont {Rusponi}, \citenamefont {Piamonteze}, \citenamefont
  {di~Gennaro}, \citenamefont {Khare}, \citenamefont {Granozio},\ and\
  \citenamefont {Filippetti}}]{Telesio2018}%
  \BibitemOpen
  \bibfield  {author} {\bibinfo {author} {\bibfnamefont {F.}~\bibnamefont
  {Telesio}}, \bibinfo {author} {\bibfnamefont {R.}~\bibnamefont {Moroni}},
  \bibinfo {author} {\bibfnamefont {I.}~\bibnamefont {Pallecchi}}, \bibinfo
  {author} {\bibfnamefont {D.}~\bibnamefont {Marré}}, \bibinfo {author}
  {\bibfnamefont {G.}~\bibnamefont {Vinai}}, \bibinfo {author} {\bibfnamefont
  {G.}~\bibnamefont {Panaccione}}, \bibinfo {author} {\bibfnamefont
  {P.}~\bibnamefont {Torelli}}, \bibinfo {author} {\bibfnamefont
  {S.}~\bibnamefont {Rusponi}}, \bibinfo {author} {\bibfnamefont
  {C.}~\bibnamefont {Piamonteze}}, \bibinfo {author} {\bibfnamefont
  {E.}~\bibnamefont {di~Gennaro}}, \bibinfo {author} {\bibfnamefont
  {A.}~\bibnamefont {Khare}}, \bibinfo {author} {\bibfnamefont {F.~M.}\
  \bibnamefont {Granozio}}, \ and\ \bibinfo {author} {\bibfnamefont
  {A.}~\bibnamefont {Filippetti}},\ }\href {\doibase 10.1088/2399-6528/aaa943}
  {\bibfield  {journal} {\bibinfo  {journal} {Journal of Physics
  Communications}\ }\textbf {\bibinfo {volume} {2}},\ \bibinfo {pages} {025010}
  (\bibinfo {year} {2018})}\BibitemShut {NoStop}%
\bibitem [{\citenamefont {Edelstein}(1990)}]{Edelstein1990}%
  \BibitemOpen
  \bibfield  {author} {\bibinfo {author} {\bibfnamefont {V.}~\bibnamefont
  {Edelstein}},\ }\href {\doibase https://doi.org/10.1016/0038-1098(90)90963-C}
  {\bibfield  {journal} {\bibinfo  {journal} {Solid State Communications}\
  }\textbf {\bibinfo {volume} {73}},\ \bibinfo {pages} {233} (\bibinfo {year}
  {1990})}\BibitemShut {NoStop}%
\bibitem [{\citenamefont {Dyakonov}\ and\ \citenamefont
  {Perel}(1971)}]{Dyakonov1971}%
  \BibitemOpen
  \bibfield  {author} {\bibinfo {author} {\bibfnamefont {M.}~\bibnamefont
  {Dyakonov}}\ and\ \bibinfo {author} {\bibfnamefont {V.}~\bibnamefont
  {Perel}},\ }\href {\doibase https://doi.org/10.1016/0375-9601(71)90196-4}
  {\bibfield  {journal} {\bibinfo  {journal} {Physics Letters A}\ }\textbf
  {\bibinfo {volume} {35}},\ \bibinfo {pages} {459} (\bibinfo {year}
  {1971})}\BibitemShut {NoStop}%
\bibitem [{\citenamefont {Sinova}\ \emph {et~al.}(2015)\citenamefont {Sinova},
  \citenamefont {Valenzuela}, \citenamefont {Wunderlich}, \citenamefont
  {Back},\ and\ \citenamefont {Jungwirth}}]{Sinova2015}%
  \BibitemOpen
  \bibfield  {author} {\bibinfo {author} {\bibfnamefont {J.}~\bibnamefont
  {Sinova}}, \bibinfo {author} {\bibfnamefont {S.~O.}\ \bibnamefont
  {Valenzuela}}, \bibinfo {author} {\bibfnamefont {J.}~\bibnamefont
  {Wunderlich}}, \bibinfo {author} {\bibfnamefont {C.~H.}\ \bibnamefont
  {Back}}, \ and\ \bibinfo {author} {\bibfnamefont {T.}~\bibnamefont
  {Jungwirth}},\ }\href {\doibase 10.1103/RevModPhys.87.1213} {\bibfield
  {journal} {\bibinfo  {journal} {Rev. Mod. Phys.}\ }\textbf {\bibinfo {volume}
  {87}},\ \bibinfo {pages} {1213} (\bibinfo {year} {2015})}\BibitemShut
  {NoStop}%
\bibitem [{\citenamefont {Qian}\ \emph {et~al.}(2014)\citenamefont {Qian},
  \citenamefont {Liu}, \citenamefont {Fu},\ and\ \citenamefont
  {Li}}]{Qian2014}%
  \BibitemOpen
  \bibfield  {author} {\bibinfo {author} {\bibfnamefont {X.}~\bibnamefont
  {Qian}}, \bibinfo {author} {\bibfnamefont {J.}~\bibnamefont {Liu}}, \bibinfo
  {author} {\bibfnamefont {L.}~\bibnamefont {Fu}}, \ and\ \bibinfo {author}
  {\bibfnamefont {J.}~\bibnamefont {Li}},\ }\href {\doibase
  10.1126/science.1256815} {\bibfield  {journal} {\bibinfo  {journal}
  {Science}\ }\textbf {\bibinfo {volume} {346}},\ \bibinfo {pages} {1344}
  (\bibinfo {year} {2014})},\ \Eprint
  {http://arxiv.org/abs/https://www.science.org/doi/pdf/10.1126/science.1256815}
  {https://www.science.org/doi/pdf/10.1126/science.1256815} \BibitemShut
  {NoStop}%
\bibitem [{\citenamefont {Nie}\ \emph {et~al.}(2015)\citenamefont {Nie},
  \citenamefont {Song}, \citenamefont {Weng},\ and\ \citenamefont
  {Fang}}]{Nie2015}%
  \BibitemOpen
  \bibfield  {author} {\bibinfo {author} {\bibfnamefont {S.~M.}\ \bibnamefont
  {Nie}}, \bibinfo {author} {\bibfnamefont {Z.}~\bibnamefont {Song}}, \bibinfo
  {author} {\bibfnamefont {H.}~\bibnamefont {Weng}}, \ and\ \bibinfo {author}
  {\bibfnamefont {Z.}~\bibnamefont {Fang}},\ }\href {\doibase
  10.1103/PhysRevB.91.235434} {\bibfield  {journal} {\bibinfo  {journal} {Phys.
  Rev. B}\ }\textbf {\bibinfo {volume} {91}},\ \bibinfo {pages} {235434}
  (\bibinfo {year} {2015})}\BibitemShut {NoStop}%
\bibitem [{\citenamefont {Safeer}\ \emph {et~al.}(2019)\citenamefont {Safeer},
  \citenamefont {Ingla-Ayn\'es}, \citenamefont {Herling}, \citenamefont
  {Garcia}, \citenamefont {Vila}, \citenamefont {Ontoso}, \citenamefont
  {Calvo}, \citenamefont {Roche}, \citenamefont {Hueso},\ and\ \citenamefont
  {Casanova}}]{Safeer2019}%
  \BibitemOpen
  \bibfield  {author} {\bibinfo {author} {\bibfnamefont {C.~K.}\ \bibnamefont
  {Safeer}}, \bibinfo {author} {\bibfnamefont {J.}~\bibnamefont
  {Ingla-Ayn\'es}}, \bibinfo {author} {\bibfnamefont {F.}~\bibnamefont
  {Herling}}, \bibinfo {author} {\bibfnamefont {J.~H.}\ \bibnamefont {Garcia}},
  \bibinfo {author} {\bibfnamefont {M.}~\bibnamefont {Vila}}, \bibinfo {author}
  {\bibfnamefont {N.}~\bibnamefont {Ontoso}}, \bibinfo {author} {\bibfnamefont
  {M.~R.}\ \bibnamefont {Calvo}}, \bibinfo {author} {\bibfnamefont
  {S.}~\bibnamefont {Roche}}, \bibinfo {author} {\bibfnamefont {L.~E.}\
  \bibnamefont {Hueso}}, \ and\ \bibinfo {author} {\bibfnamefont
  {F.}~\bibnamefont {Casanova}},\ }\href {\doibase
  10.1021/acs.nanolett.8b04368} {\bibfield  {journal} {\bibinfo  {journal}
  {Nano Letters}\ }\textbf {\bibinfo {volume} {19}},\ \bibinfo {pages} {1074}
  (\bibinfo {year} {2019})}\BibitemShut {NoStop}%
\bibitem [{\citenamefont {Benítez}\ \emph {et~al.}(2020)\citenamefont
  {Benítez}, \citenamefont {Savero~Torres}, \citenamefont {Sierra},
  \citenamefont {Timmermans}, \citenamefont {Garcia}, \citenamefont {Roche},
  \citenamefont {Costache},\ and\ \citenamefont {Valenzuela}}]{Benitez2020}%
  \BibitemOpen
  \bibfield  {author} {\bibinfo {author} {\bibfnamefont {L.~A.}\ \bibnamefont
  {Benítez}}, \bibinfo {author} {\bibfnamefont {W.}~\bibnamefont
  {Savero~Torres}}, \bibinfo {author} {\bibfnamefont {J.~F.}\ \bibnamefont
  {Sierra}}, \bibinfo {author} {\bibfnamefont {M.}~\bibnamefont {Timmermans}},
  \bibinfo {author} {\bibfnamefont {J.~H.}\ \bibnamefont {Garcia}}, \bibinfo
  {author} {\bibfnamefont {S.}~\bibnamefont {Roche}}, \bibinfo {author}
  {\bibfnamefont {M.~V.}\ \bibnamefont {Costache}}, \ and\ \bibinfo {author}
  {\bibfnamefont {S.~O.}\ \bibnamefont {Valenzuela}},\ }\href {\doibase
  10.1038/s41563-019-0575-1} {\bibfield  {journal} {\bibinfo  {journal} {Nature
  Materials}\ }\textbf {\bibinfo {volume} {19}},\ \bibinfo {pages} {170}
  (\bibinfo {year} {2020})}\BibitemShut {NoStop}%
\bibitem [{\citenamefont {Safeer}\ \emph {et~al.}(2020)\citenamefont {Safeer},
  \citenamefont {Ingla-Ayn\'es}, \citenamefont {Ontoso}, \citenamefont
  {Herling}, \citenamefont {Yan}, \citenamefont {Hueso},\ and\ \citenamefont
  {Casanova}}]{Safeer2020}%
  \BibitemOpen
  \bibfield  {author} {\bibinfo {author} {\bibfnamefont {C.~K.}\ \bibnamefont
  {Safeer}}, \bibinfo {author} {\bibfnamefont {J.}~\bibnamefont
  {Ingla-Ayn\'es}}, \bibinfo {author} {\bibfnamefont {N.}~\bibnamefont
  {Ontoso}}, \bibinfo {author} {\bibfnamefont {F.}~\bibnamefont {Herling}},
  \bibinfo {author} {\bibfnamefont {W.}~\bibnamefont {Yan}}, \bibinfo {author}
  {\bibfnamefont {L.~E.}\ \bibnamefont {Hueso}}, \ and\ \bibinfo {author}
  {\bibfnamefont {F.}~\bibnamefont {Casanova}},\ }\href {\doibase doi:
  10.1021/acs.nanolett.0c01428} {\bibfield  {journal} {\bibinfo  {journal}
  {Nano Letters}\ }\textbf {\bibinfo {volume} {20}},\ \bibinfo {pages} {4573}
  (\bibinfo {year} {2020})}\BibitemShut {NoStop}%
\bibitem [{\citenamefont {Farzaneh}\ and\ \citenamefont
  {Rakheja}(2020)}]{Farzaneh2020}%
  \BibitemOpen
  \bibfield  {author} {\bibinfo {author} {\bibfnamefont {S.~M.}\ \bibnamefont
  {Farzaneh}}\ and\ \bibinfo {author} {\bibfnamefont {S.}~\bibnamefont
  {Rakheja}},\ }\href {\doibase 10.1103/PhysRevMaterials.4.114202} {\bibfield
  {journal} {\bibinfo  {journal} {Phys. Rev. Mater.}\ }\textbf {\bibinfo
  {volume} {4}},\ \bibinfo {pages} {114202} (\bibinfo {year}
  {2020})}\BibitemShut {NoStop}%
\bibitem [{\citenamefont {Trier}\ \emph {et~al.}(2020)\citenamefont {Trier},
  \citenamefont {Vaz}, \citenamefont {Bruneel}, \citenamefont {No\"el},
  \citenamefont {Fert}, \citenamefont {Vila}, \citenamefont {Attan\'e},
  \citenamefont {Barth\'el\'emy}, \citenamefont {Gabay}, \citenamefont
  {Jaffr\`es},\ and\ \citenamefont {Bibes}}]{Trier2020}%
  \BibitemOpen
  \bibfield  {author} {\bibinfo {author} {\bibfnamefont {F.}~\bibnamefont
  {Trier}}, \bibinfo {author} {\bibfnamefont {D.~C.}\ \bibnamefont {Vaz}},
  \bibinfo {author} {\bibfnamefont {P.}~\bibnamefont {Bruneel}}, \bibinfo
  {author} {\bibfnamefont {P.}~\bibnamefont {No\"el}}, \bibinfo {author}
  {\bibfnamefont {A.}~\bibnamefont {Fert}}, \bibinfo {author} {\bibfnamefont
  {L.}~\bibnamefont {Vila}}, \bibinfo {author} {\bibfnamefont {J.-P.}\
  \bibnamefont {Attan\'e}}, \bibinfo {author} {\bibfnamefont {A.}~\bibnamefont
  {Barth\'el\'emy}}, \bibinfo {author} {\bibfnamefont {M.}~\bibnamefont
  {Gabay}}, \bibinfo {author} {\bibfnamefont {H.}~\bibnamefont {Jaffr\`es}}, \
  and\ \bibinfo {author} {\bibfnamefont {M.}~\bibnamefont {Bibes}},\ }\href
  {\doibase 10.1021/acs.nanolett.9b04079} {\bibfield  {journal} {\bibinfo
  {journal} {Nano Letters}\ }\textbf {\bibinfo {volume} {20}},\ \bibinfo
  {pages} {395} (\bibinfo {year} {2020})},\ \bibinfo {note} {pMID:
  31859513}\BibitemShut {NoStop}%
\bibitem [{\citenamefont {Sinova}\ \emph {et~al.}(2004)\citenamefont {Sinova},
  \citenamefont {Culcer}, \citenamefont {Niu}, \citenamefont {Sinitsyn},
  \citenamefont {Jungwirth},\ and\ \citenamefont {MacDonald}}]{Sinova2004}%
  \BibitemOpen
  \bibfield  {author} {\bibinfo {author} {\bibfnamefont {J.}~\bibnamefont
  {Sinova}}, \bibinfo {author} {\bibfnamefont {D.}~\bibnamefont {Culcer}},
  \bibinfo {author} {\bibfnamefont {Q.}~\bibnamefont {Niu}}, \bibinfo {author}
  {\bibfnamefont {N.~A.}\ \bibnamefont {Sinitsyn}}, \bibinfo {author}
  {\bibfnamefont {T.}~\bibnamefont {Jungwirth}}, \ and\ \bibinfo {author}
  {\bibfnamefont {A.~H.}\ \bibnamefont {MacDonald}},\ }\href {\doibase
  10.1103/PhysRevLett.92.126603} {\bibfield  {journal} {\bibinfo  {journal}
  {Phys. Rev. Lett.}\ }\textbf {\bibinfo {volume} {92}},\ \bibinfo {pages}
  {126603} (\bibinfo {year} {2004})}\BibitemShut {NoStop}%
\bibitem [{\citenamefont {King}\ \emph {et~al.}(2014)\citenamefont {King},
  \citenamefont {McKeown~Walker}, \citenamefont {Tamai}, \citenamefont {de~la
  Torre}, \citenamefont {Eknapakul}, \citenamefont {Buaphet}, \citenamefont
  {Mo}, \citenamefont {Meevasana}, \citenamefont {Bahramy},\ and\ \citenamefont
  {Baumberger}}]{King2014}%
  \BibitemOpen
  \bibfield  {author} {\bibinfo {author} {\bibfnamefont {P.~D.~C.}\
  \bibnamefont {King}}, \bibinfo {author} {\bibfnamefont {S.}~\bibnamefont
  {McKeown~Walker}}, \bibinfo {author} {\bibfnamefont {A.}~\bibnamefont
  {Tamai}}, \bibinfo {author} {\bibfnamefont {A.}~\bibnamefont {de~la Torre}},
  \bibinfo {author} {\bibfnamefont {T.}~\bibnamefont {Eknapakul}}, \bibinfo
  {author} {\bibfnamefont {P.}~\bibnamefont {Buaphet}}, \bibinfo {author}
  {\bibfnamefont {S.~K.}\ \bibnamefont {Mo}}, \bibinfo {author} {\bibfnamefont
  {W.}~\bibnamefont {Meevasana}}, \bibinfo {author} {\bibfnamefont {M.~S.}\
  \bibnamefont {Bahramy}}, \ and\ \bibinfo {author} {\bibfnamefont
  {F.}~\bibnamefont {Baumberger}},\ }\href {\doibase 10.1038/ncomms4414}
  {\bibfield  {journal} {\bibinfo  {journal} {Nature Communications}\ }\textbf
  {\bibinfo {volume} {5}},\ \bibinfo {pages} {3414} (\bibinfo {year}
  {2014})}\BibitemShut {NoStop}%
\bibitem [{\citenamefont {Ohtomo}\ and\ \citenamefont
  {Hwang}(2004)}]{Ohtomo2004}%
  \BibitemOpen
  \bibfield  {author} {\bibinfo {author} {\bibfnamefont {A.}~\bibnamefont
  {Ohtomo}}\ and\ \bibinfo {author} {\bibfnamefont {H.~Y.}\ \bibnamefont
  {Hwang}},\ }\href {\doibase 10.1038/nature02308} {\bibfield  {journal}
  {\bibinfo  {journal} {Nature}\ }\textbf {\bibinfo {volume} {427}},\ \bibinfo
  {pages} {423} (\bibinfo {year} {2004})}\BibitemShut {NoStop}%
\bibitem [{\citenamefont {Chen}\ \emph {et~al.}(2011)\citenamefont {Chen},
  \citenamefont {Pryds}, \citenamefont {Kleibeuker}, \citenamefont {Koster},
  \citenamefont {Sun}, \citenamefont {Stamate}, \citenamefont {Shen},
  \citenamefont {Rijnders},\ and\ \citenamefont {Linderoth}}]{Chen2011}%
  \BibitemOpen
  \bibfield  {author} {\bibinfo {author} {\bibfnamefont {Y.}~\bibnamefont
  {Chen}}, \bibinfo {author} {\bibfnamefont {N.}~\bibnamefont {Pryds}},
  \bibinfo {author} {\bibfnamefont {J.~E.}\ \bibnamefont {Kleibeuker}},
  \bibinfo {author} {\bibfnamefont {G.}~\bibnamefont {Koster}}, \bibinfo
  {author} {\bibfnamefont {J.}~\bibnamefont {Sun}}, \bibinfo {author}
  {\bibfnamefont {E.}~\bibnamefont {Stamate}}, \bibinfo {author} {\bibfnamefont
  {B.}~\bibnamefont {Shen}}, \bibinfo {author} {\bibfnamefont {G.}~\bibnamefont
  {Rijnders}}, \ and\ \bibinfo {author} {\bibfnamefont {S.}~\bibnamefont
  {Linderoth}},\ }\href {\doibase 10.1021/nl201821j} {\bibfield  {journal}
  {\bibinfo  {journal} {Nano Letters}\ }\textbf {\bibinfo {volume} {11}},\
  \bibinfo {pages} {3774} (\bibinfo {year} {2011})}\BibitemShut {NoStop}%
\bibitem [{\citenamefont {R\"odel}\ \emph {et~al.}(2016)\citenamefont
  {R\"odel}, \citenamefont {Fortuna}, \citenamefont {Sengupta}, \citenamefont
  {Frantzeskakis}, \citenamefont {F\`evre}, \citenamefont {Bertran},
  \citenamefont {Mercey}, \citenamefont {Matzen}, \citenamefont {Agnus},
  \citenamefont {Maroutian}, \citenamefont {Lecoeur},\ and\ \citenamefont
  {Santander-Syro}}]{Rodel2016}%
  \BibitemOpen
  \bibfield  {author} {\bibinfo {author} {\bibfnamefont {T.~C.}\ \bibnamefont
  {R\"odel}}, \bibinfo {author} {\bibfnamefont {F.}~\bibnamefont {Fortuna}},
  \bibinfo {author} {\bibfnamefont {S.}~\bibnamefont {Sengupta}}, \bibinfo
  {author} {\bibfnamefont {E.}~\bibnamefont {Frantzeskakis}}, \bibinfo {author}
  {\bibfnamefont {P.~L.}\ \bibnamefont {F\`evre}}, \bibinfo {author}
  {\bibfnamefont {F.}~\bibnamefont {Bertran}}, \bibinfo {author} {\bibfnamefont
  {B.}~\bibnamefont {Mercey}}, \bibinfo {author} {\bibfnamefont
  {S.}~\bibnamefont {Matzen}}, \bibinfo {author} {\bibfnamefont
  {G.}~\bibnamefont {Agnus}}, \bibinfo {author} {\bibfnamefont
  {T.}~\bibnamefont {Maroutian}}, \bibinfo {author} {\bibfnamefont
  {P.}~\bibnamefont {Lecoeur}}, \ and\ \bibinfo {author} {\bibfnamefont
  {A.~F.}\ \bibnamefont {Santander-Syro}},\ }\href {\doibase
  https://doi.org/10.1002/adma.201505021} {\bibfield  {journal} {\bibinfo
  {journal} {Advanced Materials}\ }\textbf {\bibinfo {volume} {28}},\ \bibinfo
  {pages} {1976} (\bibinfo {year} {2016})}\BibitemShut {NoStop}%
\bibitem [{\citenamefont {Caviglia}\ \emph {et~al.}(2008)\citenamefont
  {Caviglia}, \citenamefont {Gariglio}, \citenamefont {Reyren}, \citenamefont
  {Jaccard}, \citenamefont {Schneider}, \citenamefont {Gabay}, \citenamefont
  {Thiel}, \citenamefont {Hammerl}, \citenamefont {Mannhart},\ and\
  \citenamefont {Triscone}}]{Caviglia2008}%
  \BibitemOpen
  \bibfield  {author} {\bibinfo {author} {\bibfnamefont {A.~D.}\ \bibnamefont
  {Caviglia}}, \bibinfo {author} {\bibfnamefont {S.}~\bibnamefont {Gariglio}},
  \bibinfo {author} {\bibfnamefont {N.}~\bibnamefont {Reyren}}, \bibinfo
  {author} {\bibfnamefont {D.}~\bibnamefont {Jaccard}}, \bibinfo {author}
  {\bibfnamefont {T.}~\bibnamefont {Schneider}}, \bibinfo {author}
  {\bibfnamefont {M.}~\bibnamefont {Gabay}}, \bibinfo {author} {\bibfnamefont
  {S.}~\bibnamefont {Thiel}}, \bibinfo {author} {\bibfnamefont
  {G.}~\bibnamefont {Hammerl}}, \bibinfo {author} {\bibfnamefont
  {J.}~\bibnamefont {Mannhart}}, \ and\ \bibinfo {author} {\bibfnamefont
  {J.~M.}\ \bibnamefont {Triscone}},\ }\href {\doibase 10.1038/nature07576}
  {\bibfield  {journal} {\bibinfo  {journal} {Nature}\ }\textbf {\bibinfo
  {volume} {456}},\ \bibinfo {pages} {624} (\bibinfo {year}
  {2008})}\BibitemShut {NoStop}%
\bibitem [{\citenamefont {Caviglia}\ \emph {et~al.}(2010)\citenamefont
  {Caviglia}, \citenamefont {Gabay}, \citenamefont {Gariglio}, \citenamefont
  {Reyren}, \citenamefont {Cancellieri},\ and\ \citenamefont
  {Triscone}}]{Caviglia2010}%
  \BibitemOpen
  \bibfield  {author} {\bibinfo {author} {\bibfnamefont {A.~D.}\ \bibnamefont
  {Caviglia}}, \bibinfo {author} {\bibfnamefont {M.}~\bibnamefont {Gabay}},
  \bibinfo {author} {\bibfnamefont {S.}~\bibnamefont {Gariglio}}, \bibinfo
  {author} {\bibfnamefont {N.}~\bibnamefont {Reyren}}, \bibinfo {author}
  {\bibfnamefont {C.}~\bibnamefont {Cancellieri}}, \ and\ \bibinfo {author}
  {\bibfnamefont {J.-M.}\ \bibnamefont {Triscone}},\ }\href {\doibase
  10.1103/PhysRevLett.104.126803} {\bibfield  {journal} {\bibinfo  {journal}
  {Phys. Rev. Lett.}\ }\textbf {\bibinfo {volume} {104}},\ \bibinfo {pages}
  {126803} (\bibinfo {year} {2010})}\BibitemShut {NoStop}%
\bibitem [{\citenamefont {Trama}\ \emph {et~al.}(2021)\citenamefont {Trama},
  \citenamefont {Cataudella},\ and\ \citenamefont {Perroni}}]{Trama1}%
  \BibitemOpen
  \bibfield  {author} {\bibinfo {author} {\bibfnamefont {M.}~\bibnamefont
  {Trama}}, \bibinfo {author} {\bibfnamefont {V.}~\bibnamefont {Cataudella}}, \
  and\ \bibinfo {author} {\bibfnamefont {C.~A.}\ \bibnamefont {Perroni}},\
  }\href {\doibase 10.1103/PhysRevResearch.3.043038} {\bibfield  {journal}
  {\bibinfo  {journal} {Phys. Rev. Research}\ }\textbf {\bibinfo {volume}
  {3}},\ \bibinfo {pages} {043038} (\bibinfo {year} {2021})}\BibitemShut
  {NoStop}%
\bibitem [{\citenamefont {Lesne}\ \emph {et~al.}(2016)\citenamefont {Lesne},
  \citenamefont {Fu}, \citenamefont {Oyarzun}, \citenamefont {Rojas-S\`anchez},
  \citenamefont {Vaz}, \citenamefont {Naganuma}, \citenamefont {Sicoli},
  \citenamefont {Attan\'e}, \citenamefont {Jamet}, \citenamefont {Jacquet},
  \citenamefont {George}, \citenamefont {Barth\'el\'emy}, \citenamefont
  {Jaffr\`es}, \citenamefont {Fert}, \citenamefont {Bibes},\ and\ \citenamefont
  {Vila}}]{Lesne2016}%
  \BibitemOpen
  \bibfield  {author} {\bibinfo {author} {\bibfnamefont {E.}~\bibnamefont
  {Lesne}}, \bibinfo {author} {\bibfnamefont {Y.}~\bibnamefont {Fu}}, \bibinfo
  {author} {\bibfnamefont {S.}~\bibnamefont {Oyarzun}}, \bibinfo {author}
  {\bibfnamefont {J.~C.}\ \bibnamefont {Rojas-S\`anchez}}, \bibinfo {author}
  {\bibfnamefont {D.~C.}\ \bibnamefont {Vaz}}, \bibinfo {author} {\bibfnamefont
  {H.}~\bibnamefont {Naganuma}}, \bibinfo {author} {\bibfnamefont
  {G.}~\bibnamefont {Sicoli}}, \bibinfo {author} {\bibfnamefont {J.~P.}\
  \bibnamefont {Attan\'e}}, \bibinfo {author} {\bibfnamefont {M.}~\bibnamefont
  {Jamet}}, \bibinfo {author} {\bibfnamefont {E.}~\bibnamefont {Jacquet}},
  \bibinfo {author} {\bibfnamefont {J.~M.}\ \bibnamefont {George}}, \bibinfo
  {author} {\bibfnamefont {A.}~\bibnamefont {Barth\'el\'emy}}, \bibinfo
  {author} {\bibfnamefont {H.}~\bibnamefont {Jaffr\`es}}, \bibinfo {author}
  {\bibfnamefont {A.}~\bibnamefont {Fert}}, \bibinfo {author} {\bibfnamefont
  {M.}~\bibnamefont {Bibes}}, \ and\ \bibinfo {author} {\bibfnamefont
  {L.}~\bibnamefont {Vila}},\ }\href {\doibase 10.1038/nmat4726} {\bibfield
  {journal} {\bibinfo  {journal} {Nature Materials}\ }\textbf {\bibinfo
  {volume} {15}},\ \bibinfo {pages} {1261} (\bibinfo {year}
  {2016})}\BibitemShut {NoStop}%
\bibitem [{\citenamefont {Varignon}\ \emph {et~al.}(2018)\citenamefont
  {Varignon}, \citenamefont {Vila}, \citenamefont {Barth\'el\'emy},\ and\
  \citenamefont {Bibes}}]{Varignon2018}%
  \BibitemOpen
  \bibfield  {author} {\bibinfo {author} {\bibfnamefont {J.}~\bibnamefont
  {Varignon}}, \bibinfo {author} {\bibfnamefont {L.}~\bibnamefont {Vila}},
  \bibinfo {author} {\bibfnamefont {A.}~\bibnamefont {Barth\'el\'emy}}, \ and\
  \bibinfo {author} {\bibfnamefont {M.}~\bibnamefont {Bibes}},\ }\href
  {\doibase 10.1038/s41567-018-0112-1} {\bibfield  {journal} {\bibinfo
  {journal} {Nature Physics}\ }\textbf {\bibinfo {volume} {14}},\ \bibinfo
  {pages} {322} (\bibinfo {year} {2018})}\BibitemShut {NoStop}%
\bibitem [{\citenamefont {Abanin}\ \emph {et~al.}(2009)\citenamefont {Abanin},
  \citenamefont {Shytov}, \citenamefont {Levitov},\ and\ \citenamefont
  {Halperin}}]{Abanin2009}%
  \BibitemOpen
  \bibfield  {author} {\bibinfo {author} {\bibfnamefont {D.~A.}\ \bibnamefont
  {Abanin}}, \bibinfo {author} {\bibfnamefont {A.~V.}\ \bibnamefont {Shytov}},
  \bibinfo {author} {\bibfnamefont {L.~S.}\ \bibnamefont {Levitov}}, \ and\
  \bibinfo {author} {\bibfnamefont {B.~I.}\ \bibnamefont {Halperin}},\ }\href
  {\doibase 10.1103/PhysRevB.79.035304} {\bibfield  {journal} {\bibinfo
  {journal} {Phys. Rev. B}\ }\textbf {\bibinfo {volume} {79}},\ \bibinfo
  {pages} {035304} (\bibinfo {year} {2009})}\BibitemShut {NoStop}%
\bibitem [{\citenamefont {Mishchenko}\ \emph {et~al.}(2004)\citenamefont
  {Mishchenko}, \citenamefont {Shytov},\ and\ \citenamefont
  {Halperin}}]{Mishchenko2004}%
  \BibitemOpen
  \bibfield  {author} {\bibinfo {author} {\bibfnamefont {E.~G.}\ \bibnamefont
  {Mishchenko}}, \bibinfo {author} {\bibfnamefont {A.~V.}\ \bibnamefont
  {Shytov}}, \ and\ \bibinfo {author} {\bibfnamefont {B.~I.}\ \bibnamefont
  {Halperin}},\ }\href {\doibase 10.1103/PhysRevLett.93.226602} {\bibfield
  {journal} {\bibinfo  {journal} {Phys. Rev. Lett.}\ }\textbf {\bibinfo
  {volume} {93}},\ \bibinfo {pages} {226602} (\bibinfo {year}
  {2004})}\BibitemShut {NoStop}%
\bibitem [{\citenamefont {Inoue}\ \emph {et~al.}(2004)\citenamefont {Inoue},
  \citenamefont {Bauer},\ and\ \citenamefont {Molenkamp}}]{Inoue2004}%
  \BibitemOpen
  \bibfield  {author} {\bibinfo {author} {\bibfnamefont {J.-i.}\ \bibnamefont
  {Inoue}}, \bibinfo {author} {\bibfnamefont {G.~E.~W.}\ \bibnamefont {Bauer}},
  \ and\ \bibinfo {author} {\bibfnamefont {L.~W.}\ \bibnamefont {Molenkamp}},\
  }\href {\doibase 10.1103/PhysRevB.70.041303} {\bibfield  {journal} {\bibinfo
  {journal} {Phys. Rev. B}\ }\textbf {\bibinfo {volume} {70}},\ \bibinfo
  {pages} {041303} (\bibinfo {year} {2004})}\BibitemShut {NoStop}%
\bibitem [{\citenamefont {Khaetskii}(2006)}]{Khaetskii2006}%
  \BibitemOpen
  \bibfield  {author} {\bibinfo {author} {\bibfnamefont {A.}~\bibnamefont
  {Khaetskii}},\ }\href {\doibase 10.1103/PhysRevLett.96.056602} {\bibfield
  {journal} {\bibinfo  {journal} {Phys. Rev. Lett.}\ }\textbf {\bibinfo
  {volume} {96}},\ \bibinfo {pages} {056602} (\bibinfo {year}
  {2006})}\BibitemShut {NoStop}%
\bibitem [{\citenamefont {Raimondi}\ and\ \citenamefont
  {Schwab}(2005)}]{Raimondi2004}%
  \BibitemOpen
  \bibfield  {author} {\bibinfo {author} {\bibfnamefont {R.}~\bibnamefont
  {Raimondi}}\ and\ \bibinfo {author} {\bibfnamefont {P.}~\bibnamefont
  {Schwab}},\ }\href {\doibase 10.1103/PhysRevB.71.033311} {\bibfield
  {journal} {\bibinfo  {journal} {Phys. Rev. B}\ }\textbf {\bibinfo {volume}
  {71}},\ \bibinfo {pages} {033311} (\bibinfo {year} {2005})}\BibitemShut
  {NoStop}%
\bibitem [{\citenamefont {Dimitrova}(2005)}]{Dimitrova2005}%
  \BibitemOpen
  \bibfield  {author} {\bibinfo {author} {\bibfnamefont {O.~V.}\ \bibnamefont
  {Dimitrova}},\ }\href {\doibase 10.1103/PhysRevB.71.245327} {\bibfield
  {journal} {\bibinfo  {journal} {Phys. Rev. B}\ }\textbf {\bibinfo {volume}
  {71}},\ \bibinfo {pages} {245327} (\bibinfo {year} {2005})}\BibitemShut
  {NoStop}%
\bibitem [{\citenamefont {Perroni}\ \emph {et~al.}(2019)\citenamefont
  {Perroni}, \citenamefont {Cataudella}, \citenamefont {Salluzzo},
  \citenamefont {Cuoco},\ and\ \citenamefont {Citro}}]{Perroni2019}%
  \BibitemOpen
  \bibfield  {author} {\bibinfo {author} {\bibfnamefont {C.~A.}\ \bibnamefont
  {Perroni}}, \bibinfo {author} {\bibfnamefont {V.}~\bibnamefont {Cataudella}},
  \bibinfo {author} {\bibfnamefont {M.}~\bibnamefont {Salluzzo}}, \bibinfo
  {author} {\bibfnamefont {M.}~\bibnamefont {Cuoco}}, \ and\ \bibinfo {author}
  {\bibfnamefont {R.}~\bibnamefont {Citro}},\ }\href {\doibase
  10.1103/PhysRevB.100.094526} {\bibfield  {journal} {\bibinfo  {journal}
  {Phys. Rev. B}\ }\textbf {\bibinfo {volume} {100}},\ \bibinfo {pages}
  {094526} (\bibinfo {year} {2019})}\BibitemShut {NoStop}%
\bibitem [{\citenamefont {Bistritzer}\ \emph {et~al.}(2011)\citenamefont
  {Bistritzer}, \citenamefont {Khalsa},\ and\ \citenamefont
  {MacDonald}}]{Bistritzer2011}%
  \BibitemOpen
  \bibfield  {author} {\bibinfo {author} {\bibfnamefont {R.}~\bibnamefont
  {Bistritzer}}, \bibinfo {author} {\bibfnamefont {G.}~\bibnamefont {Khalsa}},
  \ and\ \bibinfo {author} {\bibfnamefont {A.~H.}\ \bibnamefont {MacDonald}},\
  }\href {\doibase 10.1103/PhysRevB.83.115114} {\bibfield  {journal} {\bibinfo
  {journal} {Phys. Rev. B}\ }\textbf {\bibinfo {volume} {83}},\ \bibinfo
  {pages} {115114} (\bibinfo {year} {2011})}\BibitemShut {NoStop}%
\bibitem [{\citenamefont {Michaeli}\ \emph {et~al.}(2012)\citenamefont
  {Michaeli}, \citenamefont {Potter},\ and\ \citenamefont {Lee}}]{Karen2012}%
  \BibitemOpen
  \bibfield  {author} {\bibinfo {author} {\bibfnamefont {K.}~\bibnamefont
  {Michaeli}}, \bibinfo {author} {\bibfnamefont {A.~C.}\ \bibnamefont
  {Potter}}, \ and\ \bibinfo {author} {\bibfnamefont {P.~A.}\ \bibnamefont
  {Lee}},\ }\href {\doibase 10.1103/PhysRevLett.108.117003} {\bibfield
  {journal} {\bibinfo  {journal} {Phys. Rev. Lett.}\ }\textbf {\bibinfo
  {volume} {108}},\ \bibinfo {pages} {117003} (\bibinfo {year}
  {2012})}\BibitemShut {NoStop}%
\bibitem [{\citenamefont {Khalsa}\ \emph {et~al.}(2013)\citenamefont {Khalsa},
  \citenamefont {Lee},\ and\ \citenamefont {MacDonald}}]{Khalsa2013}%
  \BibitemOpen
  \bibfield  {author} {\bibinfo {author} {\bibfnamefont {G.}~\bibnamefont
  {Khalsa}}, \bibinfo {author} {\bibfnamefont {B.}~\bibnamefont {Lee}}, \ and\
  \bibinfo {author} {\bibfnamefont {A.~H.}\ \bibnamefont {MacDonald}},\ }\href
  {\doibase 10.1103/PhysRevB.88.041302} {\bibfield  {journal} {\bibinfo
  {journal} {Phys. Rev. B}\ }\textbf {\bibinfo {volume} {88}},\ \bibinfo
  {pages} {041302} (\bibinfo {year} {2013})}\BibitemShut {NoStop}%
\bibitem [{\citenamefont {Shanavas}\ \emph {et~al.}(2014)\citenamefont
  {Shanavas}, \citenamefont {Popovi\ifmmode~\acute{c}\else \'{c}\fi{}},\ and\
  \citenamefont {Satpathy}}]{Shanavas2014}%
  \BibitemOpen
  \bibfield  {author} {\bibinfo {author} {\bibfnamefont {K.~V.}\ \bibnamefont
  {Shanavas}}, \bibinfo {author} {\bibfnamefont {Z.~S.}\ \bibnamefont
  {Popovi\ifmmode~\acute{c}\else \'{c}\fi{}}}, \ and\ \bibinfo {author}
  {\bibfnamefont {S.}~\bibnamefont {Satpathy}},\ }\href {\doibase
  10.1103/PhysRevB.90.165108} {\bibfield  {journal} {\bibinfo  {journal} {Phys.
  Rev. B}\ }\textbf {\bibinfo {volume} {90}},\ \bibinfo {pages} {165108}
  (\bibinfo {year} {2014})}\BibitemShut {NoStop}%
\bibitem [{\citenamefont {Borge}\ \emph {et~al.}(2014)\citenamefont {Borge},
  \citenamefont {Gorini}, \citenamefont {Vignale},\ and\ \citenamefont
  {Raimondi}}]{Borge2014}%
  \BibitemOpen
  \bibfield  {author} {\bibinfo {author} {\bibfnamefont {J.}~\bibnamefont
  {Borge}}, \bibinfo {author} {\bibfnamefont {C.}~\bibnamefont {Gorini}},
  \bibinfo {author} {\bibfnamefont {G.}~\bibnamefont {Vignale}}, \ and\
  \bibinfo {author} {\bibfnamefont {R.}~\bibnamefont {Raimondi}},\ }\href
  {\doibase 10.1103/PhysRevB.89.245443} {\bibfield  {journal} {\bibinfo
  {journal} {Phys. Rev. B}\ }\textbf {\bibinfo {volume} {89}},\ \bibinfo
  {pages} {245443} (\bibinfo {year} {2014})}\BibitemShut {NoStop}%
\bibitem [{\citenamefont {Rashba}(2004)}]{Rashba2004}%
  \BibitemOpen
  \bibfield  {author} {\bibinfo {author} {\bibfnamefont {E.~I.}\ \bibnamefont
  {Rashba}},\ }\href {\doibase 10.1103/PhysRevB.70.201309} {\bibfield
  {journal} {\bibinfo  {journal} {Phys. Rev. B}\ }\textbf {\bibinfo {volume}
  {70}},\ \bibinfo {pages} {201309} (\bibinfo {year} {2004})}\BibitemShut
  {NoStop}%
\bibitem [{\citenamefont {Shytov}\ \emph {et~al.}(2006)\citenamefont {Shytov},
  \citenamefont {Mishchenko}, \citenamefont {Engel},\ and\ \citenamefont
  {Halperin}}]{Shytov2006}%
  \BibitemOpen
  \bibfield  {author} {\bibinfo {author} {\bibfnamefont {A.~V.}\ \bibnamefont
  {Shytov}}, \bibinfo {author} {\bibfnamefont {E.~G.}\ \bibnamefont
  {Mishchenko}}, \bibinfo {author} {\bibfnamefont {H.-A.}\ \bibnamefont
  {Engel}}, \ and\ \bibinfo {author} {\bibfnamefont {B.~I.}\ \bibnamefont
  {Halperin}},\ }\href {\doibase 10.1103/PhysRevB.73.075316} {\bibfield
  {journal} {\bibinfo  {journal} {Phys. Rev. B}\ }\textbf {\bibinfo {volume}
  {73}},\ \bibinfo {pages} {075316} (\bibinfo {year} {2006})}\BibitemShut
  {NoStop}%
\bibitem [{\citenamefont {Vivek}\ \emph {et~al.}(2017)\citenamefont {Vivek},
  \citenamefont {Goerbig},\ and\ \citenamefont {Gabay}}]{Vivek2017}%
  \BibitemOpen
  \bibfield  {author} {\bibinfo {author} {\bibfnamefont {M.}~\bibnamefont
  {Vivek}}, \bibinfo {author} {\bibfnamefont {M.~O.}\ \bibnamefont {Goerbig}},
  \ and\ \bibinfo {author} {\bibfnamefont {M.}~\bibnamefont {Gabay}},\ }\href
  {\doibase 10.1103/PhysRevB.95.165117} {\bibfield  {journal} {\bibinfo
  {journal} {Phys. Rev. B}\ }\textbf {\bibinfo {volume} {95}},\ \bibinfo
  {pages} {165117} (\bibinfo {year} {2017})}\BibitemShut {NoStop}%
\bibitem [{\citenamefont {Huijben}\ \emph {et~al.}(2017)\citenamefont
  {Huijben}, \citenamefont {Hassink}, \citenamefont {Stehno}, \citenamefont
  {Liao}, \citenamefont {Rijnders}, \citenamefont {Brinkman},\ and\
  \citenamefont {Koster}}]{Huijben2017}%
  \BibitemOpen
  \bibfield  {author} {\bibinfo {author} {\bibfnamefont {M.}~\bibnamefont
  {Huijben}}, \bibinfo {author} {\bibfnamefont {G.~W.~J.}\ \bibnamefont
  {Hassink}}, \bibinfo {author} {\bibfnamefont {M.~P.}\ \bibnamefont {Stehno}},
  \bibinfo {author} {\bibfnamefont {Z.~L.}\ \bibnamefont {Liao}}, \bibinfo
  {author} {\bibfnamefont {G.}~\bibnamefont {Rijnders}}, \bibinfo {author}
  {\bibfnamefont {A.}~\bibnamefont {Brinkman}}, \ and\ \bibinfo {author}
  {\bibfnamefont {G.}~\bibnamefont {Koster}},\ }\href {\doibase
  10.1103/PhysRevB.96.075310} {\bibfield  {journal} {\bibinfo  {journal} {Phys.
  Rev. B}\ }\textbf {\bibinfo {volume} {96}},\ \bibinfo {pages} {075310}
  (\bibinfo {year} {2017})}\BibitemShut {NoStop}%
\bibitem [{\citenamefont {Joshua}\ \emph {et~al.}(2012)\citenamefont {Joshua},
  \citenamefont {Pecker}, \citenamefont {Ruhman}, \citenamefont {Altman},\ and\
  \citenamefont {Ilani}}]{Joshua2012}%
  \BibitemOpen
  \bibfield  {author} {\bibinfo {author} {\bibfnamefont {A.}~\bibnamefont
  {Joshua}}, \bibinfo {author} {\bibfnamefont {S.}~\bibnamefont {Pecker}},
  \bibinfo {author} {\bibfnamefont {J.}~\bibnamefont {Ruhman}}, \bibinfo
  {author} {\bibfnamefont {E.}~\bibnamefont {Altman}}, \ and\ \bibinfo {author}
  {\bibfnamefont {S.}~\bibnamefont {Ilani}},\ }\href {\doibase
  10.1038/ncomms2116} {\bibfield  {journal} {\bibinfo  {journal} {Nature
  Communications}\ }\textbf {\bibinfo {volume} {3}},\ \bibinfo {pages} {1}
  (\bibinfo {year} {2012})}\BibitemShut {NoStop}%
\bibitem [{\citenamefont {Johansson}\ \emph {et~al.}(2021)\citenamefont
  {Johansson}, \citenamefont {G\"obel}, \citenamefont {Henk}, \citenamefont
  {Bibes},\ and\ \citenamefont {Mertig}}]{Johansson2021}%
  \BibitemOpen
  \bibfield  {author} {\bibinfo {author} {\bibfnamefont {A.}~\bibnamefont
  {Johansson}}, \bibinfo {author} {\bibfnamefont {B.}~\bibnamefont {G\"obel}},
  \bibinfo {author} {\bibfnamefont {J.}~\bibnamefont {Henk}}, \bibinfo {author}
  {\bibfnamefont {M.}~\bibnamefont {Bibes}}, \ and\ \bibinfo {author}
  {\bibfnamefont {I.}~\bibnamefont {Mertig}},\ }\href {\doibase
  10.1103/PhysRevResearch.3.013275} {\bibfield  {journal} {\bibinfo  {journal}
  {Phys. Rev. Research}\ }\textbf {\bibinfo {volume} {3}},\ \bibinfo {pages}
  {013275} (\bibinfo {year} {2021})}\BibitemShut {NoStop}%
\bibitem [{\citenamefont {Nachawaty}\ \emph {et~al.}(2018)\citenamefont
  {Nachawaty}, \citenamefont {Yang}, \citenamefont {Nanot}, \citenamefont
  {Kazazis}, \citenamefont {Yakimova}, \citenamefont {Escoffier},\ and\
  \citenamefont {Jouault}}]{Nachawaty2018}%
  \BibitemOpen
  \bibfield  {author} {\bibinfo {author} {\bibfnamefont {A.}~\bibnamefont
  {Nachawaty}}, \bibinfo {author} {\bibfnamefont {M.}~\bibnamefont {Yang}},
  \bibinfo {author} {\bibfnamefont {S.}~\bibnamefont {Nanot}}, \bibinfo
  {author} {\bibfnamefont {D.}~\bibnamefont {Kazazis}}, \bibinfo {author}
  {\bibfnamefont {R.}~\bibnamefont {Yakimova}}, \bibinfo {author}
  {\bibfnamefont {W.}~\bibnamefont {Escoffier}}, \ and\ \bibinfo {author}
  {\bibfnamefont {B.}~\bibnamefont {Jouault}},\ }\href {\doibase
  10.1103/PhysRevB.98.045403} {\bibfield  {journal} {\bibinfo  {journal} {Phys.
  Rev. B}\ }\textbf {\bibinfo {volume} {98}},\ \bibinfo {pages} {045403}
  (\bibinfo {year} {2018})}\BibitemShut {NoStop}%
\bibitem [{\citenamefont {Tagliacozzo}\ \emph {et~al.}(2019)\citenamefont
  {Tagliacozzo}, \citenamefont {Campagnano}, \citenamefont {Giuliano},
  \citenamefont {Lucignano},\ and\ \citenamefont {Jouault}}]{Tagliacozzo2019}%
  \BibitemOpen
  \bibfield  {author} {\bibinfo {author} {\bibfnamefont {A.}~\bibnamefont
  {Tagliacozzo}}, \bibinfo {author} {\bibfnamefont {G.}~\bibnamefont
  {Campagnano}}, \bibinfo {author} {\bibfnamefont {D.}~\bibnamefont
  {Giuliano}}, \bibinfo {author} {\bibfnamefont {P.}~\bibnamefont {Lucignano}},
  \ and\ \bibinfo {author} {\bibfnamefont {B.}~\bibnamefont {Jouault}},\ }\href
  {\doibase 10.1103/PhysRevB.99.155417} {\bibfield  {journal} {\bibinfo
  {journal} {Phys. Rev. B}\ }\textbf {\bibinfo {volume} {99}},\ \bibinfo
  {pages} {155417} (\bibinfo {year} {2019})}\BibitemShut {NoStop}%
\bibitem [{\citenamefont {Diez}\ \emph {et~al.}(2015)\citenamefont {Diez},
  \citenamefont {Monteiro}, \citenamefont {Mattoni}, \citenamefont {Cobanera},
  \citenamefont {Hyart}, \citenamefont {Mulazimoglu}, \citenamefont {Bovenzi},
  \citenamefont {Beenakker},\ and\ \citenamefont {Caviglia}}]{caviglia_tras}%
  \BibitemOpen
  \bibfield  {author} {\bibinfo {author} {\bibfnamefont {M.}~\bibnamefont
  {Diez}}, \bibinfo {author} {\bibfnamefont {A.}~\bibnamefont {Monteiro}},
  \bibinfo {author} {\bibfnamefont {G.}~\bibnamefont {Mattoni}}, \bibinfo
  {author} {\bibfnamefont {E.}~\bibnamefont {Cobanera}}, \bibinfo {author}
  {\bibfnamefont {T.}~\bibnamefont {Hyart}}, \bibinfo {author} {\bibfnamefont
  {E.}~\bibnamefont {Mulazimoglu}}, \bibinfo {author} {\bibfnamefont
  {N.}~\bibnamefont {Bovenzi}}, \bibinfo {author} {\bibfnamefont
  {C.}~\bibnamefont {Beenakker}}, \ and\ \bibinfo {author} {\bibfnamefont
  {A.}~\bibnamefont {Caviglia}},\ }\href {\doibase
  10.1103/PhysRevLett.115.016803} {\bibfield  {journal} {\bibinfo  {journal}
  {Phys. Rev. Lett.}\ }\textbf {\bibinfo {volume} {115}},\ \bibinfo {pages}
  {016803} (\bibinfo {year} {2015})}\BibitemShut {NoStop}%
\bibitem [{\citenamefont {Trama}\ \emph {et~al.}(2022)\citenamefont {Trama},
  \citenamefont {Cataudella}, \citenamefont {Perroni}, \citenamefont {Romeo},\
  and\ \citenamefont {Citro}}]{mattia_new}%
  \BibitemOpen
  \bibfield  {author} {\bibinfo {author} {\bibfnamefont {M.}~\bibnamefont
  {Trama}}, \bibinfo {author} {\bibfnamefont {V.}~\bibnamefont {Cataudella}},
  \bibinfo {author} {\bibfnamefont {C.~A.}\ \bibnamefont {Perroni}}, \bibinfo
  {author} {\bibfnamefont {F.}~\bibnamefont {Romeo}}, \ and\ \bibinfo {author}
  {\bibfnamefont {R.}~\bibnamefont {Citro}},\ }\href {\doibase
  10.3390/nano12142494} {\bibfield  {journal} {\bibinfo  {journal}
  {Nanomaterials}\ }\textbf {\bibinfo {volume} {12}},\ \bibinfo {pages} {2494}
  (\bibinfo {year} {2022})}\BibitemShut {NoStop}%
\bibitem [{\citenamefont {Trier}\ \emph {et~al.}(2022)\citenamefont {Trier},
  \citenamefont {No{\"e}l}, \citenamefont {Kim}, \citenamefont {Attan{\'e}},
  \citenamefont {Vila},\ and\ \citenamefont {Bibes}}]{Trier2022}%
  \BibitemOpen
  \bibfield  {author} {\bibinfo {author} {\bibfnamefont {F.}~\bibnamefont
  {Trier}}, \bibinfo {author} {\bibfnamefont {P.}~\bibnamefont {No{\"e}l}},
  \bibinfo {author} {\bibfnamefont {J.-V.}\ \bibnamefont {Kim}}, \bibinfo
  {author} {\bibfnamefont {J.-P.}\ \bibnamefont {Attan{\'e}}}, \bibinfo
  {author} {\bibfnamefont {L.}~\bibnamefont {Vila}}, \ and\ \bibinfo {author}
  {\bibfnamefont {M.}~\bibnamefont {Bibes}},\ }\href {\doibase
  10.1038/s41578-021-00395-9} {\bibfield  {journal} {\bibinfo  {journal}
  {Nature Reviews Materials}\ }\textbf {\bibinfo {volume} {7}},\ \bibinfo
  {pages} {258} (\bibinfo {year} {2022})}\BibitemShut {NoStop}%
\bibitem [{\citenamefont {Schwab}\ and\ \citenamefont
  {Raimondi}(2002)}]{Schwab2002}%
  \BibitemOpen
  \bibfield  {author} {\bibinfo {author} {\bibfnamefont {P.}~\bibnamefont
  {Schwab}}\ and\ \bibinfo {author} {\bibfnamefont {R.}~\bibnamefont
  {Raimondi}},\ }\href {\doibase 10.1140/epjb/e20020054} {\bibfield  {journal}
  {\bibinfo  {journal} {The European Physical Journal B - Condensed Matter and
  Complex Systems}\ }\textbf {\bibinfo {volume} {25}},\ \bibinfo {pages} {483}
  (\bibinfo {year} {2002})}\BibitemShut {NoStop}%
\bibitem [{\citenamefont {Kane}\ and\ \citenamefont {Mele}(2005)}]{Kane2005}%
  \BibitemOpen
  \bibfield  {author} {\bibinfo {author} {\bibfnamefont {C.~L.}\ \bibnamefont
  {Kane}}\ and\ \bibinfo {author} {\bibfnamefont {E.~J.}\ \bibnamefont
  {Mele}},\ }\href {\doibase 10.1103/PhysRevLett.95.146802} {\bibfield
  {journal} {\bibinfo  {journal} {Phys. Rev. Lett.}\ }\textbf {\bibinfo
  {volume} {95}},\ \bibinfo {pages} {146802} (\bibinfo {year}
  {2005})}\BibitemShut {NoStop}%
\bibitem [{\citenamefont {Lepori}\ \emph {et~al.}(2021)\citenamefont {Lepori},
  \citenamefont {Giuliano}, \citenamefont {Nava},\ and\ \citenamefont
  {Perroni}}]{Lepori2021}%
  \BibitemOpen
  \bibfield  {author} {\bibinfo {author} {\bibfnamefont {L.}~\bibnamefont
  {Lepori}}, \bibinfo {author} {\bibfnamefont {D.}~\bibnamefont {Giuliano}},
  \bibinfo {author} {\bibfnamefont {A.}~\bibnamefont {Nava}}, \ and\ \bibinfo
  {author} {\bibfnamefont {C.~A.}\ \bibnamefont {Perroni}},\ }\href {\doibase
  10.1103/PhysRevB.104.134509} {\bibfield  {journal} {\bibinfo  {journal}
  {Phys. Rev. B}\ }\textbf {\bibinfo {volume} {104}},\ \bibinfo {pages}
  {134509} (\bibinfo {year} {2021})}\BibitemShut {NoStop}%
\end{thebibliography}%

 \end{document}